\documentclass[preprint,12pt]{elsarticle}

\usepackage{amssymb}% The amssymb package provides various useful mathematical symbols
\usepackage[nonumberlist]{glossaries-extra} % Glossary package, see glossaries.tex for definitions 
\usepackage{siunitx} % Easy formatting of SI units \SI{#}[\unit]
\usepackage{subcaption} % Allows subfigures
\usepackage{url} % Stop url links from breaking the document
\usepackage{booktabs} % toprule midrule bottomrule
\usepackage{tikz}% Tikz Image Processing (matlab2tikz)
\usepackage{pgfplots} %pgf... allows nice matlab vector plots
\usepackage{setspace} % Allows single spacing of nomenclature
\usepackage{moresize} % Reduce text sizing of optimal cleaning schedule table
\usepackage{hyperref}

\newglossary{latin}{latin}{latin}{List of Latin Variables}
\newglossary{greek}{greek}{greek}{List of Greek Variables}
\newglossary{acronyms}{acronyms}{acronyms}{List of Acronyms}
\newglossary{subscript}{subscripts}{subscripts}{List of Subscripts}

\newglossaryentry{arealoss}{type={latin},name={\ensuremath{A^\text{loss}}},description={area lost due to soiling}}
\newglossaryentry{areashade}{type={latin},name={\ensuremath{A^\text{shade}}},description={area lost due to shading}}
\newglossaryentry{areablock}{type={latin},name={\ensuremath{A^\text{block}}},description={area lost due to blocking}}
\newglossaryentry{areaoverlap}{type={latin},name={\ensuremath{A^\text{overlap}}},description={overlapped area of shading and blocking regions}}
\newglossaryentry{areatotal}{type={latin},name={\ensuremath{A^{\text{total}}}},description={total heliostat area}}
\newglossaryentry{areamirror}{type={latin},name={\ensuremath{A^{\text{mirror}}}},description={heliostat mirror area}}
\newglossaryentry{areaparticle}{type={latin},name={\ensuremath{a}},description={area loss due to a particle}}
\newglossaryentry{areafield}{type={latin},name={\ensuremath{A^\text{field}}},description={total heliostat-field area}}
\newglossaryentry{areanorm}{type={latin},name={\ensuremath{A^\text{norm}}},description={area loss at normal incidence}}
\newglossaryentry{ccl}{type={latin},name={\ensuremath{\text{C}^\text{cl}}},description={cleaning cost}}
\newglossaryentry{cdeg}{type={latin},name={\ensuremath{\text{C}^\text{deg}}},description={degradation cost}}
\newglossaryentry{costoam}{type={latin},name={\ensuremath{\text{C}^\text{o\&m}}},description={operation and maintenance costs}}
\newglossaryentry{costwf}{type={latin},name={\ensuremath{\text{C}^\text{w\&f}}},description={water and fuel costs}}
\newglossaryentry{costop}{type={latin},name={\ensuremath{\text{C}^\text{salary}}},description={operator salary}}
\newglossaryentry{costoperational}{type={latin},name={\ensuremath{\text{C}^\text{op}}},description={operation cost}}
\newglossaryentry{costtruck}{type={latin},name={\ensuremath{\text{C}^\text{truck}}},description={truck cost}}
\newglossaryentry{costmaint}{type={latin},name={\ensuremath{\text{C}^\text{maint}}},description={truck maintenance cost}}
\newglossaryentry{cleandecision}{type={latin},name={\ensuremath{\mathcal{D}}},description={cleaning action time}}
\newglossaryentry{cleanliness}{type={latin},name={\ensuremath{\zeta}},description={cleanliness}}
\newglossaryentry{depreciation}{type={latin},name={\ensuremath{b}},description={asset depreciation time}}
\newglossaryentry{geofactor}{type={latin},name={\ensuremath{g}},description={geometry factor}}
\newglossaryentry{glassthick}{type={latin},name={\ensuremath{t_{g}}},description={glass thickness}}
\newglossaryentry{hrz0}{type={latin},name={\ensuremath{h_{r}/z_{o}}},description={surface roughness ratio}}
\newglossaryentry{ndp}{type={latin},name={\ensuremath{n}},description={particle number size distribution deposition rate}}
\newglossaryentry{pdfn}{type={latin},name={\ensuremath{N}},description={cumulative particle number distribution}}
\newglossaryentry{ncleans}{type={latin},name={\ensuremath{F}},description={annual field cleanings per heliostat}}

\newglossaryentry{ntrucks}{type={latin},name={\ensuremath{M}},description={number of trucks}}
\newglossaryentry{overlaplength}{type={latin},name={\ensuremath{x_o}},description={overlap length}}
\newglossaryentry{particlediameter}{type={latin},name={\ensuremath{D}},description={particle diameter}}
\newglossaryentry{pel}{type={latin},name={\ensuremath{P}},description={price of electricity}}
\newglossaryentry{qmax}{type={latin},name={\ensuremath{Q^\text{max}}},description={maximum allowable receiver energy}}
\newglossaryentry{qloss}{type={latin},name={\ensuremath{Q^\text{loss}}},description={constant receiver thermal energy loss}}
\newglossaryentry{qrated}{type={latin},name={\ensuremath{Q^\text{rated}}},description={rated power-block thermal energy usage}}
\newglossaryentry{qtes}{type={latin},name={\ensuremath{Q^\text{tes}}},description={thermal energy storage output}}
\newglossaryentry{qrec}{type={latin},name={\ensuremath{Q^\text{rec}}},description={total receiver energy output}}
\newglossaryentry{qarray}{type={latin},name={\ensuremath{Q^\text{array}}},description={receiver array energy output}}
\newglossaryentry{qin}{type={latin},name={\ensuremath{Q^\text{field}}},description={field-array reflected energy at receiver}}
\newglossaryentry{qpb}{type={latin},name={\ensuremath{Q^\text{pb}}},description={power block thermal energy input}}
\newglossaryentry{sectorscleanedpertruck}{type={latin},name={\ensuremath{R}},description={sectors cleaned per crew}}
\newglossaryentry{storagecapacity}{type={latin},name={\ensuremath{s}},description={thermal storage charge}}
\newglossaryentry{maxstoragecapacity}{type={latin},name={\ensuremath{s^{\text{max}}}},description={maximum storage charge}}
\newglossaryentry{temperature}{type={latin},name={\ensuremath{T}},description={ambient temperature}}
\newglossaryentry{uptimeratio}{type={latin},name={\ensuremath{U}},description={uptime ratio}}
\newglossaryentry{windspeed}{type={latin},name={\ensuremath{v}},description={wind speed}}
\newglossaryentry{pbwork}{type={latin},name={\ensuremath{w}},description={work produced}}
\newglossaryentry{workclean}{type={latin},name={\ensuremath{w^\text{clean}}},description={work produced with clean field}}
\newglossaryentry{worksoil}{type={latin},name={\ensuremath{w^\text{soil}}},description={work produced with soiled field}}
\newglossaryentry{workmax}{type={latin},name={\ensuremath{w^\text{dayC}}},description={work produced with clean field and daytime cleaning}}
\newglossaryentry{pbrated}{type={latin},name={\ensuremath{w^\text{rated}}},description={rated power block work}}
\newglossaryentry{deltat}{type={greek},name={\ensuremath{\Delta t}},description={time between epochs}}
\newglossaryentry{deltatd}{type={greek},name={\ensuremath{\Delta t_{d}}},description={sampling periods in a day}}
\newglossaryentry{deltak}{type={greek},name={\ensuremath{\Delta k}},description={days between successive field cleanings}}
\newglossaryentry{eccentricanomaly}{type={greek},name={\ensuremath{\bar{t}}},description={eccentric anomaly}}
\newglossaryentry{etapb}{type={greek},name={\ensuremath{\eta_\text{pb}}},description={power block efficiency}}
\newglossaryentry{etacl}{type={greek},name={\ensuremath{\eta_\text{cl}}},description={cleaning efficiency}}
\newglossaryentry{opticalefficiency}{type={greek},name={\ensuremath{\eta^\text{opt}}},description={optical efficiency}}
\newglossaryentry{incidenceangle}{type={greek},name={\ensuremath{\theta}},description={incidence angle}}
\newglossaryentry{tiltangle}{type={greek},name={\ensuremath{\phi}},description={tilt angle}}
\newglossaryentry{refractedangle}{type={greek},name={\ensuremath{\psi}},description={refracted angle}}
\newglossaryentry{measuredreflectance}{type={greek},name={\ensuremath{\rho}},description={measured reflectance}}
\newglossaryentry{predictedreflectance}{type={greek},name={\ensuremath{\hat\rho}},description={reflectance prediction}}

\newglossaryentry{idtime}{type={subscript},name={\ensuremath{k}},description={time index}}
\newglossaryentry{ndays}{type={subscript},name={\ensuremath{K}},description={total number of time epochs}}
\newglossaryentry{idreceiver}{type={subscript},name={\ensuremath{\ell}},description={target receiver index}}
\newglossaryentry{nreceivers}{type={subscript},name={\ensuremath{L}},description={number of receivers}}
\newglossaryentry{idsect}{type={subscript},name={\ensuremath{j}},description={heliostat sector index}}
\newglossaryentry{nsects}{type={subscript},name={\ensuremath{J}},description={total number of heliostat sectors}}

\newglossaryentry{arma}{type={acronyms},name={ARMA},description={autoregressive-moving-average model}}
\newglossaryentry{ar}{type={acronyms},name={AR},description={autoregressive}}
\newglossaryentry{ann}{type={acronyms},name={ANN},description={artificial neural network}}
\newglossaryentry{carf}{type={acronyms},name={CARF},description={Central Analytical Research Facility}}
\newglossaryentry{csp}{type={acronyms},name={CSP},description={concentrated solar power}}
\newglossaryentry{dni}{type={acronyms},name={\text{DNI}},description={direct normal irradiance}}
\newglossaryentry{dsm}{type={acronyms},name={DSM},description={dry-deposition soiling model}}
\newglossaryentry{fftbcs}{type={acronyms},name={FFTBCS},description={fixed-frequency-time-based cleaning schedule}}
\newglossaryentry{mse}{type={acronyms},name={MSE},description={mean squared error}}
\newglossaryentry{milp}{type={acronyms},name={MILP},description={mixed integer linear programming}}
\newglossaryentry{nwqhpp}{type={acronyms},name={NWQHPP},description={North-West Queensland Hybrid Power Plant}}
\newglossaryentry{oam}{type={acronyms},name={O\&M},description={operation and maintenance}}
\newglossaryentry{pv}{type={acronyms},name={PV},description={photovoltaic}}
\newglossaryentry{tcc}{type={acronyms},name={\text{TCC}},description={total cleaning cost}}
\newglossaryentry{tes}{type={acronyms},name={TES},description={thermal energy storage}}
\newglossaryentry{tracs}{type={acronyms},name={TraCs},description={Tracking Cleanliness Sensor}}
\newglossaryentry{tsp}{type={acronyms},name={TSP},description={total suspended particles}}
\newglossaryentry{soilingrate}{type={acronyms},name={SR},description={soiling rate}}
\newglossaryentry{pm10}{type={acronyms},name={\ensuremath{\text{PM}}},description={particulate matter concentration below \SI{10}{\micro\metre}}}
\newglossaryentry{qut}{type={acronyms},name={QUT},description={Queensland University of Technology}}
\newglossaryentry{xrd}{type={acronyms},name={XRD},description={X\-ray diffraction}}
\newglossaryentry{sza}{type={acronyms},name={\text{SZA}},description={solar zenith angle}}

\makenoidxglossaries
\setglossarystyle{list}
\glsdisablehyper % Disables hyeprreferencing for glossary entries
\DeclareSIUnit\year{yr} % Define year as yr in SIunitx package
\pgfplotsset{compat=newest}
\pgfplotsset{plot coordinates/math parser=false}
\newlength\fwidth
\newlength\fheight
\def\orcid#1{\kern .08em\href{https://orcid.org/#1}{\includegraphics[keepaspectratio,width=0.7em]{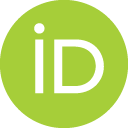}}}

\renewcommand{\glossarysection}[2][]{} % Remove bookmark titles of glossaries
\newcommand{\ta}[1]{\glsname{#1}} % Shortcut for glossary symbols
\newcommand{\als}[1]{\glsdesc{#1}, \glsname{#1}} % shortcut for creating [acronym long], [acronym short] style
\newcommand{\Als}[1]{\Glsdesc{#1}, \Glsname{#1}}
\journal{Applied Energy}
\begin{document}

\begin{frontmatter}
\title{Heliostat-field soiling predictions and cleaning resource optimization for solar tower plants}

\author[QUT,PoliMi]{Cody B. Anderson
                    \corref{correspondingAuthor}} % Corresponding author star symbol
                    \cortext[correspondingAuthor]{Corresponding author Cody B. Anderson \orcid{0000-0002-0153-3883}} % Footnote header
                    \ead{cb.anderson@qut.edu.au} % Corresponding author email address
\author[QUT]{Giovanni Picotti}
\author[QUT]{Michael E. Cholette}
\author[Vast]{Bruce Leslie}
\author[QUT]{Theodore A. Steinberg}
\author[PoliMi]{Giampaolo Manzolini}

\affiliation[QUT]{organization={School of Mechanical, Medical and Process Engineering, Queensland University of Technology}, %Department and Organization
            addressline={2 George Street}, 
            city={Brisbane},
            postcode={4000}, 
            state={QLD},
            country={Australia}}
\affiliation[PoliMi]{organization={Dipartimento di Energia, Politecnico di Milano},
            addressline={Via Lambruschini, 4},
            city={Milano}, 
            postcode={20156},
            state={Lombardy},
            country={Italy}}
\affiliation[Vast]{organization={Vast},
            addressline={225 Brisbane Terrace},
            city={Ipswich},
            postcode={4300},
            state={QLD},
            country={Australia}}

%% Front Matter 
% Abstract
\begin{abstract}
One of the primary cost drivers for concentrated solar power operations and maintenance is the soiling of heliostats in the solar field. Although this loss can be largely recovered through cleaning operations, the associated costs are substantial and require careful consideration in relation to productivity gains.

This paper presents a novel methodology for characterizing soiling losses through experimental measurements. Soiling predictions were obtained by calibrating a soiling model based on field measurements from a 50 MW modular solar tower project in Mount Isa, Australia. The study found that the mean predicted soiling rate for horizontally fixed mirrors was \SI{0.12}{pp\per\day} during low dust seasons and \SI{0.22}{pp\per\day} during high seasons. Autoregressive time series models were employed to extend two years of onsite meteorological measurements to a 10-year period, enabling the prediction of heliostat-field soiling rates. A fixed-frequency cleaning heuristic was applied to optimise the cleaning resources for various operational policies by balancing direct cleaning resource costs against the expected lost production, which was computed by averaging multiple simulated soiling loss trajectories. Analysis of resource usage showed that the cost of fuel and operator salaries contributed \SI{42}{\percent} and \SI{35}{\percent} respectively towards the cleaning cost. In addition, stowing heliostats in the horizontal position at night increased daily soiling rates by \SI{114}{\percent} and the total cleaning costs by \SI{51}{\percent} relative to vertically stowed heliostat-field. Under a simplified night-time-only power production configuration, the oversized solar field effectively charged the thermal storage during the day, despite reduced mirror reflectance due to soiling. These findings suggest that the plant can maintain efficient operation even with a reduced cleaning rate. Finally, it was observed that performing cleaning operations during the day led to a \SI{7}{\percent} increase in the total cleaning cost compared to a night-time cleaning policy. This was primarily attributed to the need to park operational heliostats for cleaning.
\end{abstract}
\begin{highlights}
    \item Cleaning resources evaluated in the design phase of a modular solar tower power plant.
    \item Soiling prediction methodology allows multiple soiling trajectories to be evaluated.
    \item Parking heliostats horizontally at night reduces plant productivity.
    \item CSP night-time power generation dispatch policies can tolerate higher soiling loads.
    \item Cleaning during day reduces production for a negligible reflectance based production gain.
\end{highlights}

% Keywords
\end{frontmatter}

%% Main matter - see sections for chapter script
\section{Introduction}
\label{sec:introduction}
% Problem statement - Soiling
The transition towards renewable energy is underway globally to mitigate the effects of climate change. Throughout 2019, Oceania was the fastest growing region in the world with an \SI{18.4}{\percent} renewable capacity increase to \SI{40}{\giga\watt}~\citep{IRENA2020}. Australia has the highest average solar radiation per square metre of any continent in the world~\citep{Bahadori2013}. This potential has thus far been exploited via \gls{pv}. However, \gls{csp} remains a promising technology owing to the economic advantages of its use with \gls{tes}, which is competitive with similarly-sized \gls{pv} installations with electrochemical storage~\citep{Feldman2016}. 

One of the key challenges for \gls{csp} is the \gls{oam} costs, which can be significantly higher than competing technologies~\citep{Picotti2018c}. A unique cost driver is the accumulation of soil on heliostats, which can quickly diminish the field specular reflectance and thus the productivity of the plant \citep{Kolb2012,Berg1978,Roth1980,Heimsath2019,Picotti2021a,Zhu2022}. Changes in reflectance with time (i.e. soiling rate) can vary widely in time and between sites. For example, \citet{Hachicha2019} report a soiling rate of 2.13 percentage points per day (\SI{}{pp\per\day}) in the United Arab Emirates, and \citet{Azouzoute2020} reported losses of \SI{0.35}{pp\per\day} for a \SI{25}{\degree} fix tilt mirror in Morocco. A long term Madrid soiling campaign by \citet{Conceicao2023} measured soiling rates between \SI{0.13}{pp\per\day} to \SI{0.58}{pp\per\day} depending upon the season and long-range transports of Saharan desert dust. Moreover, the key factors influencing this variation (location, field and plant design, seasonality, etc.) are not well understood. This poor understanding of soiling rates increases productivity revenue uncertainty and operational costs. This is particularity problematic during plant design. A better understanding of site soiling characteristics early in the plant design would enable more informed decisions on cleaning policy, site suitability, and possible design modifications (e.g. over-sizing the field)~\citep{Zhu2022}. 

Predictive models for soiling rates have been the subject of a large number of studies in both the \gls{pv} and CSP literature. These models can be separated into two broad classes: regression/empirical models (e.g. \gls{ann}, \gls{arma}, linear regression)~\citep{Bouaddi2015,Conceicao2018a,Dehghan2022} and physical models, which are usually based on atmospheric deposition models~\citep{Picotti2018,Coello2019,Sengupta2020}. 

In one study by \citet{Elboujdaini2022} trained an \gls{ann} to predict soiling losses of a \gls{tracs}~\citep{Wolfertstetter2014} soiling station in Eastern Morocco over a 20 week period. Alternatively, \citet{Ballestrin2022} predicted soiling losses of a \gls{pv} system by training an \gls{arma} model using 8 months of electrical loss measurements due to soiling. The fitted \gls{arma} model was then able to predict 38 days of soiling losses with a mean relative error of \SI{3.12}{\percent} between predictions and actual measurements. The main drawbacks to these empirical models are their lack of physical interpretations of the parameters, poor portability to different sites, and the need for large amounts of data for reliable predictions. 

Other studies, have pursued physical models as \citet{Picotti2018} who developed a physical \glsfirst{dsm} and validated for a fixed tilt angle solar mirror. The model utilises meteorological measurements and the solar field design to predict the particle deposition velocity and eventually soiling rates with a relative error of \SI{14}{\percent} between the model and measurements. Similarly, \citet{Sengupta2020} adapted a particle deposition velocity and wind removal model to predict the transmittance and the dust deposition density of an outdoor \gls{pv} module over 15 days. The deviation between measured and predicted dust deposition density was found to be \SI{4.84}{\percent}. A weakness in these studies is the specular reflectance loss model --- which is either not needed (i.e. for \gls{pv} studies) or was based on a first-surface reflector (while most CSP mirrors are second-surface). 

Once the site soiling is understood, some researchers have aimed to optimize cleaning strategies for fixed field designs. For instance, \citet{Picotti2020} conducted a study comparing two time-based cleaning optimization strategies based on daily soiling rate predictions obtained from a physical soiling model \citep{Picotti2018}. The results demonstrated that simple heuristic approaches performed as effectively as more complex and theoretically optimal planning methodologies. Specifically, the simple heuristic achieved near-optimal performance, with only a marginal deviation, while requiring significantly lower computational costs.

Similarly, \citet{Wales2021} explored a similar formulation, considering a fixed monthly soiling rate for the solar field to optimize the allocation of permanent and seasonal cleaning workers over a ten-year period.

Alternatively, reflectance-based policies have been applied to the optimization of heliostat cleaning \citep{Wolfertstetter2018,Truong2017} whereby the decision to clean is triggered when a heliostat's reflectance falls below a threshold value. \citet{Truong-Ba2020} utilised a sectorial condition-based cleaning policy to optimize cleaning resources for a hypothetical \gls{csp} plant over a ten-year horizon. While all policies resulted in the same number of cleaning crews being deployed, the sectorial condition policy exhibited a cost-saving of \SI{2.08}{\percent} relative to a time-based cleaning policy.

Although some research has explored loss models and cleaning optimization, there remains a gap in the literature regarding reflectance loss in Australian locations relevant to \gls{csp}. Consequently, typical loss rates specific to these regions are not yet established. Furthermore, existing studies on cleaning optimization often assume the availability of long-term soiling loss models (e.g., loss dynamics over several months), either by treating them as constant \citep{Wales2021} or by calibrating physical models for future plant/sites. While it is clear that the constant models have serious limitations, it is not yet clear how to develop long-term loss models from limited site data --- particularity when historical data from weather stations is not sufficiently close to the site (a likely case for the remote locations typically considered \gls{csp} candidate sites). Lastly, current cleaning approaches are not adequately tailored for soiling planning during the design phase.

%Novelty Statement
% Updated geometry model/imperfect cleaning/ consideration of receiver and thermal energy storage constraints, introduction of ARMA models to extend meteorologcal database to 10 years of simulation time allowing for averaged yearly objective function values. Consideration of different operation properties
This study details a new methodology used to characterize soiling losses through experimental measurements and predict ongoing soiling losses using an extended \glsfirst{dsm}~\citep{Picotti2018} for a \SI{50}{\mega\watt} modular solar tower site in Australia. The previously developed \gls{dsm} is extended to include a second-surface loss model to better represent the mirrors that are to be used in the future plant. An experimental procedure for collecting soiling-related data for a candidate site is detailed, and data analysis techniques are developed to characterize the airborne dust and other meteorological parameters to predict particle deposition rates and their statistical variability. Cleaning resources for the future plant are optimized using a stochastic extension adapted from \citet{Picotti2020} minimizing the \textit{average} total cleaning costs (sum of productivity losses and cleaning costs). The novelty of this study extends beyond the soiling characterization and optimization methodology. For the first time, an assessment of cost implications related to operational choices is undertaken. This includes factors such as daytime cleaning, stowing positions, and constraints linked to the receiver and storage systems.

The remainder of the paper is organized as follows. Section~\ref{sec:methodology} introduces the soiling model and cleaning resource heuristic. Section~\ref{sec:results} presents the developed methodology as part of a case study to optimize the cleaning resource needs of a candidate \gls{csp} hybrid solar plant with conclusions drawn in section~\ref{sec:conclusion}. 
% % Section 2: Methodology
\section{Methodology}
\label{sec:methodology}
\begin{figure*} 
    \centering
    \includegraphics[width=0.85\textwidth,keepaspectratio]{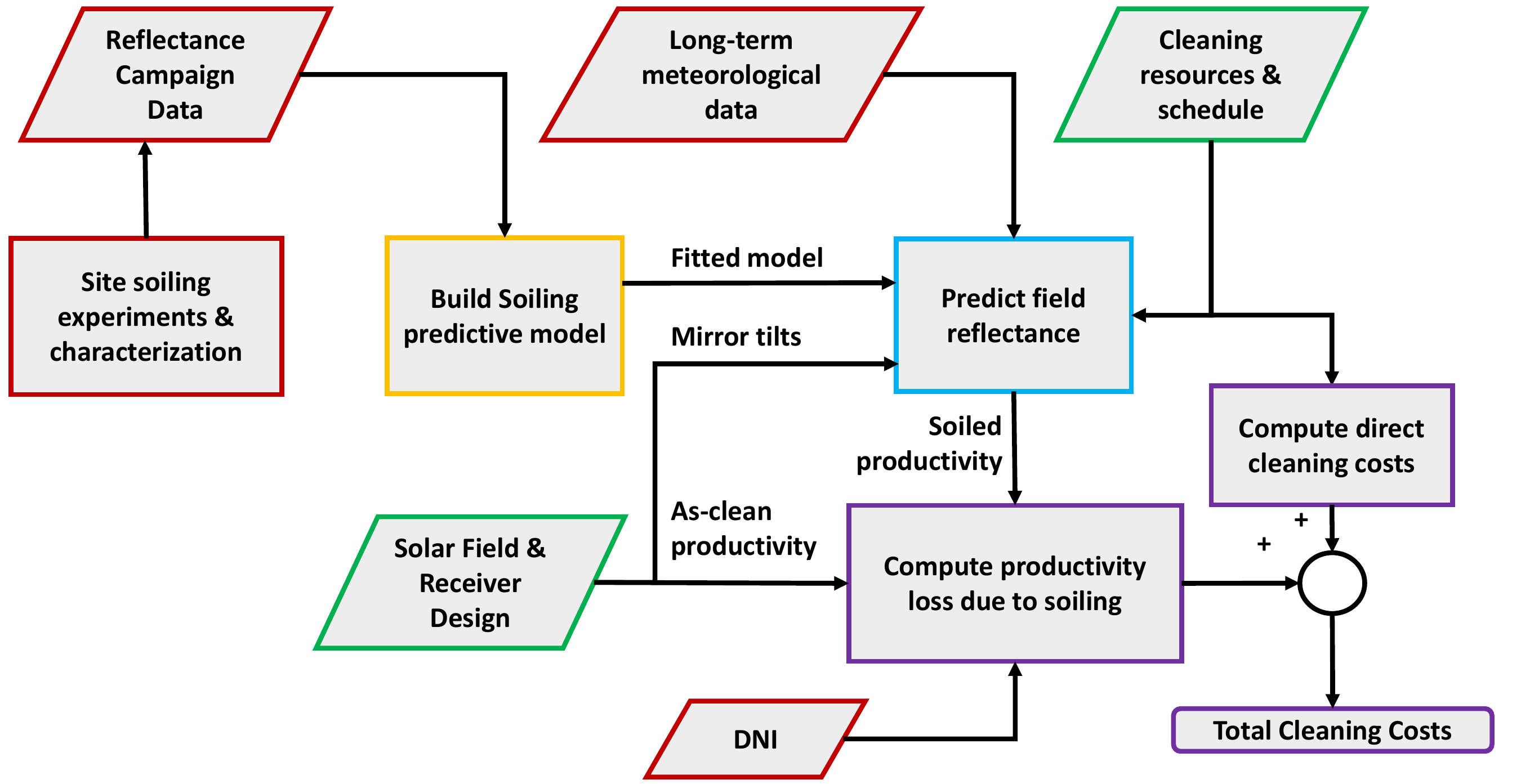}
    \caption{Overview of Cleaning Resource Optimisation Procedure}
    \label{fig:cleaningheuristic}
\end{figure*}

An overview of the soiling prediction and cleaning optimization methodology is shown in Fig.~\ref{fig:cleaningheuristic}. An optimal strategy will minimize the total cleaning cost that is calculated as the sum of the direct cleaning costs\footnote{Direct cleaning costs include salaries, truck depreciation, fuel, and water} and revenue losses due to soiling induced productivity loss. Solar field reflectance predictions are computed using a fixed cleaning schedule, corresponding to a number of deployed cleaning crews and cleaning frequency. During a cleaning event a reflectance recovery model is used to calculate the number of particles remaining depending upon the cleaning technology used. Solar field soiling rates are created using a site calibrated \glsfirst{dsm} for several simulation trajectories using ambient temperature, wind speed, and \glsfirst{tsp}. The \gls{dsm} is tuned to the \gls{csp} site through a site soiling experimental and characterization, where local soiling properties have been experimentally determined. The proposed methodology detailed below allows for the optimisation of cleaning resources under stochastic soiling trajectories early within the site selection and \gls{csp} design stage of a power plant.  

    \subsection{Soiling Predictive Model and Second Surface Geometry Factor}
        The cumulative number of particles deposited onto the surface of a heliostat is computed as
        \begin{equation}
            \label{eq:numbertotal}
            \ta{pdfn}[_{\ta{idtime}}] \left( \ta{particlediameter} \right) = \sum^{\ta{idtime}}_{i=1} \ta{ndp}_{i} \left( \ta{particlediameter} \right) \cdot \ta{deltat}
        \end{equation}
        where \ta{particlediameter} is the particle diameter (assumed spherical), \ta{deltat} is the sample period and $\ta{ndp}[_{i}]\left( \ta{particlediameter}\right)$ is the rate of deposition during the sample period, as determined from the \gls{dsm}\cite{Picotti2018c}. Typically, \ta{deltat} is the sampling period of the meteorological measurements (wind speed, airborne dust, etc.) which is constant.

        The cumulative number of particles adhered to a heliostat at time \ta{idtime} is computed as an effective reflective area loss due to soiling with:
        \begin{equation}
            \label{eq:arealoss}
            \ta{arealoss}[_{\ta{idtime}}] \left(\ta{incidenceangle}[_{\ta{idtime}}] \right) = \int^{\infty}_{0} \ta{pdfn}[_{\ta{idtime}}] (\ta{particlediameter}) \cdot \ta{areaparticle}(\ta{incidenceangle}[_{\ta{idtime}}] ,\ta{particlediameter}) \, d\ta{particlediameter}
        \end{equation}
        Where $\ta{areaparticle}\left( \ta{particlediameter}\right)$, is the area lost to specular reflectance for a particle of diameter, \ta{particlediameter} at incidence angle, \ta{incidenceangle}. This factor depends both on the nature of the reflector and the density of the particles on the surface. Assuming that particles on the surface are well separated (i.e. do not block/shade the same area)
        \begin{equation}
            \label{eq:geometryfactor}
            \ta{areaparticle} (\ta{incidenceangle},\ta{particlediameter}) = \ta{areashade} (\ta{incidenceangle},\ta{particlediameter}) + \ta{areablock} (\ta{incidenceangle},\ta{particlediameter}) - \ta{areaoverlap} (\ta{incidenceangle},\ta{particlediameter})
        \end{equation}
        The first two terms on the right-hand side are the shaded and blocked area of the reflector, which are elliptical with major axis $\frac{D}{\cos{\ta{incidenceangle}}}$ and minor axis \ta{particlediameter}, so the area can be computed as
        \begin{equation}
            \label{eq:areaobstructed}
            \ta{areashade} (\ta{incidenceangle},\ta{particlediameter}) = \ta{areablock} (\ta{incidenceangle},\ta{particlediameter}) =\frac{\pi \ta{particlediameter}^{2}}{4 \cos \ta{incidenceangle}}
        \end{equation}
        The overlap area $\ta{areaoverlap}\left( \ta{incidenceangle},\ta{particlediameter}\right)$ is the area that is both blocked and shaded by a single particle. This is derived in \ref{apx:a} calculated analytically as
        \begin{equation}
            \label{eq:areaoverlap}
            \ta{areaoverlap} (\ta{incidenceangle},\ta{particlediameter}) = \frac{ \ta{particlediameter}^{2}}{ 2 \cos \ta{incidenceangle}} \cdot \left( \ta{eccentricanomaly} - \sin \ta{eccentricanomaly} \cdot \cos \ta{eccentricanomaly} \right)
        \end{equation}
        where \ta{eccentricanomaly} is:
        \begin{equation}
            \label{eq:eccentricanomalyshort}
            \ta{eccentricanomaly} = \cos^{-1}{ \left( \frac{ \ta{particlediameter} - 2 \ta{overlaplength}\cos{\ta{incidenceangle}}}{\ta{particlediameter}} \right)}
        \end{equation}
        Where \ta{overlaplength} is the length of overlap of the blocking and shading ellipse (see Fig.~\ref{fig:shadeblockoverlaparea}). This length can be computed (also in \ref{apx:a}):
        \begin{equation}
            \label{eq:overlaplengthshort}
            \ta{overlaplength} = \frac{\ta{particlediameter}}{2} \tan{ \left( \frac{\pi}{4} - \frac{\ta{incidenceangle}}{2} \right)} - \ta{glassthick} \tan{ \left( \sin^{-1}{ \left(\frac{n_{\text{air}}}{n_{\text{glass}}}\cdot \sin{ \ta{incidenceangle}} \right)} \right)}
        \end{equation}
        where $n_\text{air}$ and $n_\text{glass}$ are the refractive indices of air and glass, respectively, and $\ta{glassthick}$ is the thickness of the glass between the particles and the reflective surface.
        \begin{figure*}[ht]
            \centering
            \begin{subfigure}[c]{0.49\textwidth}
                \resizebox{\linewidth}{!}{
\definecolor{mycolor1}{rgb}{0.00000,0.44700,0.74100}%
\definecolor{mycolor2}{rgb}{0.85000,0.32500,0.09800}%
\definecolor{mycolor3}{rgb}{0.92900,0.69400,0.12500}%
\definecolor{mycolor4}{rgb}{0.49400,0.18400,0.55600}%
\begin{tikzpicture}[%
trim axis left, trim axis right
]

\begin{axis}[%
width=0.974\fwidth,
height=\fheight,
at={(0\fwidth,0\fheight)},
scale only axis,
xmode=log,
xmin=0.001,
xmax=100.000,
xminorticks=true,
xlabel style={font=\color{white!15!black}},
xlabel={Incidence Angle, \ta{incidenceangle} [\SI{}{\degree}]},
ymin=1.000,
ymax=4.000,
ylabel style={font=\color{white!15!black}},
ylabel={Geometry Factor, \ta{geofactor} [-]},
axis background/.style={fill=white},
title style={font=\bfseries},
title={Geometry Factor with 2 mm Glass},
xmajorgrids,
xminorgrids,
ymajorgrids,
legend pos=north west,
legend style={legend cell align=left, align=left, draw=white!15!black}
]
\addplot [color=mycolor1, line width=2.0pt]
  table[row sep=crcr]{%
0.001	1.059\\
0.001	1.071\\
0.001	1.085\\
0.002	1.100\\
0.002	1.116\\
0.002	1.134\\
0.003	1.153\\
0.003	1.173\\
0.003	1.195\\
0.004	1.218\\
0.004	1.243\\
0.005	1.270\\
0.005	1.299\\
0.006	1.329\\
0.006	1.362\\
0.007	1.397\\
0.008	1.435\\
0.008	1.476\\
0.009	1.520\\
0.010	1.569\\
0.011	1.622\\
0.012	1.682\\
0.014	1.755\\
0.019	1.940\\
0.020	1.969\\
0.021	1.986\\
0.021	1.996\\
0.022	2.000\\
0.022	2.000\\
1.938	2.001\\
3.436	2.004\\
4.884	2.007\\
6.308	2.012\\
7.716	2.018\\
9.113	2.026\\
10.498	2.034\\
11.876	2.044\\
13.246	2.055\\
14.611	2.067\\
15.973	2.080\\
17.333	2.095\\
18.694	2.111\\
20.060	2.129\\
21.432	2.149\\
22.816	2.170\\
24.212	2.193\\
25.626	2.218\\
27.061	2.246\\
28.518	2.276\\
30.002	2.309\\
31.513	2.346\\
33.054	2.386\\
34.627	2.431\\
36.232	2.479\\
37.870	2.534\\
39.541	2.593\\
41.243	2.660\\
42.976	2.734\\
44.738	2.816\\
46.527	2.907\\
48.340	3.009\\
50.173	3.123\\
52.024	3.250\\
53.887	3.393\\
55.759	3.554\\
57.634	3.736\\
59.507	3.941\\
60.000	4.000\\
};
\addlegendentry{$\text{D =  1 }\mu\text{m}$}

\addplot [color=mycolor2, dotted, line width=2.0pt]
  table[row sep=crcr]{%
0.001	1.001\\
0.004	1.003\\
0.010	1.006\\
0.017	1.010\\
0.026	1.016\\
0.038	1.023\\
0.051	1.031\\
0.066	1.040\\
0.083	1.051\\
0.103	1.062\\
0.124	1.075\\
0.147	1.089\\
0.173	1.105\\
0.200	1.122\\
0.230	1.140\\
0.262	1.159\\
0.296	1.180\\
0.334	1.202\\
0.373	1.226\\
0.416	1.251\\
0.462	1.279\\
0.511	1.308\\
0.565	1.339\\
0.622	1.373\\
0.684	1.409\\
0.752	1.448\\
0.827	1.490\\
0.909	1.535\\
1.001	1.585\\
1.106	1.640\\
1.231	1.703\\
1.399	1.783\\
1.757	1.925\\
1.870	1.960\\
1.957	1.981\\
2.024	1.994\\
2.074	2.000\\
2.104	2.001\\
3.595	2.004\\
5.039	2.008\\
6.461	2.013\\
7.868	2.019\\
9.263	2.026\\
10.648	2.035\\
12.024	2.045\\
13.394	2.056\\
14.759	2.068\\
16.120	2.082\\
17.480	2.097\\
18.842	2.113\\
20.208	2.131\\
21.582	2.151\\
22.966	2.172\\
24.365	2.196\\
25.781	2.221\\
27.217	2.249\\
28.677	2.280\\
30.164	2.313\\
31.678	2.350\\
33.223	2.391\\
34.800	2.436\\
36.408	2.485\\
38.050	2.540\\
39.724	2.600\\
41.430	2.668\\
43.167	2.742\\
44.932	2.825\\
46.723	2.917\\
48.538	3.021\\
50.374	3.136\\
52.226	3.265\\
54.090	3.410\\
55.962	3.573\\
57.837	3.757\\
59.710	3.965\\
60.000	4.000\\
};
\addlegendentry{$\text{D =100 }\mu\text{m}$}

\addplot [color=mycolor3, dashed, line width=2.0pt]
  table[row sep=crcr]{%
0.001	2.000\\
1.936	2.001\\
3.435	2.004\\
4.883	2.007\\
6.307	2.012\\
7.715	2.018\\
9.111	2.026\\
10.497	2.034\\
11.875	2.044\\
13.245	2.055\\
14.610	2.067\\
15.972	2.080\\
17.332	2.095\\
18.693	2.111\\
20.058	2.129\\
21.431	2.149\\
22.814	2.170\\
24.210	2.193\\
25.625	2.218\\
27.059	2.246\\
28.516	2.276\\
29.999	2.309\\
31.511	2.346\\
33.053	2.386\\
34.625	2.430\\
36.231	2.479\\
37.869	2.534\\
39.539	2.593\\
41.241	2.660\\
42.974	2.734\\
44.736	2.816\\
46.525	2.907\\
48.338	3.009\\
50.171	3.123\\
52.021	3.250\\
53.885	3.393\\
55.756	3.554\\
57.632	3.736\\
59.505	3.941\\
60.000	4.000\\
};
\addlegendentry{$\text{2/cos(}\theta\text{)}$}

\addplot [color=mycolor4, dash dot, line width=2.0pt]
  table[row sep=crcr]{%
0.001	1.000\\
0.071	1.001\\
0.214	1.004\\
0.427	1.007\\
0.707	1.012\\
1.052	1.019\\
1.461	1.026\\
1.931	1.034\\
2.460	1.044\\
3.047	1.055\\
3.689	1.067\\
4.385	1.080\\
5.132	1.094\\
5.929	1.109\\
6.773	1.126\\
7.662	1.144\\
8.596	1.163\\
9.573	1.183\\
10.592	1.204\\
11.653	1.227\\
12.753	1.252\\
13.894	1.278\\
15.076	1.305\\
16.297	1.334\\
17.559	1.365\\
18.863	1.398\\
20.207	1.434\\
21.593	1.471\\
23.022	1.511\\
24.492	1.554\\
26.005	1.600\\
27.560	1.650\\
29.157	1.703\\
30.796	1.760\\
32.474	1.822\\
34.192	1.888\\
35.947	1.960\\
37.737	2.038\\
39.560	2.123\\
41.413	2.215\\
43.293	2.316\\
45.197	2.426\\
47.120	2.546\\
49.059	2.679\\
51.010	2.825\\
52.967	2.986\\
54.926	3.164\\
56.882	3.363\\
58.829	3.585\\
60.763	3.834\\
61.928	4.000\\
};
\addlegendentry{$\text{1+sin(}\theta\text{)/cos(}\theta\text{)}$}

\end{axis}

\begin{axis}[%
width=1.257\fwidth,
height=1.257\fheight,
at={(-0.163\fwidth,-0.163\fheight)},
scale only axis,
xmin=0.000,
xmax=1.000,
ymin=0.000,
ymax=1.000,
axis line style={draw=none},
ticks=none,
axis x line*=bottom,
axis y line*=left
]
\end{axis}
\end{tikzpicture}%
                }
            \end{subfigure}
            \begin{subfigure}[c]{0.49\textwidth}
                \resizebox{\linewidth}{!}{
                    \input{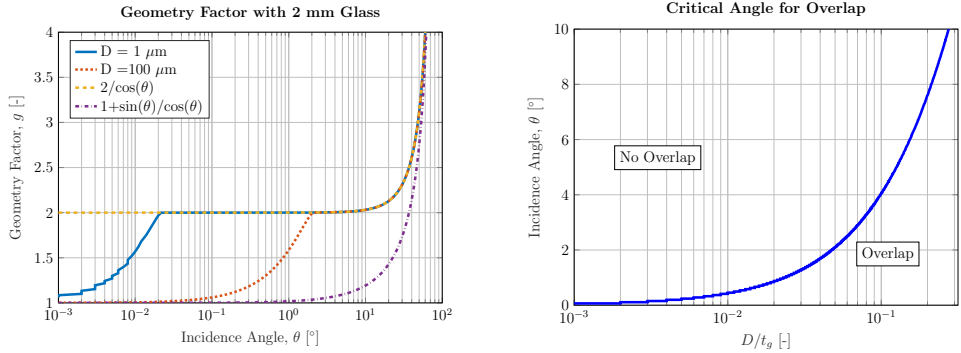}
                }
            \end{subfigure}
            \caption{Left: Geometry factor for particle diameters and simplified first and second geometry models. Right: Incidence angles and particle size/glass thickness required for blocking and shading overlap.}
            \label{fig:overlapcheck}
         \end{figure*}
        Examining these equations, it is clear that $\ta{incidenceangle} \rightarrow 0$ yields an overlap area that is equal to the shading and blocking area, i.e. the area loss $\ta{areaparticle}(\ta{incidenceangle},\ta{particlediameter}) =\frac{\pi \ta{particlediameter}^2}{4}$ is simply equal to the cross-sectional are of the particle. In the other extreme, $\ta{incidenceangle} \rightarrow \frac{\pi}{2}$ yields no overlap. Thus, there is a critical incidence angle where overlap will occur that will be a function of \ta{particlediameter} and \ta{glassthick}. This critical angle is shown in Fig.~\ref{fig:overlapcheck} right. 
        
        Figure~\ref{fig:overlapcheck} left compares a small and large particle's geometry factor which is calculated as
        \begin{equation}
            \label{eq:geometryfactorfig}
            \ta{geofactor} = \ta{areaparticle}(\ta{incidenceangle},\ta{particlediameter}) \cdot \frac{4}{\pi \ta{particlediameter}^2}
        \end{equation}
        i.e. it is the ratio of reflective area loss and the cross-sectional area of the particle. In the case of no overlap $\left(\ta{areaoverlap}=0\right)$, this yields
        \begin{equation}
            \label{eq:geometryfactornooverlap}
            g = \frac{\pi \ta{particlediameter}^2}{2 \cos{\ta{incidenceangle}}} \cdot \frac{4}{\pi \ta{particlediameter}^2} = \frac{2}{\cos{\ta{incidenceangle}}}
        \end{equation}
        This approximation is compared with the actual geometry factor and the first surface geometry factor in the left-hand side of Fig.~\ref{fig:overlapcheck} for $\ta{glassthick}=\SI{2}{\milli\metre}$. The vast majority of airborne dust particles will typically be below \SI{100}{\micro\metre}, so it can be said that the geometry factor approximation of \eqref{eq:geometryfactornooverlap} is valid for incidence angles above \SI{1}{\degree}, which encompasses the vast majority of incidence angles in a CSP tower plant~\citep{Sutter2018}. This approximation yields the following expression for the area loss:
        \begin{equation}
            \label{eq:aproxsecondsurface}
            \ta{arealoss}[_{\ta{idtime}}](\ta{incidenceangle}) = \frac{2}{\cos{\ta{incidenceangle}}} \cdot \ta{areanorm}[_{\ta{idtime}}]\left( \ta{particlediameter} \right)
        \end{equation}
        where 
        \begin{equation}
        \label{eq:areanormalcalc}
            \ta{areanorm}[_{\ta{idtime}}]=\frac{\pi}{4}\int^{\infty}_{0}\ta{particlediameter}^{2}\ta{pdfn}[_{\ta{idtime}}](\ta{particlediameter}) \, d\ta{particlediameter}
        \end{equation}
        is the total area loss with a normal incidence angle. Eventually, the reflectance is computed as
        \begin{equation}
            \label{eq:arealosstoreflectance}
            \ta{predictedreflectance}[_{\ta{idtime}}](\ta{incidenceangle}[_{\ta{idtime}}]) = \ta{measuredreflectance}_{0} \left( 1 - \frac{2}{\cos{\ta{incidenceangle}[_{\ta{idtime}}]}} \cdot \frac{\ta{areanorm}[_{\ta{idtime}}]}{\ta{areamirror}} \right)
        \end{equation}
        where \ta{areamirror} is the total mirror reflective area and $\ta{measuredreflectance}_0$ is the as-clean reflectance.
        
    \subsubsection{Site Soiling Characterization}
    \label{subsubsec:characterization}
    For accurate heliostat-field soiling predictions, the \glspl{dsm} free parameter \als{hrz0} should be estimated from site data using solar mirror reflectance values measured on site~\citep{Picotti2018}. Using the \gls{dsm}, reflectance predictions throughout the measurement campaign may be made using the measured meteorological variables. A least mean squared fit between measured and predicted reflectance values may then be used to determine an appropriate \gls{hrz0} as shown in \eqref{eq:parameterfitting}:
        \begin{subequations}
            \label{eq:parameterfitting}
            \begin{align}
                \ta{hrz0} = \arg \min_{\text{h}} \quad & \sum_{\ta{idtime}=1}^{\ta{ndays}} \left( \ta{measuredreflectance}[_{\ta{idtime}}] - \ta{predictedreflectance}[_{\ta{idtime}}] \right) ^{2} \\
                \text{s.t.} \quad & \ta{predictedreflectance}_{\ta{idtime}+1} = f \left( \ta{predictedreflectance}[_{\ta{idtime}}], \ta{temperature}[_{\ta{idtime}}],\ta{windspeed}[_{\ta{idtime}}], \ta{tsp}[_{\ta{idtime}}], \ta{tiltangle}[_{\ta{idtime}}]; h \right) \label{eq:hrz0fitfunction}\\
                & \ta{predictedreflectance}_{0}=\ta{measuredreflectance}_{0} (\text{measured})
            \end{align}
        \end{subequations}
    where measured reflectance, $\ta{measuredreflectance}$ is measured during the experimental campaign and reflectance prediction, $\ta{predictedreflectance}$ is determined from the \gls{dsm} and the relevant mirror properties and weather inputs in \eqref{eq:hrz0fitfunction} -- the \als{tiltangle} between  horizon and solar mirror; \als{windspeed}, \als{tsp}, and \als{temperature} as measured from the weather station for each measurement, $\ta{idtime}=1,2,\ldots,\ta{ndays}$. 

    In the absence of long term meteorological data, time series analysis can be used to extend existing meteorological data~\cite{Seckin2017,Zhang2018} and explore several yearly trajectories. Since \gls{tsp} data is non-normal (a key assumption for time series modelling) a common transform is employed:
    \begin{equation}
        y_{k}=
            \begin{cases}
                \log \left( \ta{tsp}[_{\ta{idtime}}] + \omega_{\ta{idtime}} \right) & \ta{tsp} > 0\\
                \log \left( \ta{tsp}[_{\ta{idtime}}] + \epsilon_{\ta{idtime}} \right) & \ta{tsp} = 0
            \end{cases}
            \label{eq:tsptransform}
        \end{equation}
        where $\ta{tsp}[_{\ta{idtime}}]$ is the \gls{tsp} measurement at time \ta{idtime}, s is the resolution of the \gls{tsp} measurements, $\omega_{\ta{idtime}} \sim \mathcal{U} \left( 0,\frac{\alpha}{2} \right)$, and $\epsilon_{\ta{idtime}} \sim \mathcal{U} \left( \frac{-s}{2},\frac{\alpha}{2} \right)$. This choice acknowledges that due to the resolution of the instrument, if a measurement has $\tau_{\ta{idtime}}$ then the true value is in the range $\left[ \tau_{\ta{idtime}}-\frac{\alpha}{2},\tau_{\ta{idtime}}+\frac{\alpha}{2} \right)$ if $\tau_{\ta{idtime}} > 0$ and in the range $\left[ 0,\tau_{\ta{idtime}} + \frac{\alpha}{2} \right)$ if $\tau_{\ta{idtime}} =0$.
    
        Using the transformed time series $y_{\ta{idtime}}$, an autoregressive model of the form:
        \begin{equation}
            y_{\ta{idtime}}=\sum^{n_{a}}_{i=1} \phi_{i} y_{\ta{idtime}-i} + \kappa_{\ta{idtime}}
            \label{eq:armodel}
        \end{equation}
        is estimated from the data, where the noise terms $\kappa_{\ta{idtime}}$ are assumed to be normal, independent and identically distributed (i.i.d.)  with variance $\sigma_a^2$. The parameters are estimated via least squares for $n_{a} \in \left[1,50 \right]$ and the model with the minimal Akaike's Information Criterion is selected as the best model.

    A similar approach was adopted for the wind speed. However, due to the diurnal non-stationarity of wind speeds, a Z-score normalization technique~\cite{Altan2021a} was applied to standardize the data
        \begin{equation}
            u_{k}=\frac{\ta{windspeed}[_{\ta{idtime}}]-\Bar{\ta{windspeed}}_{\text{tod}(\ta{idtime})}}{\sigma_{\text{tod}(\ta{idtime})}}
            \label{eq:normalizedwindspeed}
        \end{equation}
        where $\ta{windspeed}[_{\ta{idtime}}]$ is the nominal wind speed, and $\bar{\ta{windspeed}}_{\text{tod}(\ta{idtime})}$ and $\sigma_{\text{tod}(\ta{idtime})}$ are the average and standard deviation, respectively, of the wind speed during the same time of the day as $\ta{idtime}$ in the historical record (i.e. the wind speed statistics are assumed to be 24-hour-periodic).
    
\subsection{Cleaning Resource Optimizer}
Once \gls{hrz0} is estimated from the experimental site data, the \gls{dsm} can be used to predict soiling rates for a hypothetical operating solar field using only meteorological data and solar field design conditions. The predicted soiling rates are utilised with a cleaning resource optimiser to determine a cost effective cleaning resource plan. Decisions of when to clean must consider the economic balance of cleaning costs and plant productivity. A \glsfirst{fftbcs} policy was employed here, which has been shown to yield near-optimal cleaning schedules for the case where trucks are owned~\citep{Picotti2020}.

The \gls{fftbcs} policy is described as follows. Firstly, the field is sectorized into $\ta{idsect}=1,2,\ldots,\ta{nsects}$ sectors, each of which has a representative heliostat for which the soiling losses are simulated by the \gls{dsm}. All heliostats in the sector are assumed to have the same soiled area and incidence angle. There are \ta{ntrucks} cleaning crews of which can each clean $\ta{sectorscleanedpertruck}$ sectors in one day of cleaning and the number of full-field cleans per year is $\ta{ncleans}\in0,\text{floor}\frac{365\cdot\ta{ntrucks}\cdot\ta{sectorscleanedpertruck}}{\ta{nsects}}$. Since the sectors are cleaned in sequential order, the set of time indices where section $\ta{idsect} = 1$ is cleaned is
\begin{equation}
    \label{eq:cleaningstart}
    \ta{cleandecision}_1 = \left\{ 1,\lceil\ta{deltak} \rceil, \lceil2 \ta{deltak} \rceil, ..., \lceil(\ta{ncleans}-1)\cdot \ta{deltak} \rceil \right\}  
\end{equation}
where $\lceil\cdot \rceil$ denotes a rounding up operation and the number of days in between cleans, $\ta{deltak}$ is
\begin{equation}
    \label{eq:cleaninginterval}
    \ta{deltak} = \frac{365\cdot\ta{deltatd}}{\ta{ncleans}}
\end{equation}
where $\ta{deltatd}=\frac{24}{\ta{deltat}}$ (\ta{deltat} in hours) is the number of sampling periods in a day. The cleaning time for any sector $\ta{idsect}=2,3,\ldots,\ta{nsects}$ are:
\begin{equation}
    \label{eq:cleaningrotation}
    \ta{cleandecision}_j = \left\{ \left. d+\mathrm{floor}\left( \frac{\ta{idsect}-1}{\ta{ntrucks}\cdot \ta{sectorscleanedpertruck}} \right) \, \right| \, d\in\ta{cleandecision}_1, \, \ta{idsect}>1\right\}
\end{equation}

The parameters of this model are (\ta{nsects}, \ta{sectorscleanedpertruck}, \ta{ntrucks}, \ta{ncleans}). The parameter \ta{nsects} is selected based on the desired spatial resolution of the soiling predictions, while \ta{sectorscleanedpertruck} is a function of the technology and number of people in a crew. Once these parameters are fixed, the \gls{fftbcs} policy is thus parameterized by \ta{ntrucks} and \ta{ncleans}.

The optimal (\ta{ntrucks}, \ta{ncleans}) are calculated by minimizing the total annual cleaning cost:
\begin{equation}
    \label{eq:tcc}
    \ta{tcc} (\ta{ntrucks},\ta{ncleans}) = \ta{ccl} + \ta{cdeg}
\end{equation}
% Cleanign Cost
where \ta{ccl} is an annualized sum of all direct cleaning related costs (i.e. fixed yearly costs and variable costs with number of field cleanings):
\begin{equation}
\label{eq:ccl}
    \ta{ccl}(\ta{ntrucks},\ta{ncleans}) = \left( \frac{\ta{costtruck}}{\ta{depreciation}} + \ta{costop} + \ta{costmaint} \right) \ta{ntrucks} \\
     + \sum_{\ta{idsect}=1}^{\ta{nsects}} \ta{costwf} \cdot \ta{areatotal}[_{\ta{idsect}}] \cdot \ta{ncleans}
\end{equation}
where \glsname{costtruck} is the cost of a cleaning vehicle, \glsname{depreciation} is the useful life of the truck, \glsname{costop} is the cleaning vehicle operator salary, \glsname{ntrucks} is the number of cleaning crews, water and fuel costs, \glsname{costwf} represent the cleaning truck consumable costs per square metre of heliostat cleaned, and $\ta{areatotal}[_{\ta{idsect}}]$ is the heliostat area of sector \ta{idsect}.

The degradation costs, \ta{cdeg} are associated with the loss in thermal energy due to soiling and is calculated as the difference in revenue that would have been obtained from an always clean heliostat field and the predicted soiled reflectance values for the given cleaning schedule, as shown in \eqref{eq:cdeg}.
\begin{equation}
    \label{eq:cdeg}
    \ta{cdeg}(\ta{ntrucks},\ta{ncleans}) = \left( \ta{pel} - \ta{costoam} \right) \sum^{\ta{ndays}}_{\ta{idtime}=1}{ \left( \ta{workclean}[_{\ta{idtime}}] - \ta{worksoil}[_{\ta{idtime}}] \right) \ta{deltat}}
\end{equation}
where \glsname{pel} is the sale price of electricity, \glsname{costoam} is the cost of non-cleaning operation and maintenance tasks per unit of power production\footnote{Thus, higher soiling losses lead to lower non-cleaning operation and maintenance costs, which is common in energy production economic analyses}, \ta{deltat} is the time between two epochs (usually the sampling frequency of the meteorology data), $\ta{workclean}[_{\ta{idtime}}]$ and $\ta{worksoil}[_{\ta{idtime}}]$ is the electrical work generated from a solar field in clean $\left( \ta{predictedreflectance}=\ta{measuredreflectance}_{0} \right)$ and predicted soiled conditions respectively at time \ta{idtime}, and \ta{ndays} is the total number of samplings for the simulation trajectory. The amount of work produced for both the clean and soiled conditions depends on the plant design and operating strategy. The plant under consideration is one with thermal energy storage (no direct steam capability) and a number of receivers, each with their own dedicated solar field. A simplified dispatch strategy is adopted where the power block is run at rated load when the thermal storage is sufficient and shutdown otherwise.

When the energy from the storage is sufficient to run the power block at rated loads for the full duration (i.e. $\ta{qpb}=\ta{qtes}[_{\ta{idtime}}]$) the amount of work produced is:
\begin{equation}
\label{eq:qtes}
    \ta{pbwork}[_{\ta{idtime}}] =\frac{\ta{qtes}[_{\ta{idtime}}]}{\ta{etapb}}
\end{equation}
where
\begin{equation}
    \label{eq:dischargeconstraints}
    \ta{qtes}[_{\ta{idtime}}] = 
    \begin{cases}
        \ta{qpb} & \frac{\ta{storagecapacity}_{\ta{idtime}-1}}{\ta{deltat}} + \ta{qrec}[_{\ta{idtime}}] > \ta{qpb}\\
        \frac{\ta{storagecapacity}_{\ta{idtime}-1}}{\ta{deltat}} + \ta{qrec}[_{\ta{idtime}}] & 0 < \frac{\ta{storagecapacity}_{\ta{idtime}-1}}{\ta{deltat}} + \ta{qrec}[_{\ta{idtime}}] \leq \ta{qpb}\\
        0 & \text{otherwise}
    \end{cases}
\end{equation}
and \ta{qpb} is the thermal input required to run the power block at rated load, $\ta{storagecapacity}[_{\ta{idtime}}]$ is the stored thermal energy, and $\ta{qrec}[_{\ta{idtime}}]$ is the thermal energy coming from the receiver(s). The thermal storage charge is calculated as
\begin{equation} 
    \label{eq:storagecapacity}
    \ta{storagecapacity}[_{\ta{idtime}}] = \ta{storagecapacity}_{\ta{idtime}-1} + \left( \ta{qrec}[_{\ta{idtime}}]- \ta{qtes} \right) \cdot \ta{deltat}
\end{equation} 
The charge state is constrained to the \als{maxstoragecapacity} of the \gls{tes}. When the maximum charge is reached the heliostat-field will partially or fully defocus to limit the amount of energy into the thermal storage depending upon its current and maximum capacity, as shown in \eqref{eq:receivertesconditions}:
\begin{equation}
    \label{eq:receivertesconditions}
    \ta{qrec}[_{\ta{idtime}}] = 
    \begin{cases}
        \ta{qrec}[_{\ta{idtime}}]& \frac{\ta{storagecapacity}_{\ta{idtime}-1}}{\ta{deltat}} + \ta{qrec}[_{\ta{idtime}}] < \frac{\ta{maxstoragecapacity}}{\ta{deltat}}\\
        \frac{\ta{maxstoragecapacity} - \ta{storagecapacity}_{\ta{idtime}-1}}{\ta{deltat}} & \frac{\ta{storagecapacity}_{\ta{idtime}-1}}{\ta{deltat}} + \ta{qrec}[_{\ta{idtime}}] > \frac{\ta{maxstoragecapacity}}{\ta{deltat}}\\
        0 & \frac{\ta{storagecapacity}_{\ta{idtime}-1}}{\ta{deltat}} \geq \frac{\ta{maxstoragecapacity}}{\ta{deltat}}
    \end{cases}
\end{equation}

Thermal energy from the receiver is the sum of several arrays each with there own heliostat field and receiver system. Let there be \ta{nreceivers} arrays and $\ta{idreceiver}=1,2,...,\ta{nreceivers}$ be the array index for all solar towers. The total receiver energy output from all arrays is therefore calculated as the sum from each array:
\begin{equation}
    \ta{qrec}[_{\ta{idtime}}] = \sum^{\ta{nreceivers}}_{\ta{idreceiver} = 1} \ta{qarray}[_{\ta{idtime},\ta{idreceiver}}]
\end{equation}
\Als{qarray} is subjected to a \als{qloss} as such the contribution from each tower array is subjected to:
\begin{equation}
    \label{eq:recpower}
    \ta{qarray}[_{\ta{idtime},\ta{idreceiver}}] = \ta{qin}[_{\ta{idtime},\ta{idreceiver}}] - \ta{qloss}
\end{equation}

Receiver saturation and deprivation conditional requirements are introduced to improve the receiver model. Saturation conditions occur when the incident receiver thermal energy is greater then the operating capacity, for modelling purposes the receiver power is capped at the maximum design point, \ta{qmax}. Receiver deprivation conditions are met when the incident thermal power is less then the constant thermal losses of the receiver, the receiver is turned off and the energy output is set to zero during these conditions. It is assumed that when the receiver is operating in a restricted capacity either from storage constraints, receiver saturation or depravation the heliostat fields will defocus by adjusting their azimuth angle to redirect \gls{dni} off the receiver while maintaining the correct tilt angles. In addition to the constraints of Eqs~\ref{eq:receivertesconditions} the aforementioned constraints must also comply for each array:
\begin{equation}
\label{eq:recconditions}
    \ta{qarray}[_{\ta{idtime},\ta{idreceiver}}] = 
        \begin{cases}
            \ta{qin}[_{\ta{idtime},\ta{idreceiver}}] - \ta{qloss} & \ta{qloss} < \ta{qarray}[_{\ta{idtime},\ta{idreceiver}}] \leq \ta{qmax}\\
            \ta{qmax} - \ta{qloss}  & \ta{qarray}[_{\ta{idtime},\ta{idreceiver}}] > \ta{qmax}\\
            0   & \ta{qin}[_{\ta{idtime},\ta{idreceiver}}] \leq \ta{qloss}
        \end{cases}
\end{equation}

Reflected \gls{dni} received by each array from their respective heliostat fields are calculated with:
\begin{equation}
    \label{eq:recinpower}
    \ta{qin}[_{\ta{idtime},\ta{idreceiver}}] = \ta{dni}[_{\ta{idtime}}] \sum_{\ta{idsect} \in \mathcal{J}_{\ta{idreceiver}}}{\ta{areatotal}[_{\ta{idsect}}] \cdot \ta{opticalefficiency}[_{\ta{idtime},\ta{idsect}}]} \cdot \ta{predictedreflectance}[_{\ta{idtime},\ta{idsect}}]
\end{equation}
where $\ta{dni}[_{\ta{idtime}}]$ is the incident \gls{dni}, $\mathcal{J}_{\ta{idreceiver}}$ is the set of representative heliostats (i.e.) sector that are targeted at receiver \ta{idreceiver}, $\ta{areatotal}[_{\ta{idsect}}]$ is the sum of reflective area for heliostat sector \ta{idsect}, and the $\ta{opticalefficiency}[_{\ta{idtime},\ta{idsect}}]$ is the mean product of heliostat cosine, attenuation, blocking, shading, and image intercept efficiencies of all heliostats in sector \ta{idsect} at time index \ta{idtime}.

\subsubsection{Simulating Heliostat Reflectance with Cleaning}
 For each time step, the soiling state of each representative mirror is updated by adding the current soiling rate to the cumulative particle number distribution. When a cleaning event occurs a \textit{reflectance recovery} (i.e. cleaning) model is utilised to determine the total number of particles removed from the cleaned heliostat. Let $\ta{pdfn}[_{\ta{idtime},\ta{idsect}}](\ta{particlediameter})$ be the cumulative number of particles on the mirror before a cleaning action and $\ta{pdfn}[_{\ta{idtime}+1}](\ta{particlediameter})$ be the remaining particles just after a cleaning event. The reflectance recovery model is given as:
\begin{equation}
    \label{eq:restorationclean}
    \ta{pdfn}[_{\ta{idtime}+1,\ta{idsect}}] \left( \ta{particlediameter} \right) = 
    \begin{cases}
        \ta{ndp}[_{\ta{idtime},\ta{idsect}}] \left( \ta{particlediameter} \right) + \ta{pdfn}[_{\ta{idtime}\ta{idsect}}] \left( \ta{particlediameter} \right) \left( 1 - \ta{etacl}(\ta{particlediameter}) \right) & \ta{cleandecision}[_{\ta{idtime}+1,\ta{idsect}}] = 1\\
        \ta{ndp}[_{\ta{idtime},\ta{idsect}}] \left( \ta{particlediameter} \right) + \ta{pdfn}[_{\ta{idtime},\ta{idsect}}] \left(\ta{particlediameter} \right) & \ta{cleandecision}[_{\ta{idtime}+1,\ta{idsect}}] = 0\\
    \end{cases}
\end{equation}

where $0< \ta{etacl}\leq1$ is the cleaning efficiency (i.e. the fraction of particles that are removed at \ta{particlediameter}) for a particular cleaning technology and \ta{cleandecision} is the cleaning decision that is given for a particular cleaning schedule.

Initial reflectances of each representative heliostat should be specified. It is desirable to set these reflectances so that the initial and final conditions are the same --- thus producing a ``steady-state'' reflectance trajectory. This steady-state condition ensures that the initial soiling state of the heliostats are aligned with typical soiling status of a plant at the start of the year (not perfectly clean). To achieve this, a boundary condition check for each representative heliostat is conducted after running the cleaning model. If there is a substantial difference between $\ta{pdfn}[_{1,\ta{idsect}}]$ and $\ta{pdfn}[_{\ta{ndays},\ta{idsect}}]$ the cleaning model is repeated using the initial conditions of $\ta{pdfn}[_{1,\ta{idsect}}]=\ta{ndp}[_{1,\ta{idsect}}]\cdot\ta{deltat}+\ta{pdfn}[_{\ta{ndays},\ta{idsect}}]$ for the given cleaning schedule. After a set tolerance is reached, predicted reflectance values for each heliostat sector are computed using \eqref{eq:areanormalcalc} and \eqref{eq:arealosstoreflectance}. 
\section{Case Study: A 50~MW Modular CSP Plant}
\label{sec:results}
    \gls{nwqhpp} (lat =\SI{-20.732}{\degree} lon = \SI{139.384}{\degree}) is a prospective \SI{50}{\mega\watt} solar hybrid baseload power plant integrating gas turbine, \gls{pv}, modular \gls{csp} + \gls{tes} technology to provide continuous utility scale power production for Queensland's rural North West Power System grid. Cleaning resources for the heliostat field have been optimised using the presented \gls{fftbcs} heuristic and modified \gls{dsm} for an early design of the \gls{csp} plant. The degradation costs were computed as the average of multiple weather trajectories to estimate the \emph{expected} degradation costs. This averaging will prevent over-specialization of the cleaning schedule to one single dust trajectory and instead prefer those that lower the average over a range of plausible conditions.
    
    \subsection{Experimental Measurement Campaign}
    \label{subsec:experimentalresults}
\begin{figure*}[h!]
        \centering
        \includegraphics[width=0.8\textwidth,keepaspectratio]{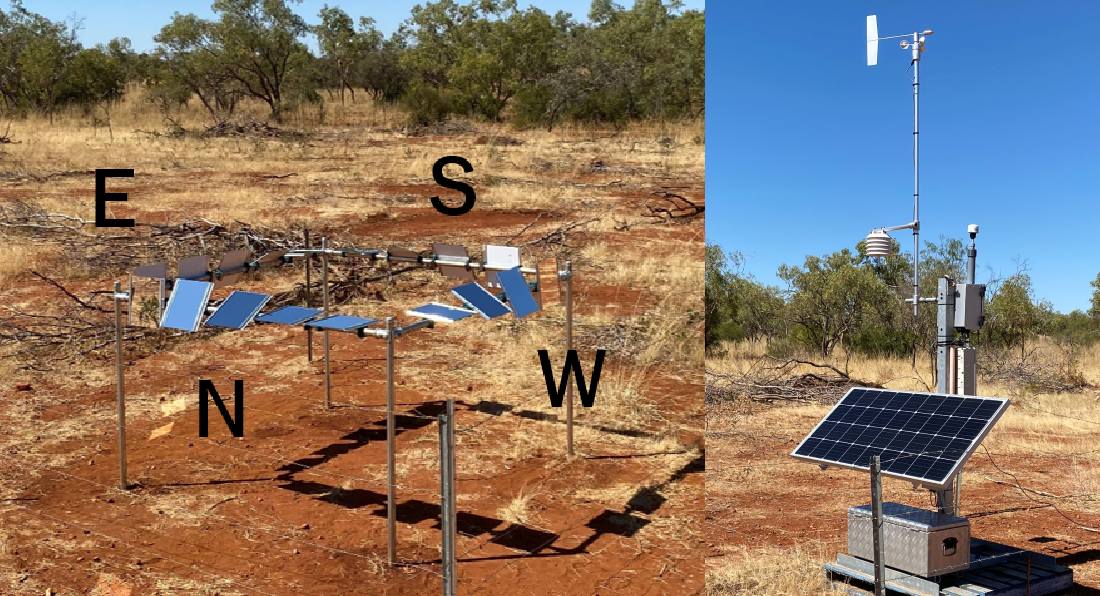}
        \caption{(Left) Mirror test-Rig; and (Right) \gls{tsp}, temperature, relative humidity, and wind speed weather station}
        \label{fig:testrig}
    \end{figure*}

    The experimental set-up is comprised of a weather monitoring station and a mirror test rig, as shown in Fig.~\ref{fig:testrig}. The weather station measured key environmental parameters required to predict soiling losses using the \gls{dsm}. \gls{tsp}, \glsdesc{windspeed} and \glsdesc{temperature} were recorded from the weather station over a two-year period. A solar mirror test rig was assembled near the weather station (within \SI{20}{\metre}) and was used to fasten solar mirrors at a fixed tilt angle and aid in the retrieval of reflectance measurements over a period of time. The test rig was comprised of 18 mirror samples each tilted at various fixed angles (\SI{5}{\degree}, \SI{30}{\degree}, \SI{60}{\degree}, and \SI{85}{\degree} with respect to the horizon) and repeated for each orientation (North, East, South, West) in addition to a vertical East facing and a horizontal mirror on the the North arm of the test rig.

        \subsubsection{Airborne Dust Characteristics}
        The \gls{tsp} was measured using an Ecotech eSampler with a one-second sampling rate, and the data was averaged every five minutes over a two-year period. The instrument employs a forward light scatter laser Nephelometer to measure \gls{tsp} concentrations ranging from 0 to \SI{65}{\milli\gram\per\metre\cubed}, with a reported precision of \SI{3}{\micro\gram\per\metre\cubed} or \SI{2}{\percent} (whichever is greater) and a resolution of \SI{1}{\micro\gram\per\metre\cubed}.

\begin{figure*}
            \centering
            \begin{subfigure}[c]{0.49\textwidth}
                \resizebox{\linewidth}{!}{            
\definecolor{mycolor1}{rgb}{0.00000,0.44700,0.74100}%
\definecolor{mycolor3}{rgb}{0.92900,0.69400,0.12500}%
\begin{tikzpicture}[%
trim axis left, trim axis right
]

\begin{axis}[%
width=0.951\fwidth,
height=\fheight,
at={(0\fwidth,0\fheight)},
scale only axis,
xtick scale label code/.code={},
xmin=738000.000,
xmax=738900.000,
xtick={738000.000,738100.000,738200.000,738300.000,738400.000,738500.000,738600.000,738700.000,738800.000,738900.000},
xticklabels={{20-07},{20-11},{21-02},{21-05},{21-09},{21-12},{22-03},{22-06},{22-10},{23-01}},
xticklabel style={rotate=30},
ymin=-5.000,
ymax=30.000,
ylabel style={font=\color{white!15!black}},
ylabel={$\text{TSP [}\mu\text{.g/m}^\text{3}\text{]}$},
axis background/.style={fill=white},
title style={font=\bfseries},
title={Site Monthly TSP},
xmajorgrids,
ymajorgrids,
legend pos=north east,
legend style={legend cell align=left, align=left, draw=white!15!black}
]
\addplot [color=mycolor1, line width=2.0]
  table[row sep=crcr]{%
738004.000	4.590\\
738035.000	8.444\\
738065.000	8.564\\
738096.000	12.342\\
738126.000	8.169\\
738157.000	7.257\\
738188.000	5.972\\
738216.000	4.979\\
738247.000	4.731\\
738277.000	7.218\\
738308.000	7.093\\
738338.000	8.906\\
738369.000	13.421\\
738400.000	16.079\\
738430.000	23.659\\
738461.000	13.345\\
738491.000	12.332\\
738522.000	8.369\\
738553.000	8.833\\
738581.000	8.731\\
738612.000	8.962\\
738642.000	5.890\\
738673.000	6.444\\
738703.000	5.419\\
738734.000	7.339\\
738765.000	9.387\\
738795.000	10.315\\
738826.000	12.844\\
738856.000	4.455\\
};
\addlegendentry{Avg}

\addplot [color=red, dashed, line width=2.0pt]
  table[row sep=crcr]{%
738004.000	6.984\\
738157.000	6.984\\
738188.000	7.723\\
738216.000	8.410\\
738247.000	9.357\\
738277.000	10.027\\
738308.000	10.243\\
738338.000	10.462\\
738369.000	10.628\\
738400.000	10.903\\
738430.000	11.236\\
738461.000	11.357\\
738491.000	11.275\\
738522.000	11.102\\
738553.000	10.704\\
738581.000	10.171\\
738612.000	9.336\\
738642.000	8.760\\
738673.000	8.411\\
738856.000	8.411\\
};
\addlegendentry{13-MA}

\addplot [color=mycolor3, dotted, line width=2.0pt]
  table[row sep=crcr]{%
738004.000	-0.280\\
738035.000	2.481\\
738065.000	5.247\\
738096.000	3.870\\
738126.000	-0.627\\
738157.000	-1.286\\
738188.000	-1.867\\
738216.000	-2.491\\
738247.000	-2.556\\
738277.000	-2.895\\
738308.000	-2.613\\
738338.000	-2.329\\
738369.000	-0.280\\
738400.000	2.481\\
738430.000	5.247\\
738461.000	3.870\\
738491.000	-0.627\\
738522.000	-1.286\\
738553.000	-1.867\\
738581.000	-2.491\\
738612.000	-2.556\\
738642.000	-2.895\\
738673.000	-2.613\\
738703.000	-2.329\\
738734.000	-0.280\\
738765.000	2.481\\
738795.000	5.247\\
738826.000	3.870\\
738856.000	-0.627\\
};
\addlegendentry{Seasonal}

\end{axis}

\begin{axis}[%
width=1.227\fwidth,
height=1.227\fheight,
at={(-0.16\fwidth,-0.135\fheight)},
scale only axis,
xmin=0.000,
xmax=1.000,
ymin=0.000,
ymax=1.000,
axis line style={draw=none},
ticks=none,
axis x line*=bottom,
axis y line*=left
]
\end{axis}
\end{tikzpicture}%
                }
            \end{subfigure}
            \hfill
            \begin{subfigure}[c]{0.49\textwidth}
                \resizebox{\linewidth}{!}{
\definecolor{mycolor1}{rgb}{0.00000,0.44700,0.74100}%
\begin{tikzpicture}[%
trim axis left, trim axis right
]

\begin{axis}[%
width=0.951\fwidth,
height=\fheight,
at={(0\fwidth,0\fheight)},
scale only axis,
xmin=0.000,
xmax=25.000,
xlabel style={font=\color{white!15!black}},
xlabel={Local Solar Time Hour of Day},
ymin=6.000,
ymax=14.000,
ylabel style={font=\color{white!15!black}},
ylabel={$\text{TSP [}\mu\text{g/m}^\text{3}\text{]}$},
axis background/.style={fill=white},
title style={font=\bfseries},
title={Site Time Synchronous Average TSP},
grid=both,
grid style={line width=.1pt, draw=gray!25},
major grid style={line width=.2pt,draw=gray!50},
]
\addplot [color=mycolor1, forget plot, line width=2.0pt]
  table[row sep=crcr]{%
1.000	8.237\\
2.000	8.083\\
3.000	7.886\\
4.000	7.880\\
5.000	8.571\\
6.000	8.911\\
7.000	10.247\\
8.000	11.227\\
9.000	12.784\\
10.000	12.585\\
11.000	12.375\\
12.000	11.389\\
13.000	9.733\\
14.000	9.166\\
15.000	9.251\\
16.000	8.725\\
17.000	8.794\\
18.000	8.011\\
19.000	8.272\\
20.000	8.151\\
21.000	8.368\\
22.000	7.964\\
23.000	8.014\\
24.000	6.552\\
};
\end{axis}

\begin{axis}[%
width=1.227\fwidth,
height=1.227\fheight,
at={(-0.16\fwidth,-0.135\fheight)},
scale only axis,
xmin=0.000,
xmax=1.000,
ymin=0.000,
ymax=1.000,
axis line style={draw=none},
ticks=none,
axis x line*=bottom,
axis y line*=left
]
\end{axis}
\end{tikzpicture}%
                }
            \end{subfigure}
            \caption{Left: Monthly averaged site TSP with 2x12 moving average (MA) filter and stable seasonal component. Right: Solar time synchronous average total suspended particle count.}
            \label{fig:seasonalcomponentcheck}
        \end{figure*}
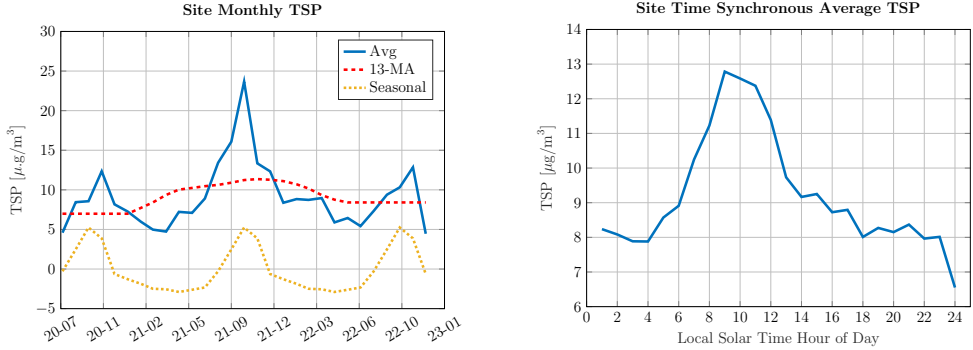

The measurements revealed a mean \gls{tsp} concentration of \SI{9.1}{\micro\gram\per\metre\cubed}, with a standard deviation of \SI{14.9}{\micro\gram\per\metre\cubed} and a peak of \SI{1075}{\micro\gram\per\metre\cubed}. Analyzing the two years of data using an additive decomposition time series model~\cite{Cleveland1990}, evidence of cyclic seasonality was observed. The mean monthly \gls{tsp} concentrations showed an increase of approximately \SI{5}{\micro\gram\per\metre\cubed} in October and a decrease of \SI{3}{\micro\gram\per\metre\cubed} in May each year, as shown in Fig.~\ref{fig:seasonalcomponentcheck} (left).

Additionally, Fig.~\ref{fig:seasonalcomponentcheck} (right) presents the intraday solar time synchronous average \gls{tsp}. The intraday \gls{tsp} levels were higher during \gls{csp} operational times (7 to 19) and decreased during the night. 

        \subsubsection{Cleanliness Measurements}
        Mirrors were cleaned with demineralized water and a squeegee to measure the nominal clean reflectance values using a D\&S Portable Specular Reflectometer (15R-USB, S/N: 713). Four reflectance measurements were made for each mirror, one at the centre of each quadrant, with measurements of each mirror being recorded. Following the initial set of measurements, the reflectance of all mirrors were measured twice a day over a seven day period in September 2020. The reflectometer was configured to measure specular reflectance with the instruments white LED at an \glsdesc{incidenceangle} of \SI{15}{\degree} with an acceptance half angle of \SI{12.5}{\milli\radian}. For the duration of the measurement campaign there was no additional mirror cleaning from either manual labor or rain. All reflectance measurements were converted to a \gls{cleanliness} ratio using \eqref{eq:cleanliness}, where $\rho_{0,\ta{idsect}}$ is the mean initial clean reflectance of mirror \ta{idsect}.
        \begin{equation}
            \label{eq:cleanliness}
            \ta{cleanliness} = \dfrac{\rho_{k,j}}{\rho_{0,j}}
        \end{equation}
        
        Fig.~\ref{fig:tiltcleanlinessmeasures} shows the mean cleanliness for each tilt angle when considering all orientations. Results show that the soiling rate\footnote{Soiling rate is the change in cleanliness as a percentage over a period of time} decreases with increasing tilt angle, where the horizontal mirror (\SI{0}{\degree}) experienced higher soiling rates of \SI{0.4}{pp\per\day} relative to the near vertical tilted (\SI{85}{\degree}) mirrors of \SI{0.02}{pp\per\day}. Mean cleanliness values for each orientation are shown in Fig.~\ref{fig:orientationcleanlinessmeasures}, the mean is calculated using the cleanliness values from the 5, 30, 60, and \SI{85}{\degree} tilted solar mirrors of each orientation. Soiling rates for each orientation are similar with the means being within the standard deviations of each other, as such for soiling rate predictions it is assumed that soiling rates are independent from heliostat azimuth angles.
        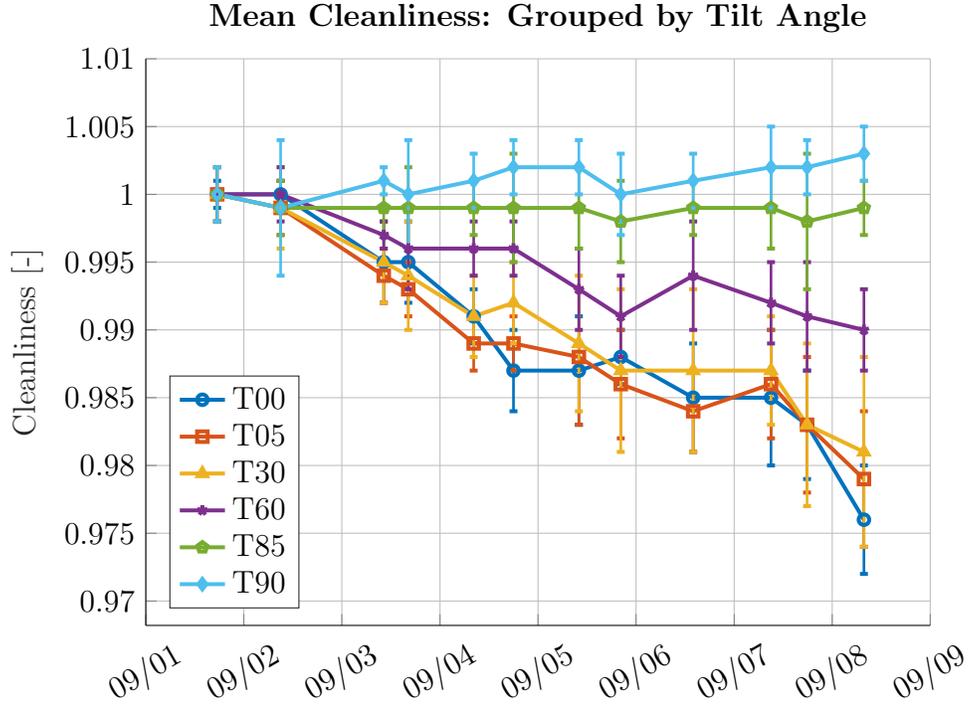
\begin{figure}[h!]
            \centering
\definecolor{mycolor1}{rgb}{0.00000,0.44700,0.74100}%
\definecolor{mycolor2}{rgb}{0.85000,0.32500,0.09800}%
\definecolor{mycolor3}{rgb}{0.92900,0.69400,0.12500}%
\definecolor{mycolor4}{rgb}{0.49400,0.18400,0.55600}%
\definecolor{mycolor5}{rgb}{0.46600,0.67400,0.18800}%
\definecolor{mycolor6}{rgb}{0.30100,0.74500,0.93300}%
\begin{tikzpicture}[%
trim axis left, trim axis right
]

\begin{axis}[%
width=0.951\fwidth,
height=\fheight,
at={(0\fwidth,0\fheight)},
scale only axis,
xtick scale label code/.code={},
xmin=738035.000,
xmax=738043.000,
xtick={738035.000,738036.000,738037.000,738038.000,738039.000,738040.000,738041.000,738042.000,738043.000},
xticklabels={{09/01},{09/02},{09/03},{09/04},{09/05},{09/06},{09/07},{09/08},{09/09}},
xticklabel style={rotate=30},
xlabel style={font=\color{white!15!black}},=
ymin=0.97,
ymax=1.01,
y tick label style={/pgf/number format/fixed, /pgf/number format/precision=5},
ylabel style={font=\color{white!15!black}},
ylabel={$\text{Cleanliness [-]}$},
axis background/.style={fill=white},
title style={font=\bfseries},
title={Mean Cleanliness: Grouped by Tilt Angle},
axis x line*=bottom,
axis y line*=left,
xmajorgrids,
ymajorgrids,
legend pos=south west,
legend style={legend cell align=left, align=left, draw=white!15!black}
]
\addplot [color=mycolor1, line width=1.5pt, mark=o, mark options={solid, mycolor1}]
 plot [error bars/.cd, y dir=both, y explicit, error bar style={line width=1pt}, error mark options={line width=1.5pt, mark size=1.5pt, rotate=90}]
 table[row sep=crcr, y error plus index=2, y error minus index=3]{%
738035.729	1.000	0.001	0.001\\
738036.375	1.000	0.001	0.001\\
738037.427	0.995	0.003	0.003\\
738037.677	0.995	0.003	0.003\\
738038.344	0.991	0.002	0.002\\
738038.750	0.987	0.003	0.003\\
738039.417	0.987	0.004	0.004\\
738039.844	0.988	0.002	0.002\\
738040.583	0.985	0.004	0.004\\
738041.375	0.985	0.005	0.005\\
738041.743	0.983	0.004	0.004\\
738042.323	0.976	0.004	0.004\\
};
\addlegendentry{T00}

\addplot [color=mycolor2, line width=1.5pt, mark=square, mark options={solid, mycolor2}]
 plot [error bars/.cd, y dir=both, y explicit, error bar style={line width=1pt}, error mark options={line width=1.5pt, mark size=1.5pt, rotate=90}]
 table[row sep=crcr, y error plus index=2, y error minus index=3]{%
738035.729	1.000	0.002	0.002\\
738036.375	0.999	0.002	0.002\\
738037.427	0.994	0.002	0.002\\
738037.677	0.993	0.002	0.002\\
738038.344	0.989	0.002	0.002\\
738038.750	0.989	0.002	0.002\\
738039.417	0.988	0.005	0.005\\
738039.844	0.986	0.004	0.004\\
738040.583	0.984	0.003	0.003\\
738041.375	0.986	0.004	0.004\\
738041.743	0.983	0.005	0.005\\
738042.323	0.979	0.005	0.005\\
};
\addlegendentry{T05}

\addplot [color=mycolor3, line width=1.5pt, mark=triangle, mark options={solid, mycolor3}]
 plot [error bars/.cd, y dir=both, y explicit, error bar style={line width=1pt}, error mark options={line width=1.5pt, mark size=1.5pt, rotate=90}]
 table[row sep=crcr, y error plus index=2, y error minus index=3]{%
738035.729	1.000	0.002	0.002\\
738036.375	0.999	0.003	0.003\\
738037.427	0.995	0.003	0.003\\
738037.677	0.994	0.004	0.004\\
738038.344	0.991	0.003	0.003\\
738038.750	0.992	0.003	0.003\\
738039.417	0.989	0.005	0.005\\
738039.844	0.987	0.006	0.006\\
738040.583	0.987	0.006	0.006\\
738041.375	0.987	0.004	0.004\\
738041.743	0.983	0.006	0.006\\
738042.323	0.981	0.007	0.007\\
};
\addlegendentry{T30}

\addplot [color=mycolor4, line width=1.5pt, mark=star, mark options={solid, mycolor4}]
 plot [error bars/.cd, y dir=both, y explicit, error bar style={line width=1pt}, error mark options={line width=1.5pt, mark size=1.5pt, rotate=90}]
 table[row sep=crcr, y error plus index=2, y error minus index=3]{%
738035.729	1.000	0.002	0.002\\
738036.375	1.000	0.002	0.002\\
738037.427	0.997	0.001	0.001\\
738037.677	0.996	0.003	0.003\\
738038.344	0.996	0.002	0.002\\
738038.750	0.996	0.002	0.002\\
738039.417	0.993	0.003	0.003\\
738039.844	0.991	0.003	0.003\\
738040.583	0.994	0.004	0.004\\
738041.375	0.992	0.003	0.003\\
738041.743	0.991	0.004	0.004\\
738042.323	0.990	0.003	0.003\\
};
\addlegendentry{T60}

\addplot [color=mycolor5, line width=1.5pt, mark=pentagon, mark options={solid, mycolor5}]
 plot [error bars/.cd, y dir=both, y explicit, error bar style={line width=1pt}, error mark options={line width=1.5pt, mark size=1.5pt, rotate=90}]
 table[row sep=crcr, y error plus index=2, y error minus index=3]{%
738035.729	1.000	0.002	0.002\\
738036.375	0.999	0.002	0.002\\
738037.427	0.999	0.002	0.002\\
738037.677	0.999	0.003	0.003\\
738038.344	0.999	0.002	0.002\\
738038.750	0.999	0.004	0.004\\
738039.417	0.999	0.003	0.003\\
738039.844	0.998	0.003	0.003\\
738040.583	0.999	0.002	0.002\\
738041.375	0.999	0.003	0.003\\
738041.743	0.998	0.005	0.005\\
738042.323	0.999	0.002	0.002\\
};
\addlegendentry{T85}

\addplot [color=mycolor6, line width=1.5pt, mark=diamond, mark options={solid, mycolor6}]
 plot [error bars/.cd, y dir=both, y explicit, error bar style={line width=1pt}, error mark options={line width=1.5pt, mark size=1.5pt, rotate=90}]
 table[row sep=crcr, y error plus index=2, y error minus index=3]{%
738035.729	1.000	0.002	0.002\\
738036.375	0.999	0.005	0.005\\
738037.427	1.001	0.001	0.001\\
738037.677	1.000	0.004	0.004\\
738038.344	1.001	0.002	0.002\\
738038.750	1.002	0.002	0.002\\
738039.417	1.002	0.002	0.002\\
738039.844	1.000	0.003	0.003\\
738040.583	1.001	0.002	0.002\\
738041.375	1.002	0.003	0.003\\
738041.743	1.002	0.002	0.002\\
738042.323	1.003	0.002	0.002\\
};
\addlegendentry{T90}

\end{axis}

\begin{axis}[%
width=1.322\fwidth,
height=1.322\fheight,
at={(-0.196\fwidth,-0.246\fheight)},
scale only axis,
xmin=0.000,
xmax=1.000,
ymin=0.000,
ymax=1.000,
axis line style={draw=none},
ticks=none,
axis x line*=bottom,
axis y line*=left
]
\end{axis}
\end{tikzpicture}%
            \caption{Mean cleanliness of all orientations with respect to mirror tilt angle (T)}
            \label{fig:tiltcleanlinessmeasures}
        \end{figure}
        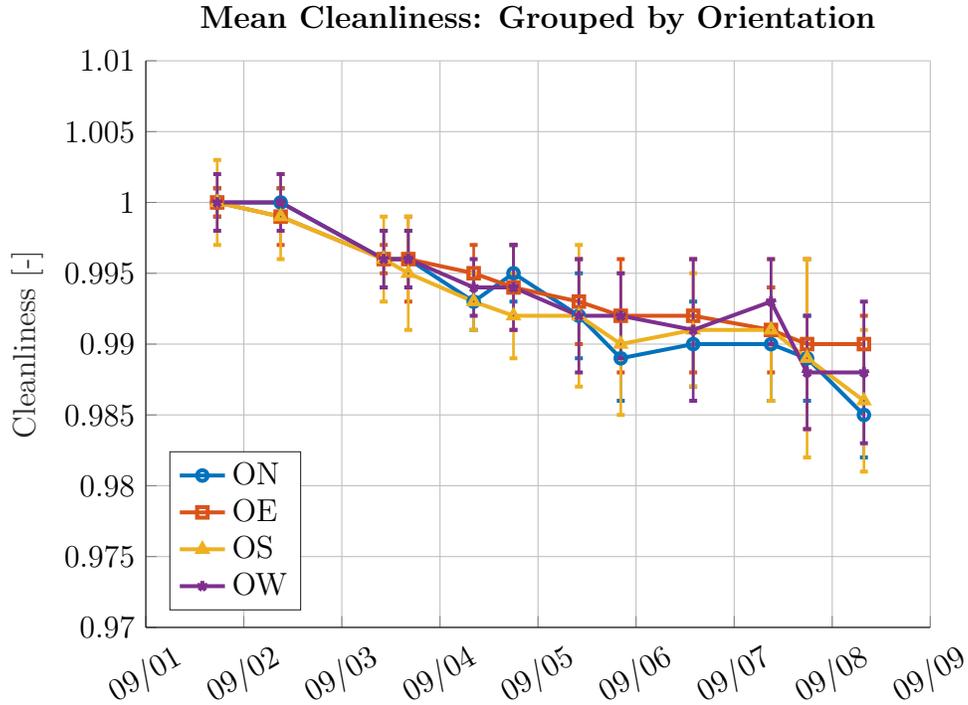
\begin{figure}[h!]
            \centering
\definecolor{mycolor1}{rgb}{0.00000,0.44700,0.74100}%
\definecolor{mycolor2}{rgb}{0.85000,0.32500,0.09800}%
\definecolor{mycolor3}{rgb}{0.92900,0.69400,0.12500}%
\definecolor{mycolor4}{rgb}{0.49400,0.18400,0.55600}%
\begin{tikzpicture}[%
trim axis left, trim axis right
]

\begin{axis}[%
width=0.951\fwidth,
height=\fheight,
at={(0\fwidth,0\fheight)},
scale only axis,
xtick scale label code/.code={},
xmin=738035.000,
xmax=738043.000,
xtick={738035.000,738036.000,738037.000,738038.000,738039.000,738040.000,738041.000,738042.000,738043.000},
xticklabels={{09/01},{09/02},{09/03},{09/04},{09/05},{09/06},{09/07},{09/08},{09/09}},
xticklabel style={rotate=30},
ymin=0.97,
ymax=1.01, 
y tick label style={/pgf/number format/fixed, /pgf/number format/precision=5},
ylabel style={font=\color{white!15!black}},
ylabel={$\text{Cleanliness [-]}$},
axis background/.style={fill=white},
title style={font=\bfseries},
title={Mean Cleanliness: Grouped by Orientation},
axis x line*=bottom,
axis y line*=left,
xmajorgrids,
ymajorgrids,
legend pos = south west,
legend style={legend cell align=left, align=left, draw=white!15!black}
]
\addplot [color=mycolor1, line width=1.5pt, mark=o, mark options={solid, mycolor1}]
 plot [error bars/.cd, y dir=both, y explicit, error bar style={line width=1.0pt}, error mark options={line width=1.5pt, mark size=1.5pt, rotate=90}]
 table[row sep=crcr, y error plus index=2, y error minus index=3]{%
738035.729	1.000	0.001	0.001\\
738036.375	1.000	0.001	0.001\\
738037.427	0.996	0.002	0.002\\
738037.677	0.996	0.002	0.002\\
738038.344	0.993	0.002	0.002\\
738038.750	0.995	0.002	0.002\\
738039.417	0.992	0.003	0.003\\
738039.844	0.989	0.003	0.003\\
738040.583	0.990	0.003	0.003\\
738041.375	0.990	0.004	0.004\\
738041.743	0.989	0.003	0.003\\
738042.323	0.985	0.003	0.003\\
};
\addlegendentry{ON}

\addplot [color=mycolor2, line width=1.5pt, mark=square, mark options={solid, mycolor2}]
 plot [error bars/.cd, y dir=both, y explicit, error bar style={line width=1.0pt}, error mark options={line width=1.5pt, mark size=1.5pt, rotate=90}]
 table[row sep=crcr, y error plus index=2, y error minus index=3]{%
738035.729	1.000	0.001	0.001\\
738036.375	0.999	0.002	0.002\\
738037.427	0.996	0.001	0.001\\
738037.677	0.996	0.003	0.003\\
738038.344	0.995	0.002	0.002\\
738038.750	0.994	0.003	0.003\\
738039.417	0.993	0.003	0.003\\
738039.844	0.992	0.004	0.004\\
738040.583	0.992	0.004	0.004\\
738041.375	0.991	0.003	0.003\\
738041.743	0.990	0.006	0.006\\
738042.323	0.990	0.002	0.002\\
};
\addlegendentry{OE}

\addplot [color=mycolor3, line width=1.5pt, mark=triangle, mark options={solid, mycolor3}]
 plot [error bars/.cd, y dir=both, y explicit, error bar style={line width=1.0pt}, error mark options={line width=1.5pt, mark size=1.5pt, rotate=90}]
 table[row sep=crcr, y error plus index=2, y error minus index=3]{%
738035.729	1.000	0.003	0.003\\
738036.375	0.999	0.003	0.003\\
738037.427	0.996	0.003	0.003\\
738037.677	0.995	0.004	0.004\\
738038.344	0.993	0.002	0.002\\
738038.750	0.992	0.003	0.003\\
738039.417	0.992	0.005	0.005\\
738039.844	0.990	0.005	0.005\\
738040.583	0.991	0.004	0.004\\
738041.375	0.991	0.005	0.005\\
738041.743	0.989	0.007	0.007\\
738042.323	0.986	0.005	0.005\\
};
\addlegendentry{OS}

\addplot [color=mycolor4, line width=1.5pt, mark=star, mark options={solid, mycolor4}]
 plot [error bars/.cd, y dir=both, y explicit, error bar style={line width=1.0pt}, error mark options={line width=1.5pt, mark size=1.5pt, rotate=90}]
 table[row sep=crcr, y error plus index=2, y error minus index=3]{%
738035.729	1.000	0.002	0.002\\
738036.375	1.000	0.002	0.002\\
738037.427	0.996	0.002	0.002\\
738037.677	0.996	0.002	0.002\\
738038.344	0.994	0.002	0.002\\
738038.750	0.994	0.003	0.003\\
738039.417	0.992	0.004	0.004\\
738039.844	0.992	0.003	0.003\\
738040.583	0.991	0.005	0.005\\
738041.375	0.993	0.003	0.003\\
738041.743	0.988	0.004	0.004\\
738042.323	0.988	0.005	0.005\\
};
\addlegendentry{OW}

\end{axis}

\begin{axis}[%
width=1.322\fwidth,
height=1.322\fheight,
at={(-0.196\fwidth,-0.246\fheight)},
scale only axis,
xmin=0.000,
xmax=1.000,
ymin=0.000,
ymax=1.000,
axis line style={draw=none},
ticks=none,
axis x line*=bottom,
axis y line*=left
]
\end{axis}
\end{tikzpicture}%   
            \caption{Mean cleanliness of tilt angles $ \left( 5,30,60,85 \right) $ with respect to orientation (North - ON, East - OE, South - OS, West - OW)}
            \label{fig:orientationcleanlinessmeasures}
        \end{figure}
            
        \subsubsection{Ground soil composition}
        Main components of the local soil were determined by collecting two samples of dust from the ground in proximity of the mirror test-rig (as a proxy to the airborne dust components). The samples were analysed with an \gls{xrd} technique at the \gls{carf} within \gls{qut}. Results showed the main crystallline phase is represented by Quartz ($\text{SiO}_{2}$) which accounts for 40 to 47 \% by weight in addition to 12 to 21 \% by weight of amorphous content. This is in accordance with the assumption adopted in the \gls{dsm} that dust is assumed to be silica with a density of \SI{2000}{\kilo\gram\per\metre\cubed}. 
        
        \subsubsection{Soiling Model Parameter Fitting and Predictions}
        \label{subsec:soilingpredictions}
        The reflectance measurements and weather station measurements were used to estimate the free parameter \gls{hrz0} of the \gls{dsm}. It should be noted that the \SI{0}{\degree} and \SI{90}{\degree} tilted mirrors have been excluded from the fitting procedure as they only exist on the North and East axis respectively. Table~\ref{tab:mseandsoilingrates} shows the \gls{mse} and soiling rates for all tested fixed tilt angles. The fitted \gls{mse} shows excellent agreement between predicted and measured reflectance values, although this is facilitated by the accurate knowledge of the initial reflectance value which may not be available for every heliostat in the field. 
        
        \begin{table}[htp]
            \centering
            \caption{Mean squared error, predicted and experimental soiling rates, and fitted \ta{hrz0} for all tilt angles}
            \label{tab:mseandsoilingrates}
            \begin{tabular}{ccccc}
                \toprule
                \multicolumn{1}{p{0.05\columnwidth}}{\centering \textbf{Tilt} (\SI{}{\degree})} & \multicolumn{1}{p{0.1\columnwidth}}{\centering \textbf{MSE} (\SI{}{\percent})} & \multicolumn{1}{p{0.25\columnwidth}}{ \centering \textbf{Predicted SR (\SI{}{pp\per\day})}} & \multicolumn{1}{p{0.3\columnwidth}}{\centering \textbf{Experimental SR} (\SI{}{pp\per\day})} & \textbf{\ta{hrz0}}\\
                \midrule
                0 & 3.4 & 0.35 & 0.40 & -\\
                5 & 2.3 & 0.33 & 0.33 & 6.6\\
                30 & 1.2 & 0.29 & 0.30 & 5.9\\
                60 & 1.3 & 0.17 & 0.17 & 5.7\\
                85 & 0.26 & 0.003 & 0.002 & 5.0\\
                \bottomrule
            \end{tabular}
        \end{table}
        Table~\ref{tab:mseandsoilingrates} also shows the fitted \gls{hrz0} values using the mean reflectance for each of the tested tilt angles. According to literature \cite{McRae1982a} the surface roughness for a scrub or open area is $z_0 \approx \SI{0.2}{\metre}$ and $z_0 \approx \SI{1}{\metre}$ for tree covered areas (which is representative of the surrounding environment at the Mount Isa site). Considering the reference wind measurement height of $h_r = \SI{3}{\metre}$, \gls{hrz0} should be within $3-15$. Thus the fitted and mean \gls{hrz0} values reported are consistent with this back-of-the-envelope analysis and future predictions will use the mean \gls{hrz0} of $5.8$. Though there is a modest trend between tilt angle and \gls{hrz0} values it will be assumed to be constant and the topic of future studies. 
        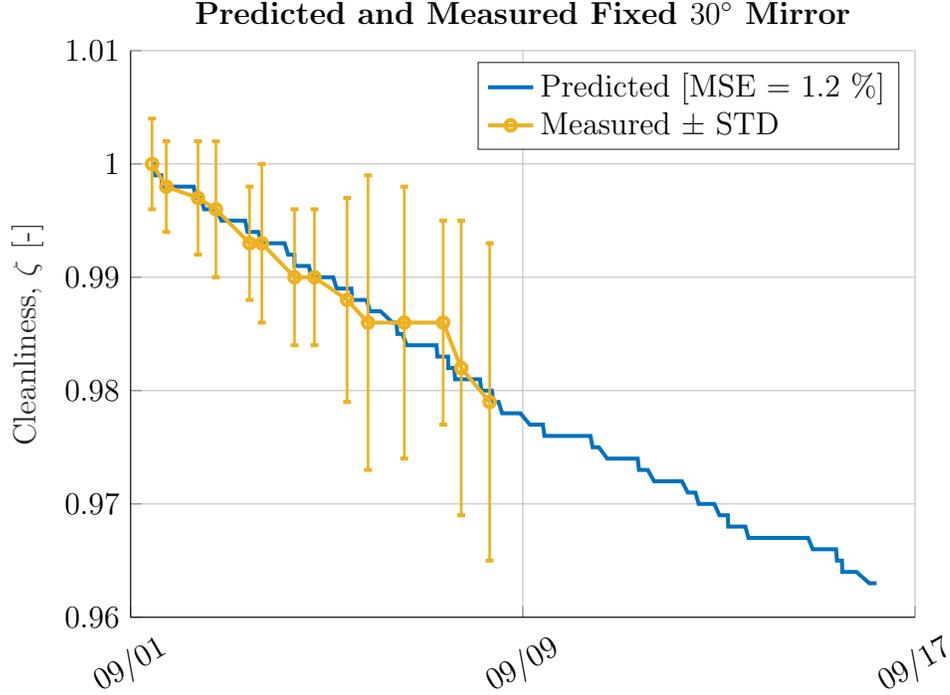
\begin{figure}[h]
            \centering
\definecolor{mycolor1}{rgb}{0.00000,0.44700,0.74100}%
\definecolor{mycolor2}{rgb}{0.92900,0.69400,0.12500}%
\begin{tikzpicture}[%
trim axis left, trim axis right
]

\begin{axis}[%
width=0.951\fwidth,
height=\fheight,
at={(0\fwidth,0\fheight)},
scale only axis,
xtick scale label code/.code={},
xmin=738035.000,
xmax=738051.000,
xtick={738035.000,738043.000,738051.000},
xticklabels={{09/01},{09/09},{09/17}},
xticklabel style={rotate=30},
ymin=0.960,
ymax=1.010,
ytick={0.960,0.970,0.980,0.990,1.000,1.010},
ylabel style={font=\color{white!15!black}},
ylabel={Cleanliness, \ta{cleanliness} [-]},
axis background/.style={fill=white},
title style={font=\bfseries},
title={$\text{Predicted and Measured Fixed \SI{30}{\degree} Mirror}$},
axis x line*=bottom,
axis y line*=left,
xmajorgrids,
ymajorgrids,
legend style={legend cell align=left, align=left, draw=white!15!black}
]
\addplot [color=mycolor1, line width=1.5pt]
  table[row sep=crcr]{%
738035.438	1.000\\
738035.444	1.000\\
738035.451	1.000\\
738035.479	1.000\\
738035.486	1.000\\
738035.500	0.999\\
738035.552	0.999\\
738035.556	0.999\\
738035.559	0.999\\
738035.566	0.999\\
738035.583	0.999\\
738035.590	0.999\\
738035.594	0.999\\
738035.625	0.999\\
738035.628	0.999\\
738035.642	0.998\\
738035.646	0.998\\
738035.656	0.998\\
738035.670	0.998\\
738035.736	0.998\\
738035.847	0.998\\
738036.083	0.998\\
738036.222	0.998\\
738036.274	0.998\\
738036.288	0.998\\
738036.326	0.997\\
738036.344	0.997\\
738036.347	0.997\\
738036.399	0.997\\
738036.427	0.997\\
738036.500	0.996\\
738036.552	0.996\\
738036.608	0.996\\
738036.719	0.996\\
738036.743	0.996\\
738036.747	0.996\\
738036.851	0.995\\
738036.920	0.995\\
738037.052	0.995\\
738037.177	0.995\\
738037.264	0.995\\
738037.337	0.995\\
738037.385	0.994\\
738037.434	0.994\\
738037.444	0.994\\
738037.465	0.994\\
738037.517	0.994\\
738037.601	0.994\\
738037.604	0.994\\
738037.681	0.993\\
738037.826	0.993\\
738037.958	0.993\\
738038.024	0.993\\
738038.073	0.993\\
738038.142	0.993\\
738038.212	0.992\\
738038.229	0.992\\
738038.306	0.992\\
738038.337	0.992\\
738038.340	0.992\\
738038.344	0.991\\
738038.392	0.991\\
738038.549	0.991\\
738038.639	0.991\\
738038.684	0.990\\
738038.826	0.990\\
738038.861	0.990\\
738038.955	0.990\\
738039.132	0.990\\
738039.201	0.989\\
738039.302	0.989\\
738039.434	0.989\\
738039.465	0.989\\
738039.479	0.989\\
738039.503	0.989\\
738039.517	0.988\\
738039.552	0.988\\
738039.566	0.988\\
738039.622	0.988\\
738039.684	0.988\\
738039.750	0.988\\
738039.816	0.988\\
738039.882	0.987\\
738039.927	0.987\\
738040.010	0.987\\
738040.097	0.987\\
738040.344	0.986\\
738040.392	0.986\\
738040.413	0.986\\
738040.424	0.986\\
738040.441	0.985\\
738040.444	0.985\\
738040.451	0.985\\
738040.462	0.985\\
738040.503	0.985\\
738040.542	0.985\\
738040.635	0.984\\
738040.653	0.984\\
738040.729	0.984\\
738041.240	0.984\\
738041.257	0.983\\
738041.299	0.983\\
738041.340	0.983\\
738041.368	0.983\\
738041.372	0.983\\
738041.385	0.983\\
738041.389	0.983\\
738041.434	0.983\\
738041.444	0.983\\
738041.476	0.983\\
738041.479	0.982\\
738041.503	0.982\\
738041.507	0.982\\
738041.531	0.982\\
738041.538	0.982\\
738041.542	0.982\\
738041.559	0.982\\
738041.576	0.982\\
738041.594	0.982\\
738041.601	0.982\\
738041.618	0.981\\
738041.642	0.981\\
738041.760	0.981\\
738041.764	0.981\\
738041.771	0.981\\
738041.778	0.981\\
738041.795	0.981\\
738042.111	0.981\\
738042.125	0.981\\
738042.128	0.981\\
738042.153	0.980\\
738042.212	0.980\\
738042.302	0.980\\
738042.309	0.980\\
738042.316	0.980\\
738042.319	0.980\\
738042.333	0.980\\
738042.337	0.980\\
738042.358	0.980\\
738042.368	0.980\\
738042.403	0.979\\
738042.413	0.979\\
738042.427	0.979\\
738042.441	0.979\\
738042.479	0.979\\
738042.507	0.979\\
738042.576	0.978\\
738042.583	0.978\\
738042.615	0.978\\
738042.660	0.978\\
738042.677	0.978\\
738042.792	0.978\\
738042.795	0.978\\
738042.858	0.978\\
738042.951	0.978\\
738043.142	0.977\\
738043.330	0.977\\
738043.337	0.977\\
738043.354	0.977\\
738043.406	0.977\\
738043.410	0.977\\
738043.441	0.976\\
738043.521	0.976\\
738043.601	0.976\\
738043.962	0.976\\
738043.997	0.976\\
738044.212	0.976\\
738044.326	0.976\\
738044.333	0.976\\
738044.375	0.976\\
738044.378	0.976\\
738044.424	0.975\\
738044.427	0.975\\
738044.431	0.975\\
738044.465	0.975\\
738044.479	0.975\\
738044.490	0.975\\
738044.528	0.975\\
738044.531	0.975\\
738044.549	0.975\\
738044.722	0.974\\
738044.806	0.974\\
738045.149	0.974\\
738045.215	0.974\\
738045.253	0.974\\
738045.323	0.974\\
738045.344	0.974\\
738045.368	0.973\\
738045.389	0.973\\
738045.497	0.973\\
738045.517	0.973\\
738045.521	0.973\\
738045.549	0.973\\
738045.677	0.972\\
738045.913	0.972\\
738046.104	0.972\\
738046.247	0.972\\
738046.365	0.971\\
738046.375	0.971\\
738046.396	0.971\\
738046.413	0.971\\
738046.479	0.971\\
738046.521	0.971\\
738046.587	0.970\\
738046.628	0.970\\
738046.771	0.970\\
738046.854	0.970\\
738046.875	0.970\\
738046.906	0.970\\
738047.010	0.969\\
738047.031	0.969\\
738047.059	0.969\\
738047.181	0.969\\
738047.188	0.968\\
738047.194	0.968\\
738047.219	0.968\\
738047.340	0.968\\
738047.424	0.968\\
738047.441	0.968\\
738047.538	0.968\\
738047.601	0.967\\
738047.642	0.967\\
738047.688	0.967\\
738048.330	0.967\\
738048.378	0.967\\
738048.385	0.967\\
738048.427	0.967\\
738048.483	0.967\\
738048.601	0.967\\
738048.698	0.967\\
738048.819	0.967\\
738048.913	0.966\\
738049.052	0.966\\
738049.319	0.966\\
738049.326	0.966\\
738049.337	0.966\\
738049.389	0.966\\
738049.406	0.965\\
738049.413	0.965\\
738049.438	0.965\\
738049.472	0.965\\
738049.476	0.965\\
738049.479	0.965\\
738049.490	0.965\\
738049.507	0.965\\
738049.510	0.965\\
738049.517	0.964\\
738049.521	0.964\\
738049.542	0.964\\
738049.628	0.964\\
738049.653	0.964\\
738049.660	0.964\\
738049.691	0.964\\
738049.694	0.964\\
738049.806	0.964\\
738050.069	0.963\\
738050.115	0.963\\
738050.215	0.963\\
};
\addlegendentry{Predicted [MSE = 1.2 \%]}

\addplot [color=mycolor2, line width=1.5pt, mark=o, mark options={solid, mycolor2}]
 plot [error bars/.cd, y dir=both, y explicit, error bar style={line width=1pt}, error mark options={line width=1.5pt, mark size=1.5pt, rotate=90}]
 table[row sep=crcr, y error plus index=2, y error minus index=3]{%
738035.438	1.000	0.004	0.004\\
738035.729	0.998	0.004	0.004\\
738036.375	0.997	0.005	0.005\\
738036.740	0.996	0.006	0.006\\
738037.427	0.993	0.005	0.005\\
738037.677	0.993	0.007	0.007\\
738038.344	0.990	0.006	0.006\\
738038.750	0.990	0.006	0.006\\
738039.417	0.988	0.009	0.009\\
738039.844	0.986	0.013	0.013\\
738040.583	0.986	0.012	0.012\\
738041.375	0.986	0.009	0.009\\
738041.743	0.982	0.013	0.013\\
738042.323	0.979	0.014	0.014\\
};
\addlegendentry{Measured ± STD}

\end{axis}

\begin{axis}[%
width=1.227\fwidth,
height=1.227\fheight,
at={(-0.16\fwidth,-0.135\fheight)},
scale only axis,
xmin=0.000,
xmax=1.000,
ymin=0.000,
ymax=1.000,
axis line style={draw=none},
ticks=none,
axis x line*=bottom,
axis y line*=left
]
\end{axis}
\end{tikzpicture}%
            \caption{Predicted and measured cleanliness of a \SI{30}{\degree} tilted mirror using \gls{hrz0}$ = 5.8$ and site meteorological data}
            \label{fig:predictedcleanliness}
        \end{figure}
        With the fitted \gls{hrz0} the \gls{dsm} can be used to predict soiling losses for not only the fixed tilt mirrors at the Mount Isa site, but also for hypothetical heliostats that follow a receiver aiming strategy. An example of future soiling loss predictions of a fixed \SI{30}{\degree} mirror is given in Fig.~\ref{fig:predictedcleanliness}. 

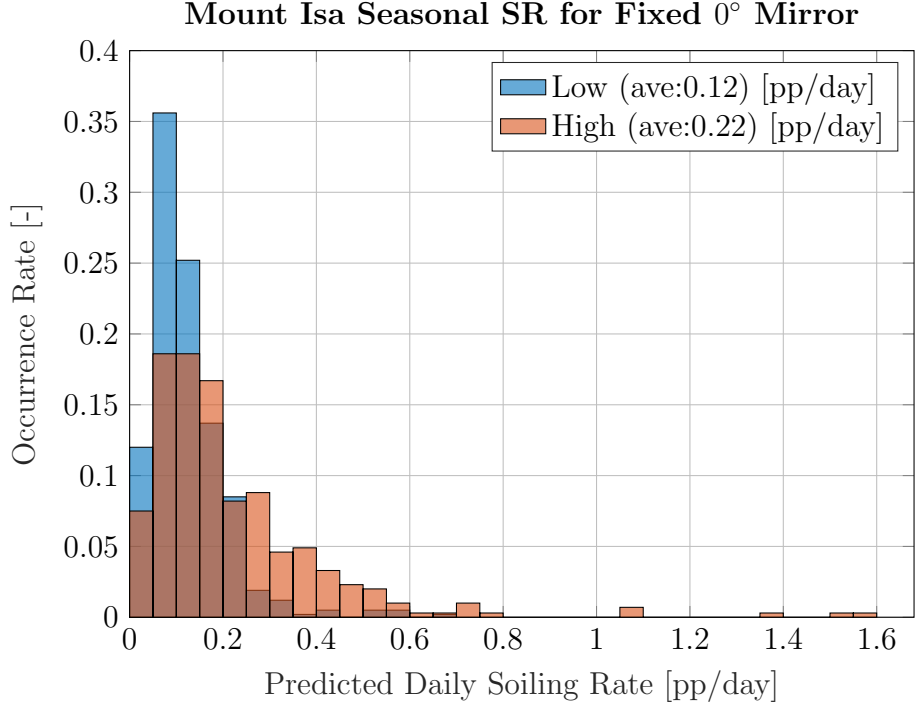
\begin{figure}[h]
            \centering
\definecolor{mycolor1}{rgb}{0.00000,0.44700,0.74100}%
\definecolor{mycolor2}{rgb}{0.85000,0.32500,0.09800}%
\begin{tikzpicture}[%
trim axis left, trim axis right
]

\begin{axis}[%
width=0.951\fwidth,
height=\fheight,
at={(0\fwidth,0\fheight)},
scale only axis,
xtick scale label code/.code={},
xmin=0,
xmax=1.680,
xlabel style={font=\color{white!15!black}},
xlabel={Predicted Daily Soiling Rate [pp/day]},
ymin=0.000,
ymax=0.400,
ylabel style={font=\color{white!15!black}},
ylabel={Occurrence Rate [-]},
tick label style={/pgf/number format/fixed},
axis background/.style={fill=white},
title style={font=\bfseries},
title={Mount Isa Seasonal SR for Fixed \SI{0}{\degree} Mirror},
xmajorgrids,
xminorgrids,
ymajorgrids,
yminorgrids,
legend style={legend cell align=left, align=left, draw=white!15!black}
]
\addplot[ybar interval, fill=mycolor1, fill opacity=0.600, draw=black, area legend] table[row sep=crcr] {%
x	y\\
0.000	0.120\\
0.050	0.356\\
0.100	0.252\\
0.150	0.137\\
0.200	0.085\\
0.250	0.019\\
0.300	0.012\\
0.350	0.002\\
0.400	0.005\\
0.450	0.000\\
0.500	0.005\\
0.550	0.005\\
0.600	0.000\\
0.650	0.002\\
0.700	0.002\\
};
\addlegendentry{Low (ave:0.12) [pp/day]}

\addplot[ybar interval, fill=mycolor2, fill opacity=0.600, draw=black, area legend] table[row sep=crcr] {%
x	y\\
0.000	0.075\\
0.050	0.186\\
0.100	0.186\\
0.150	0.167\\
0.200	0.082\\
0.250	0.088\\
0.300	0.046\\
0.350	0.049\\
0.400	0.033\\
0.450	0.023\\
0.500	0.020\\
0.550	0.010\\
0.600	0.003\\
0.650	0.003\\
0.700	0.010\\
0.750	0.003\\
0.800	0.000\\
0.850	0.000\\
0.900	0.000\\
0.950	0.000\\
1.000	0.000\\
1.050	0.007\\
1.100	0.000\\
1.150	0.000\\
1.200	0.000\\
1.250	0.000\\
1.300	0.000\\
1.350	0.003\\
1.400	0.000\\
1.450	0.000\\
1.500	0.003\\
1.550	0.003\\
1.600	0.003\\
};
\addlegendentry{High (ave:0.22) [pp/day]}

\end{axis}

\begin{axis}[%
width=1.227\fwidth,
height=1.227\fheight,
at={(-0.16\fwidth,-0.135\fheight)},
scale only axis,
xmin=0.000,
xmax=1.000,
ymin=0.000,
ymax=1.000,
axis line style={draw=none},
ticks=none,
axis x line*=bottom,
axis y line*=left
]
\end{axis}
\end{tikzpicture}%
            \caption{Mount Isa site predicted seasonal soiling rate for a \SI{0}{\degree} fixed-tilt angle mirror with $\ta{hrz0}=5.8$, high season between August and December, and low season between January and June.}
            \label{fig:seasonalsr}
        \end{figure}

        The predicted soiling rates are shown in Fig.~\ref{fig:seasonalsr} for the low and high \gls{tsp} seasons identified in Fig.~\ref{fig:seasonalcomponentcheck} left for a \SI{0}{\degree} tilted fixed mirror. With the high season defined as the period where the mean seasonal \gls{tsp} is above $0$ and low being below that. High season soiling rates have a higher mean of 0.22~pp/d and variance relative to the $0.12$~pp/d low season. Note that predicted high seasonal soiling rates are lower then that reported in Table~\ref{tab:mseandsoilingrates} during the experimental campaign ($0.35$ pp/day) as measurements were carried near the peak of the high season. 
    \subsubsection{Extension and Sampling of Meteorological Trajectories}
    \label{subsec:meteoextension}
    % Mount Isa Site 
    % DNI from satellite In addition, 26 years of satellite imagery derived \gls{dni} was sourced from the National Climate Centre at Australia's Bureau of Meteorology . 
    % Wind Speed and Ambient Temperature
    % TSP
    Soiling rate predictions and optimization of cleaning resources require long-term meteorological data. However, the available meteorological data from the Mount Isa CSP site covers just over two years, which is insufficient for optimizing cleaning resources under various meteorological scenarios. To address this limitation and extend the \gls{tsp} and wind speed time series to a full ten years, the time series analysis approach detailed in section~\ref{subsubsec:characterization} was employed.

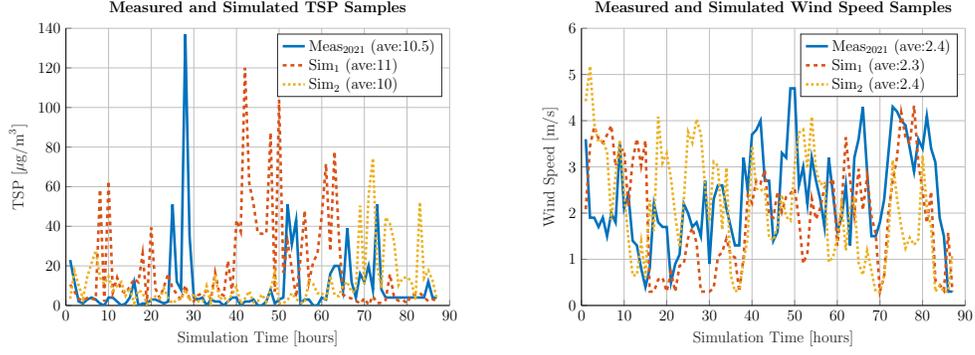
\begin{figure*}
        \centering
        \begin{subfigure}[c]{0.49\textwidth}
            \resizebox{\linewidth}{!}{            
\definecolor{mycolor1}{rgb}{0.00000,0.44700,0.74100}%
\definecolor{mycolor2}{rgb}{0.85000,0.32500,0.09800}%
\definecolor{mycolor3}{rgb}{0.92900,0.69400,0.12500}%
\begin{tikzpicture}[%
trim axis left, trim axis right
]

\begin{axis}[%
width=0.951\fwidth,
height=\fheight,
at={(0\fwidth,0\fheight)},
scale only axis,
xmin=0.000,
xmax=90.000,
xlabel style={font=\color{white!15!black}},
xlabel={Simulation Time [hours]},
ymin=0.000,
ymax=140.000,
ylabel style={font=\color{white!15!black}},
ylabel={$\text{TSP [}\mu\text{g/m}^\text{3}\text{]}$},
axis background/.style={fill=white},
title style={font=\bfseries},
title={Measured and Simulated TSP Samples},
axis x line*=bottom,
axis y line*=left,
xmajorgrids,
ymajorgrids,
legend style={legend cell align=left, align=left, draw=white!15!black}
]
\addplot [color=mycolor1, line width=2.0pt]
  table[row sep=crcr]{%
1.000	23.000\\
2.000	12.000\\
3.000	2.000\\
4.000	1.000\\
5.000	3.000\\
6.000	4.000\\
7.000	3.000\\
8.000	1.000\\
9.000	0.000\\
10.000	4.000\\
11.000	4.000\\
13.000	0.000\\
14.000	1.000\\
15.000	4.000\\
16.000	12.000\\
17.000	0.000\\
18.000	1.000\\
19.000	1.000\\
20.000	3.000\\
21.000	3.000\\
23.000	1.000\\
24.000	2.000\\
25.000	51.000\\
26.000	12.000\\
27.000	8.000\\
28.000	137.000\\
29.000	35.000\\
30.000	4.000\\
31.000	3.000\\
32.000	4.000\\
33.000	0.000\\
34.000	3.000\\
35.000	2.000\\
36.000	2.000\\
37.000	0.000\\
39.000	4.000\\
40.000	4.000\\
41.000	0.000\\
42.000	2.000\\
43.000	2.000\\
44.000	3.000\\
45.000	0.000\\
46.000	0.000\\
47.000	2.000\\
48.000	8.000\\
49.000	1.000\\
50.000	3.000\\
51.000	4.000\\
52.000	51.000\\
53.000	31.000\\
54.000	43.000\\
55.000	1.000\\
56.000	3.000\\
57.000	3.000\\
58.000	0.000\\
59.000	0.000\\
60.000	4.000\\
61.000	3.000\\
62.000	16.000\\
63.000	20.000\\
64.000	20.000\\
65.000	8.000\\
66.000	39.000\\
67.000	12.000\\
68.000	4.000\\
69.000	16.000\\
70.000	12.000\\
71.000	20.000\\
72.000	8.000\\
73.000	51.000\\
74.000	8.000\\
75.000	4.000\\
84.000	4.000\\
85.000	12.000\\
86.000	4.000\\
87.000	4.000\\
};
\addlegendentry{$\text{Meas}_{\text{2021}}\text{ (ave:10.5)}$}

\addplot [color=mycolor2, dashed, line width=2.0pt]
  table[row sep=crcr]{%
1.000	10.555\\
2.000	7.062\\
3.000	3.395\\
4.000	2.296\\
5.000	3.801\\
6.000	3.430\\
7.000	2.902\\
8.000	58.606\\
9.000	1.989\\
10.000	62.808\\
11.000	6.560\\
12.000	14.096\\
13.000	7.627\\
14.000	6.103\\
15.000	9.574\\
16.000	13.493\\
17.000	11.759\\
18.000	26.208\\
19.000	1.583\\
20.000	39.324\\
21.000	8.603\\
22.000	3.767\\
23.000	9.572\\
24.000	16.053\\
25.000	8.547\\
26.000	5.917\\
27.000	2.479\\
28.000	10.254\\
29.000	3.227\\
30.000	2.157\\
31.000	5.213\\
32.000	6.084\\
33.000	7.278\\
34.000	3.098\\
35.000	16.618\\
36.000	6.085\\
37.000	5.044\\
38.000	22.585\\
39.000	6.025\\
40.000	43.153\\
41.000	37.368\\
42.000	120.035\\
43.000	61.740\\
44.000	51.342\\
45.000	36.589\\
46.000	36.208\\
47.000	36.393\\
48.000	87.303\\
49.000	7.040\\
50.000	103.669\\
51.000	16.447\\
52.000	32.105\\
53.000	18.280\\
54.000	4.087\\
55.000	23.299\\
56.000	47.927\\
57.000	11.953\\
58.000	21.927\\
59.000	26.102\\
60.000	45.238\\
61.000	71.533\\
62.000	27.441\\
63.000	77.750\\
64.000	35.816\\
65.000	5.370\\
66.000	4.286\\
67.000	3.502\\
68.000	2.540\\
69.000	0.892\\
70.000	3.708\\
71.000	0.244\\
72.000	4.729\\
73.000	1.132\\
74.000	1.416\\
75.000	2.817\\
76.000	15.898\\
77.000	3.806\\
78.000	2.417\\
79.000	2.874\\
80.000	1.051\\
81.000	5.305\\
82.000	5.749\\
83.000	6.344\\
85.000	2.105\\
86.000	3.123\\
87.000	2.443\\
};
\addlegendentry{$\text{Sim}_\text{1}\text{ (ave:11)}$}

\addplot [color=mycolor3, dotted, line width=2.0pt]
  table[row sep=crcr]{%
1.000	3.406\\
2.000	17.448\\
3.000	10.142\\
4.000	3.419\\
5.000	14.387\\
6.000	21.244\\
7.000	27.783\\
8.000	14.126\\
9.000	10.921\\
10.000	14.708\\
11.000	7.489\\
12.000	9.777\\
13.000	11.856\\
14.000	4.954\\
15.000	9.054\\
16.000	1.104\\
17.000	7.741\\
18.000	6.473\\
19.000	0.718\\
20.000	3.032\\
21.000	4.592\\
22.000	3.602\\
23.000	8.596\\
24.000	3.655\\
25.000	1.011\\
26.000	2.846\\
27.000	3.180\\
28.000	8.044\\
29.000	2.317\\
30.000	5.872\\
31.000	2.228\\
32.000	1.891\\
33.000	9.842\\
34.000	1.472\\
35.000	15.677\\
36.000	3.086\\
37.000	4.387\\
38.000	1.928\\
39.000	5.593\\
40.000	9.621\\
41.000	0.841\\
42.000	3.568\\
43.000	7.235\\
44.000	1.509\\
45.000	2.123\\
46.000	5.553\\
47.000	1.838\\
48.000	7.941\\
49.000	16.497\\
50.000	1.107\\
51.000	1.003\\
52.000	2.830\\
53.000	6.236\\
54.000	3.276\\
55.000	1.513\\
56.000	1.501\\
57.000	12.429\\
58.000	10.826\\
59.000	6.134\\
60.000	6.501\\
61.000	2.267\\
62.000	15.024\\
63.000	5.357\\
64.000	4.708\\
65.000	10.168\\
66.000	11.987\\
67.000	6.693\\
68.000	7.259\\
69.000	50.307\\
70.000	10.949\\
71.000	55.736\\
72.000	74.404\\
73.000	19.845\\
74.000	8.749\\
75.000	44.491\\
76.000	41.258\\
77.000	31.951\\
78.000	10.206\\
79.000	13.876\\
80.000	14.273\\
81.000	12.337\\
82.000	2.554\\
83.000	52.192\\
84.000	13.830\\
85.000	17.691\\
86.000	12.505\\
87.000	3.220\\
};
\addlegendentry{$\text{Sim}_\text{2}\text{ (ave:10)}$}

\end{axis}

\begin{axis}[%
width=1.227\fwidth,
height=1.227\fheight,
at={(-0.16\fwidth,-0.135\fheight)},
scale only axis,
xmin=0.000,
xmax=1.000,
ymin=0.000,
ymax=1.000,
axis line style={draw=none},
ticks=none,
axis x line*=bottom,
axis y line*=left
]
\end{axis}
\end{tikzpicture}%
            }
        \end{subfigure}
        \begin{subfigure}[c]{0.49\textwidth}
            \resizebox{\linewidth}{!}{
\definecolor{mycolor1}{rgb}{0.00000,0.44700,0.74100}%
\definecolor{mycolor2}{rgb}{0.85000,0.32500,0.09800}%
\definecolor{mycolor3}{rgb}{0.92900,0.69400,0.12500}%
\begin{tikzpicture}[%
trim axis left, trim axis right
]

\begin{axis}[%
width=0.951\fwidth,
height=\fheight,
at={(0\fwidth,0\fheight)},
scale only axis,
xmin=0.000,
xmax=90.000,
xlabel style={font=\color{white!15!black}},
xlabel={Simulation Time [hours]},
ymin=0.000,
ymax=6.000,
ylabel style={font=\color{white!15!black}},
ylabel={Wind Speed [m/s]},
axis background/.style={fill=white},
title style={font=\bfseries},
title={Measured and Simulated Wind Speed Samples},
axis x line*=bottom,
axis y line*=left,
xmajorgrids,
ymajorgrids,
legend style={legend cell align=left, align=left, draw=white!15!black}
]
\addplot [color=mycolor1, line width=2.0pt]
  table[row sep=crcr]{%
1.000	3.600\\
2.000	1.900\\
3.000	1.900\\
4.000	1.700\\
5.000	1.900\\
6.000	1.500\\
7.000	2.000\\
8.000	1.800\\
9.000	3.500\\
10.000	2.100\\
11.000	2.400\\
12.000	1.400\\
13.000	1.300\\
14.000	0.800\\
15.000	0.400\\
16.000	0.800\\
17.000	2.200\\
18.000	1.800\\
19.000	1.700\\
20.000	1.700\\
21.000	0.500\\
22.000	0.900\\
23.000	1.100\\
24.000	2.200\\
25.000	2.000\\
26.000	1.700\\
27.000	1.800\\
28.000	1.500\\
29.000	2.700\\
30.000	0.900\\
31.000	2.300\\
32.000	2.600\\
33.000	2.600\\
34.000	2.100\\
36.000	1.300\\
37.000	1.300\\
38.000	3.200\\
39.000	2.500\\
40.000	3.700\\
41.000	3.800\\
42.000	4.000\\
43.000	2.700\\
44.000	2.700\\
45.000	1.400\\
46.000	1.600\\
47.000	3.300\\
48.000	3.200\\
49.000	4.700\\
50.000	4.700\\
51.000	2.600\\
52.000	3.000\\
53.000	2.200\\
54.000	3.200\\
55.000	2.700\\
56.000	2.300\\
57.000	1.700\\
58.000	3.200\\
60.000	1.600\\
61.000	1.900\\
62.000	2.600\\
63.000	1.300\\
64.000	3.200\\
65.000	3.600\\
66.000	4.300\\
68.000	1.500\\
69.000	1.500\\
70.000	1.800\\
71.000	2.300\\
73.000	4.300\\
74.000	4.200\\
75.000	4.000\\
76.000	3.900\\
78.000	2.900\\
79.000	3.600\\
80.000	3.400\\
81.000	4.100\\
82.000	3.400\\
83.000	3.100\\
84.000	1.900\\
85.000	1.500\\
86.000	0.300\\
87.000	0.300\\
};
\addlegendentry{$\text{Meas}_{\text{2021}}\text{ (ave:2.4)}$}

\addplot [color=mycolor2, dashed, line width=2.0pt]
  table[row sep=crcr]{%
1.000	2.084\\
2.000	3.448\\
3.000	3.873\\
4.000	3.614\\
5.000	3.521\\
6.000	3.694\\
7.000	3.897\\
8.000	3.018\\
9.000	3.362\\
10.000	3.373\\
11.000	2.141\\
12.000	3.141\\
13.000	3.608\\
14.000	3.025\\
15.000	3.572\\
16.000	0.300\\
17.000	0.300\\
18.000	0.562\\
19.000	0.603\\
20.000	0.300\\
21.000	0.717\\
22.000	0.300\\
23.000	0.858\\
24.000	0.906\\
25.000	1.461\\
26.000	1.680\\
27.000	1.352\\
28.000	0.300\\
30.000	0.300\\
31.000	0.373\\
32.000	1.052\\
33.000	1.156\\
34.000	1.448\\
35.000	1.091\\
36.000	1.040\\
37.000	0.300\\
38.000	1.029\\
39.000	2.386\\
40.000	2.215\\
41.000	2.712\\
42.000	2.641\\
43.000	2.931\\
44.000	1.955\\
45.000	1.903\\
46.000	2.465\\
47.000	2.030\\
48.000	2.349\\
49.000	2.422\\
50.000	2.524\\
51.000	1.872\\
52.000	0.905\\
53.000	1.142\\
54.000	2.310\\
55.000	1.864\\
56.000	2.110\\
57.000	2.558\\
58.000	2.612\\
59.000	2.746\\
60.000	1.299\\
61.000	2.135\\
62.000	3.641\\
63.000	2.721\\
64.000	1.967\\
65.000	2.954\\
66.000	2.066\\
67.000	2.795\\
68.000	1.936\\
69.000	1.561\\
70.000	0.300\\
71.000	0.641\\
72.000	2.018\\
73.000	2.694\\
74.000	3.199\\
75.000	4.170\\
76.000	3.644\\
77.000	3.494\\
78.000	4.326\\
79.000	3.098\\
80.000	2.476\\
81.000	1.787\\
82.000	2.467\\
83.000	0.540\\
84.000	0.454\\
85.000	0.322\\
86.000	1.572\\
87.000	0.300\\
};
\addlegendentry{$\text{Sim}_\text{1}\text{ (ave:2.3)}$}

\addplot [color=mycolor3, dotted, line width=2.0pt]
  table[row sep=crcr]{%
1.000	4.422\\
2.000	5.180\\
3.000	3.937\\
4.000	3.533\\
5.000	3.574\\
6.000	3.320\\
7.000	1.853\\
8.000	2.823\\
9.000	3.576\\
10.000	2.428\\
11.000	1.453\\
12.000	0.649\\
13.000	0.661\\
14.000	1.380\\
15.000	0.607\\
16.000	2.282\\
17.000	1.610\\
18.000	4.082\\
19.000	3.085\\
20.000	3.295\\
21.000	2.740\\
22.000	1.949\\
23.000	1.260\\
24.000	2.282\\
25.000	3.758\\
26.000	3.523\\
27.000	4.030\\
28.000	3.706\\
29.000	2.551\\
30.000	1.820\\
31.000	3.108\\
32.000	3.025\\
33.000	1.643\\
34.000	2.981\\
35.000	0.666\\
36.000	0.300\\
37.000	0.352\\
38.000	0.950\\
39.000	1.482\\
40.000	3.453\\
41.000	2.533\\
42.000	2.482\\
43.000	2.546\\
44.000	1.474\\
45.000	1.407\\
46.000	1.720\\
47.000	2.393\\
48.000	1.919\\
49.000	1.840\\
50.000	3.029\\
51.000	3.738\\
52.000	3.807\\
53.000	2.967\\
54.000	4.084\\
55.000	2.942\\
56.000	2.614\\
57.000	2.680\\
58.000	0.940\\
59.000	1.041\\
60.000	1.460\\
61.000	2.185\\
62.000	1.195\\
63.000	2.125\\
64.000	2.284\\
65.000	0.925\\
66.000	0.660\\
67.000	1.186\\
68.000	0.752\\
69.000	0.300\\
70.000	0.300\\
71.000	0.654\\
72.000	1.471\\
73.000	3.142\\
74.000	2.315\\
75.000	1.703\\
76.000	1.242\\
77.000	1.469\\
78.000	1.310\\
79.000	1.432\\
80.000	3.241\\
81.000	2.138\\
82.000	1.685\\
83.000	0.300\\
84.000	0.429\\
85.000	0.300\\
86.000	0.300\\
87.000	1.122\\
};
\addlegendentry{$\text{Sim}_\text{2}\text{ (ave:2.4)}$}

\end{axis}

\begin{axis}[%
width=1.227\fwidth,
height=1.227\fheight,
at={(-0.16\fwidth,-0.135\fheight)},
scale only axis,
xmin=0.000,
xmax=1.000,
ymin=0.000,
ymax=1.000,
axis line style={draw=none},
ticks=none,
axis x line*=bottom,
axis y line*=left
]
\end{axis}
\end{tikzpicture}%
            }
        \end{subfigure}
        \caption{Left: TSP site data and AR(24) TSP simulation snapshots. Right: Wind speed site data and AR(9) wind speed simulation snapshots.}
        \label{fig:arpredictions}
    \end{figure*}

    By utilizing Akaike's Information Criterion, an AR(24) \gls{tsp} model and an AR(9) wind speed model was selected. Fig.~\ref{fig:arpredictions} provides a preview of the simulations. Comparing the mean values and shape characteristics of the site measurements with the AR simulations revealed sufficient agreement to justify extending the meteorological trajectories.

    Temperature recordings for the ten year trajectories were made up of the original two years of measurements and extended using a Mount Isa city measurement station, as both stations recorded similar values across the same time frame. Eight years of historical temperature data was used to impute the remaining temperature trajectories. 
    
    \Gls{dni} estimates from satellite imagery between 1989 and 2015 were sourced from the Bureau of Meteorology (BOM). The data was split and grouped by day of year, such that for each day of the year there were 26 days of hourly \gls{dni} recordings. For each trajectory and simulation day of year random hourly trajectories were drawn from the grouped data to impute the simulation year. 
        
    \subsection{CSP Plant Configuration}  
    \label{subsec:plantconfiguration}
    \subsubsection{Plant Design Parameters}
    The CSP plant used for the case study is a \SI{56}{\mega\watt} facility, integrated with a 14.5-hour thermal storage system and connected to a steam turbine with a similar capacity. The plant's thermal storage is designed to be charged by 30 polar square solar field-arrays, each consisting of 2389 heliostats per array. These heliostats are strategically positioned to efficiently harness Direct Normal Irradiance (DNI) and focus it towards a $\ta{qmax}=\SI{10}{\mega\watt}$ receiver system.

    During operation, the receiver system maintains a constant thermal loss of $\ta{qloss}=\SI{0.91}{\mega\watt}$. The CSP plant operates when solar zenith angles are below \SI{80}{\degree}, and the thermal receiver power exceeds the constant thermal receiver losses. The electrical power generation is assumed to run with a steady thermal-to-electricity efficiency of $\eta_{\text{pb}}=\SI{35}{\percent}$. The parameters required for the cleaning optimization process are outlined in table~\ref{tab:cspsystemdesign}.
    
    \begin{table}
        \centering
        \caption{CSP plant model values}
        \label{tab:cspsystemdesign}
        \begin{tabular}{lcr}
            \toprule
             Parameter & Value \\
             \midrule
             \Glsdesc{pbrated}, \ta{pbrated} & \SI{56}{\mega\watt}\\
             \Glsdesc{etapb}, \ta{etapb} & \SI{35}{\percent}\\
             Max storage charge, \ta{maxstoragecapacity} & \SI{2320}{\mega\watt\hour}\\
             Max receiver energy, \ta{qmax} & \SI{10}{\mega\watt}\\
             Constant receiver loss, \ta{qloss} & \SI{0.9}{\mega\watt}\\
             Solar field arrays & 30\\
             \# Heliostats per module  & 2389 \\
             \Glsdesc{areamirror}, \ta{areamirror} & \SI{6.4}{\metre\squared}\\
             \bottomrule
        \end{tabular}
    \end{table}
    
\begin{figure}[h!]
        \centering
        \includegraphics[width=0.951\columnwidth]{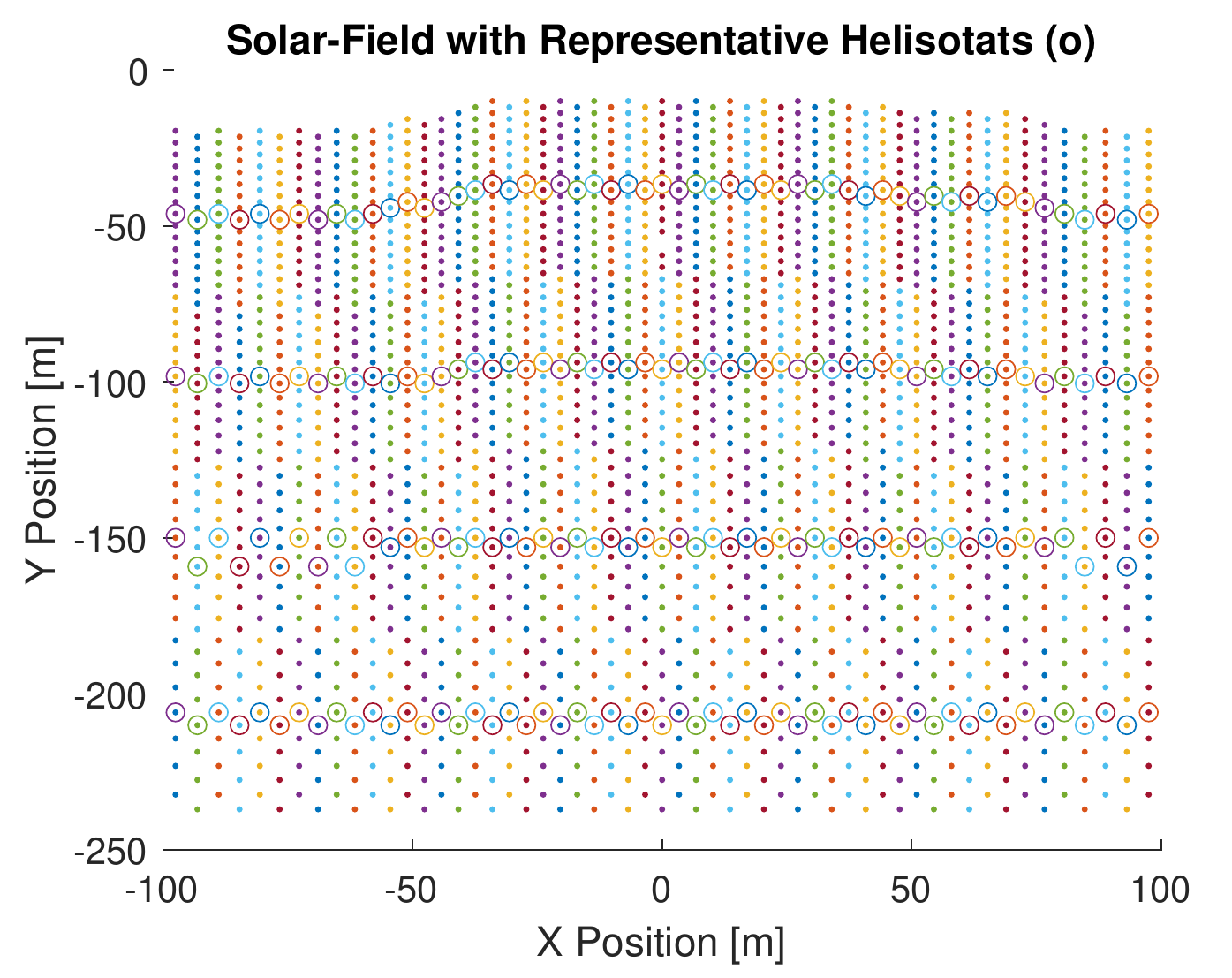}
        \caption{Solar field sectors with representative heliostats denoted as circle outlines, receiver centred at (0,0)}
        \label{fig:sectorisation}
    \end{figure}

    Each array's solar field is assumed to have the same layout as shown in Fig.~\ref{fig:sectorisation}. To reduce computational load each solar field was sectorised into a 55x4 grid with one representative heliostat per sector. Four representative heliostats were chosen for each of the fifty five cleaning corridors\footnote{Cleaning corridor is defined as the South to North route taken by cleaning crews to clean heliostats} based upon an even displacement position within the corridor. 
    
    SolarPILOT-v1.5.2 (\cite{Wagner2018}) was used to generate a optical efficiency look up table using the receiver design characteristics and heliostat field position for all feasible solar azimuth and zenith angles. Hourly optical efficiency values were interpolated for the simulated year using geographical coordinates of the site and the \citet{Michalsky1988} solar position algorithm. The optical efficiency was calculated for each heliostat of sector \ta{idsect} and simulation period \ta{idtime} and with the mean being used for the sector representative heliostat. 

    Hourly particle deposition rates, \ta{ndp} from the fitted \gls{dsm} across the ten meteorological trajectories were calculated using each representative heliostat's hourly tilt and incidence angle computed from the \citet{Guo2013} heliostat tracking algorithm. The resultant soiling rates of each sector representative were fed into the cleaning model of the cleaning optimizer to retrieve predicted reflectance trajectories for a given cleaning resource setup.
    
    \subsubsection{Cleaning Resource Parameters}   
    \begin{table}[h!]
        \centering
        \caption{Cleaning and cost model values}
        \label{tab:costcleanvalues}
        \begin{tabular}{lcr}
        \toprule
            Parameter & Value\\
        \midrule
            Tractor velocity (cleaning) &\SI{2}{\kilo\metre\per\hour}\\
            Tractor velocity (no cleaning) & \SI{10}{\kilo\metre\per\hour}\\
            Tractor work load & \SI{82}{\kilo\watt}\\
            Operator shift time & \SI{8}{\hour}\\
            Water consumption rate & \SI{0.4}{\litre\per\metre\squared}\\
            Fuel consumption rate & \SI{25}{\litre\per\hour}\\
            \Glsdesc{etacl}, \ta{etacl} & \SI{99}{\percent}\\
            \Glsdesc{pel}, \ta{pel} & \SI{140}{A\$\per\mega\watt\hour}\\
            Plant O\&M Costs, \ta{costoam} & \SI{3.5}{A\$\per\mega\watt\hour}\\
            \Glsdesc{costtruck}, \ta{costtruck} & \SI{300000}{A\$\per truck}\\
            \Glsdesc{depreciation}, \ta{depreciation} & \SI{10}{\year}\\
            \Glsdesc{costop}, \ta{costop} & \SI{80000}{A\$\per\year}\\
            \Glsdesc{costmaint}, \ta{costmaint} & \SI{15000}{A\$\per\year}\\
            \Glsdesc{costwf}, \ta{costwf} & \SI{0.0175}{A\$\per\metre\squared}\\
        \bottomrule
        \end{tabular}
    \end{table}
    
    Characteristics of the solar field geometry and operating conditions of the cleaning crew allow an estimation of the cleaning speed for a high pressure + brush operating cleaning crew. Assuming the crew cleans for \SI{8}{\hour} per shift at an average vehicle cleaning velocity of \SI{2}{\kilo\metre\per\hour} and non-cleaning travel velocity of \SI{10}{\kilo\metre\per\hour} it is expected that one solar field will be cleaned per crew per shift (heliostat area cleaning rate of \SI{2039}{\metre\squared\per\hour}) which also includes return trips for breaks and reloading fuel/water at the start of the day.
    
    For each cleaning event a crew will path through all cleaning corridors of a unique solar field, restarting the cleaning rotation once all heliostats of the power plant have been cleaned. Cleaning crews are comprised of one operator and a cleaning tractor\footnote{Fuel rate derived for \SI{82}{\kilo\watt} tractor \citep{Grisso2004}} with a high pressure spray and cleaning brush with \SI{7}{\kilo\litre} of water capacity. The high pressure wash and brush cleaning technology is assumed to remove $\ta{etacl}=\SI{99}{\percent}$ of particles for all diameters.  

    \subsubsection{Economic Assumptions}
    Determining the cleaning costs and soiling-induced-degradation costs requires economic assumptions about the cleaning crew, equipment, electricity and operation and maintenance costs for the power plant. These costs (see Table~\ref{tab:costcleanvalues}) include the tractor depreciation cost over the expected asset lifetime $(\ta{costtruck}/\ta{depreciation})$, \glsdesc{costmaint}, Operator salary. \Als{pel} of the plant has been assumed to be inline with the regions calculated cost of A\$\SI{140}{\per\mega\watt\hour} \citep{QueenslandGovernment2021}. Non-cleaning operation and maintenance costs of the plant are assumed to be A\$\SI{3.5}{\per\mega\watt\hour}. 
    
    \subsection{Operating Scenarios}
    \label{subsec:operationstrategies}
    The model parameters given in the previous section and ten meteorological trajectories have been combined with design and operating strategies to investigate the effects on \gls{tcc}. These operating scenarios include a base setup which will serve as a reference to compare against the three scenarios that change either the angle at which heliostats are stowed at night, dispatch policy, and the time at which heliostats are cleaned.
    
        \subsubsection{Base Setup}
        The base scenario is an operation strategy that assumes all thermal energy collected is converted into electricity as such the dispatch policy is the continuous production of available energy. Heliostats will have a near vertical park angle\footnote{Park angle is defined as the tilt angle of the heliostat when stowed} of \SI{85}{\degree} to limit soiling-induced reflectance loss when not in operation. In addition, all cleaning events are assumed to occur instantaneously at midnight.

        \subsubsection{Horizontal Park Setup}
        Heliostats are often parked in the horizontal position to reduce wind loads. By decreasing wind loads the costs associated with structural supports and maintenance (i.e. alignment calibration) may be lowered. To better understand the soil degradation costs associated with this operational mode a scenario has been setup to park heliostats in a horizontal position (\SI{0}{\degree} tilt) during night time stowing.
    
        \subsubsection{Night Dispatch Setup}
         The Mount Isa regional electrical grid is connected to several large mines and production facilities requiring continuous electrical demand. For the night dispatch scenario it is assumed that during the day the \gls{tes} is filled while the \gls{pv} fields provide continuous power generation below zenith angles of \SI{80}{\degree}, at which point the PV field is turned off\footnote{This is a simplification of how the PV and gas power plants will provide power. In reality a combination of thermal storage capacity and available PV power would decide when to operate the power block} and the \gls{tes} begins providing thermal energy for the power block. As such the dispatch policy of \eqref{eq:dischargeconstraints} are modified to also include \gls{sza} night time dispatch constraints:
        \begin{equation}
            \label{eq:dischargenightconstraints}
             \ta{qtes}[_{\ta{idtime}}] = 
            \begin{cases}
                \ta{qpb} & \frac{\ta{storagecapacity}_{k-1}}{\ta{deltat}} + \ta{qrec}[_{\ta{idtime}}] > \ta{qpb}~\&~\ta{sza}[_{\ta{idtime}}] > 80\\
                \frac{\ta{storagecapacity}_{k-1}}{\ta{deltat}} + \ta{qrec}[_{\ta{idtime}}] & 0 < \frac{\ta{storagecapacity}_{k-1}}{\ta{deltat}} + \ta{qrec}[_{\ta{idtime}}] \leq \ta{qpb}~\&~\ta{sza}[_{\ta{idtime}}] > 80\\
                0  & \text{otherwise}
            \end{cases}
        \end{equation}
        
        \subsubsection{Day Cleaning Setup}
        In the previous scenarios cleanings are assumed to occur instantaneous at midnight. However, heliostats are generally cleaned during the daytime when operators can maneuver cleaning vehicles in tight clearances around heliostats in visible conditions. The downside of this cleaning strategy is the lost productivity from parking the heliostats in a vertical position while they are being cleaned. This scenario has been setup to investigate the effects of day time cleaning on the optimized cleaning strategy. 
        
        With 55 corridors cleaned per shift per crew (working time from 8 A.M to 4 P.M), it would take $T_{\text{c}}=\SI{0.146}{\hour}$ to clean one corridor. When a corridor is cleaned it is assumed to be parked for two cleaning passes. The first pass to clean the corridor and the second pass to allow enough clearance to clean the neighbouring corridor. To account for the lost productivity, heliostats will be assigned an uptime or productivity ratio per hour of operation; denoting the percentage a heliostat is producing energy during an hour of cleaning. As such corridors will have an \als{uptimeratio} of $U_{k,j} = \frac{1-2\cdot T_{\text{c}}}{1}=\SI{71}{\percent}$ while being cleaned and $U_{k,j}=\SI{100}{\percent}$ when not being cleaned. 
        
        To account for productivity losses from parking the heliostats in the daytime, the reflected \gls{dni} from \eqref{eq:recinpower} is replaced with:
        \begin{equation}
            \label{eq:recoppower}
            \ta{qin}[_{\ta{idtime},\ta{idreceiver}}] = \ta{dni}[_{\ta{idtime}}] \sum_{\ta{idsect} \in \mathcal{J}_{\ta{idreceiver}}}{\ta{areatotal}[_{\ta{idsect}}] \cdot \ta{opticalefficiency}[_{\ta{idtime},\ta{idsect}}] \cdot \ta{predictedreflectance}[_{\ta{idtime},\ta{idsect}}] \cdot U_{\ta{idtime},\ta{idsect}}} 
        \end{equation}

        The \gls{costoperational} or profit loss due to parking the heliostats during the day is calculated as:
        \begin{equation}
            \label{eq:costoperational}
            \ta{costoperational}(\ta{ntrucks},\ta{ncleans}) = \left( \ta{pel} - \ta{costoam} \right) \sum^{\ta{ndays}}_{k=1}{ \left( \ta{workclean}[_{\ta{idtime}}] - \ta{workmax}[_{\ta{idtime}}]  \right) \ta{deltat}}
        \end{equation}
        where \ta{workclean} is the work produced with an always clean heliostat field and no heliostat daytime parking losses (i.e. $U=\SI{100}{\percent})$ and \ta{workmax} is the work generated with an always clean heliostat field with consideration of day time cleaning parking (i.e. $U_{k,j}=\SI{71}{\percent}$ for $\ta{cleandecision}[_{\ta{idtime},\ta{idsect}}]=1$).
        
        The total cleaning cost of \eqref{eq:tcc} is ammended to include the operational downtime cost
        \begin{equation}
            \label{eq:tccpluscop}
            \ta{tcc} = \ta{cdeg}+\ta{ccl}+\ta{costoperational}
        \end{equation}

    \subsection{Optimal Cleaning Resources}
    \label{subsec:optimalresults}
    \begin{table*}[ht]
\centering
\caption{Optimal cleaning schedules, mean work produced with a clean and soiled field, and mean costs associated for scenarios horizontal stowing (HS), night dispatching (ND) and day cleaning (DC)}
\label{table:optimalresultssummary}
\begin{tabular}{lcccccccc}
\toprule
 & \ta{ntrucks} & \ta{ncleans} & $w_{\text{rated}}^{\text{clean}}$ & $w_{\text{rated}}^{\text{soil}}$ & $C_{\text{cl}}$ & $C_{\text{deg}}$ & $C_{\text{op}}$ & $\text{TCC}$ \\
 & & & \scriptsize{$(\text{MW h/yr})$} & \scriptsize{$(\text{MW h/yr})$} & \scriptsize{$(\text{kA\$/yr})$} & \scriptsize{$(\text{kA\$/yr})$} & \scriptsize{$(\text{kA\$/yr})$} & \scriptsize{$(\text{kA\$/yr})$} \\
\midrule
 Base & 2 & 24 & 202906 & 199948 & 452 & 404 & 0 & 856 \\
HS & 3 & 36 & 202906 & 198702 & 678 & 574 & 0 & 1252 \\
ND & 2 & 24 & 198018 & 196008 & 452 & 274 & 0 & 727 \\
DC & 2 & 24 & 202408 & 199484 & 452 & 399 & 68 & 919 \\
\bottomrule
\end{tabular}
\end{table*}

    The soiling prediction and stochastic cleaning resource heuristic has been implemented using the \gls{nwqhpp} design conditions for the four operating scenarios. Table~\ref{table:optimalresultssummary} summarizes the optimal cleaning resource setup, electricity generated under clean and soiled fields and costs associated for each operating scenario. Optimal cleaning resource allocations for all near vertical stow angled scenarios occurs with two cleaning crews operating at the maximum cleaning rate of 24 cleans per year. While the optimal cleaning resources are similar, the difference in total cleaning costs is where each operating scenario differs. Contrastingly, horizontally stowed heliostats experienced the highest soil degradation cost and required the largest cleaning fleet with three crews cleaning each heliostat 36 times per year due to the increased soiling losses. Making the operating strategy one of the most costly in terms of soiling-based production loss. 

    \subsubsection{Base Scenario Evaluation}
\begin{figure}[h!]
            \centering
\definecolor{mycolor1}{rgb}{0.00000,0.44700,0.74100}%
\definecolor{mycolor2}{rgb}{0.85000,0.32500,0.09800}%
\definecolor{mycolor3}{rgb}{0.92900,0.69400,0.12500}%
\definecolor{mycolor4}{rgb}{0.49400,0.18400,0.55600}%
\begin{tikzpicture}[trim axis left, trim axis right]

\begin{axis}[%
width=0.951\fwidth,
height=\fheight,
at={(0\fwidth,0\fheight)},
scale only axis,
grid=both,
unbounded coords=jump,
xmin=0.000,
xmax=50.000,
xlabel style={font=\bfseries\color{white!15!black}},
xlabel={\ta{ncleans} [cleans/yr]},
ymin=500000.000,
ymax=2000000.000,
ylabel style={font=\bfseries\color{white!15!black}},
ylabel={TCC [A\$/yr]},
title style={font=\bfseries},
title={Base \ta{tcc}},
axis background/.style={fill=white},
axis x line*=bottom,
axis y line*=left,
grid style={line width=.1pt, draw=gray!25},
major grid style={line width=.2pt,draw=gray!50},
minor tick num=1,
legend pos=north east,
legend style={legend cell align=left, align=left, draw=white!15!black}
]
\addplot [color=mycolor1, line width=1.5pt ,mark=o, mark options={solid, mycolor1}]
 plot [error bars/.cd, y dir=both, y explicit, error bar style={line width=1.5pt}, error mark options={line width=1.5pt, mark size=1.5pt, rotate=90}]
 table[row sep=crcr, y error plus index=2, y error minus index=3]{%
1.000	9177868.881	561164.845	561164.845\\
2.000	4811056.027	482536.083	482536.083\\
3.000	3415514.039	308769.739	308769.739\\
4.000	2575058.006	257327.032	257327.032\\
5.000	2098188.169	161240.675	161240.675\\
6.000	1772336.743	127643.077	127643.077\\
7.000	1557130.075	123176.418	123176.418\\
8.000	1402748.988	110156.133	110156.133\\
9.000	1279007.406	99877.706	99877.706\\
10.000	1184295.001	84184.320	84184.320\\
11.000	1097575.785	75573.591	75573.591\\
12.000	1033298.131	69262.006	69262.006\\
13.000	nan	nan	nan\\
14.000	nan	nan	nan\\
15.000	nan	nan	nan\\
16.000	nan	nan	nan\\
17.000	nan	nan	nan\\
18.000	nan	nan	nan\\
19.000	nan	nan	nan\\
20.000	nan	nan	nan\\
21.000	nan	nan	nan\\
22.000	nan	nan	nan\\
23.000	nan	nan	nan\\
24.000	nan	nan	nan\\
25.000	nan	nan	nan\\
26.000	nan	nan	nan\\
27.000	nan	nan	nan\\
28.000	nan	nan	nan\\
29.000	nan	nan	nan\\
30.000	nan	nan	nan\\
31.000	nan	nan	nan\\
32.000	nan	nan	nan\\
33.000	nan	nan	nan\\
34.000	nan	nan	nan\\
35.000	nan	nan	nan\\
36.000	nan	nan	nan\\
37.000	nan	nan	nan\\
38.000	nan	nan	nan\\
39.000	nan	nan	nan\\
40.000	nan	nan	nan\\
41.000	nan	nan	nan\\
42.000	nan	nan	nan\\
43.000	nan	nan	nan\\
44.000	nan	nan	nan\\
45.000	nan	nan	nan\\
46.000	nan	nan	nan\\
};
\addlegendentry{M=1}

\addplot [color=mycolor2, line width=1.5pt ,mark=square, mark options={solid,mycolor2}]
 plot [error bars/.cd, y dir=both, y explicit, error bar style={line width=1.5pt}, error mark options={line width=1.5pt, mark size=2.0pt, rotate=90}]
 table[row sep=crcr, y error plus index=2, y error minus index=3]{%
1.000	9384368.174	515365.879	515365.879\\
2.000	4919227.625	473986.984	473986.984\\
3.000	3542656.891	305456.605	305456.605\\
4.000	2703378.326	267007.851	267007.851\\
5.000	2239320.186	183434.775	183434.775\\
6.000	1905818.714	137381.433	137381.433\\
7.000	1690483.355	120458.823	120458.823\\
8.000	1536114.612	108608.267	108608.267\\
9.000	1401295.647	81279.286	81279.286\\
10.000	1316233.690	80428.397	80428.397\\
11.000	1229088.158	70189.190	70189.190\\
12.000	1158630.501	68144.682	68144.682\\
13.000	1111006.267	63551.079	63551.079\\
14.000	1061805.444	67829.556	67829.556\\
15.000	1025850.414	59361.826	59361.826\\
16.000	995004.944	55765.915	55765.915\\
17.000	962561.240	49805.654	49805.654\\
18.000	939325.344	46235.119	46235.119\\
19.000	919360.218	44678.292	44678.292\\
20.000	905734.535	42599.253	42599.253\\
21.000	888249.976	41097.337	41097.337\\
22.000	875620.575	39001.561	39001.561\\
23.000	864946.696	37218.480	37218.480\\
24.000	855900.475	35321.124	35321.124\\
25.000	nan	nan	nan\\
26.000	nan	nan	nan\\
27.000	nan	nan	nan\\
28.000	nan	nan	nan\\
29.000	nan	nan	nan\\
30.000	nan	nan	nan\\
31.000	nan	nan	nan\\
32.000	nan	nan	nan\\
33.000	nan	nan	nan\\
34.000	nan	nan	nan\\
35.000	nan	nan	nan\\
36.000	nan	nan	nan\\
37.000	nan	nan	nan\\
38.000	nan	nan	nan\\
39.000	nan	nan	nan\\
40.000	nan	nan	nan\\
41.000	nan	nan	nan\\
42.000	nan	nan	nan\\
43.000	nan	nan	nan\\
44.000	nan	nan	nan\\
45.000	nan	nan	nan\\
46.000	nan	nan	nan\\
};
\addlegendentry{M=2}

\addplot [color=mycolor3, line width=1.5pt ,mark=triangle, mark options={solid, mycolor3}]
 plot [error bars/.cd, y dir=both, y explicit, error bar style={line width=1.5pt}, error mark options={line width=1.5pt, mark size=2.0pt, rotate=90}]
 table[row sep=crcr, y error plus index=2, y error minus index=3]{%
1.000	9551834.940	499573.991	499573.991\\
2.000	5040048.936	462113.906	462113.906\\
3.000	3672474.339	296781.884	296781.884\\
4.000	2832703.940	265527.424	265527.424\\
5.000	2373169.968	188854.786	188854.786\\
6.000	2035308.118	143845.593	143845.593\\
7.000	1827494.222	117196.621	117196.621\\
8.000	1666474.626	109879.232	109879.232\\
9.000	1519882.773	69708.070	69708.070\\
10.000	1437516.601	78917.402	78917.402\\
11.000	1356205.557	69676.090	69676.090\\
12.000	1283532.878	65992.274	65992.274\\
13.000	1238835.747	58797.175	58797.175\\
14.000	1191235.674	67457.783	67457.783\\
15.000	1158704.838	60626.826	60626.826\\
16.000	1122056.749	59599.498	59599.498\\
17.000	1090262.342	53004.502	53004.502\\
18.000	1060919.949	47575.997	47575.997\\
19.000	1038159.441	41177.121	41177.121\\
20.000	1031369.705	43397.882	43397.882\\
21.000	1014156.914	41458.202	41458.202\\
22.000	1001490.541	38383.541	38383.541\\
23.000	993728.749	40296.177	40296.177\\
24.000	980963.075	34930.911	34930.911\\
25.000	972361.513	32508.064	32508.064\\
26.000	965323.712	31255.973	31255.973\\
27.000	961614.909	32453.452	32453.452\\
28.000	956405.980	28405.130	28405.130\\
29.000	953159.908	30475.526	30475.526\\
30.000	951754.736	29280.042	29280.042\\
31.000	949407.720	27985.094	27985.094\\
32.000	948505.900	27352.678	27352.678\\
33.000	946334.215	26050.486	26050.486\\
34.000	947541.338	25216.062	25216.062\\
35.000	947170.538	24284.802	24284.802\\
36.000	948109.748	23781.647	23781.647\\
37.000	nan	nan	nan\\
38.000	nan	nan	nan\\
39.000	nan	nan	nan\\
40.000	nan	nan	nan\\
41.000	nan	nan	nan\\
42.000	nan	nan	nan\\
43.000	nan	nan	nan\\
44.000	nan	nan	nan\\
45.000	nan	nan	nan\\
46.000	nan	nan	nan\\
};
\addlegendentry{M=3}

\addplot [color=mycolor4, line width=1.5pt ,mark=star, mark options={solid, mycolor4}]
 plot [error bars/.cd, y dir=both, y explicit, error bar style={line width=1.5pt}, error mark options={line width=1.5pt, mark size=2.0pt, rotate=90}]
 table[row sep=crcr, y error plus index=2, y error minus index=3]{%
1.000	9686429.325	491692.388	491692.388\\
2.000	5154400.575	448049.537	448049.537\\
3.000	3799595.536	287860.974	287860.974\\
4.000	2950864.212	258177.783	258177.783\\
5.000	2496130.671	187571.224	187571.224\\
6.000	2163647.214	146336.900	146336.900\\
7.000	1957122.058	116414.489	116414.489\\
8.000	1790207.420	109131.120	109131.120\\
9.000	1644098.078	65226.527	65226.527\\
10.000	1553473.594	73156.746	73156.746\\
11.000	1480040.508	68685.944	68685.944\\
12.000	1408406.412	64669.489	64669.489\\
13.000	1364691.204	54691.507	54691.507\\
14.000	1315748.053	66691.925	66691.925\\
15.000	1285949.466	57367.321	57367.321\\
16.000	1245090.989	60959.410	60959.410\\
17.000	1217553.472	56075.742	56075.742\\
18.000	1185375.554	48643.118	48643.118\\
19.000	1160343.472	41139.005	41139.005\\
20.000	1152300.671	40669.408	40669.408\\
21.000	1140048.186	42570.638	42570.638\\
22.000	1125873.107	36948.237	36948.237\\
23.000	1119592.006	40140.849	40140.849\\
24.000	1105393.205	34647.464	34647.464\\
25.000	1096518.335	31868.011	31868.011\\
26.000	1090294.455	29988.315	29988.315\\
27.000	1088372.761	32071.712	32071.712\\
28.000	1080156.273	27517.905	27517.905\\
29.000	1077675.800	28644.810	28644.810\\
30.000	1076820.693	27476.465	27476.465\\
31.000	1075041.520	27608.237	27608.237\\
32.000	1072797.493	27304.493	27304.493\\
33.000	1070263.929	25545.828	25545.828\\
34.000	1074045.869	26074.014	26074.014\\
35.000	1073482.411	26489.333	26489.333\\
36.000	1073793.294	23932.348	23932.348\\
37.000	1073480.177	23942.036	23942.036\\
38.000	1077233.877	23647.680	23647.680\\
39.000	1078366.859	21960.998	21960.998\\
40.000	1080340.075	21348.312	21348.312\\
41.000	1082387.339	21289.480	21289.480\\
42.000	1086216.944	21398.378	21398.378\\
43.000	1088951.543	21360.860	21360.860\\
44.000	1092002.086	19749.164	19749.164\\
45.000	1095291.469	19265.708	19265.708\\
46.000	1099018.862	18843.601	18843.601\\
};
\addlegendentry{M=4}

\end{axis}
\end{tikzpicture}%
            \caption{Base scenario mean annual total cleaning cost for different quantities of cleaning crews (M)}
            \label{fig:basetcc}
        \end{figure}
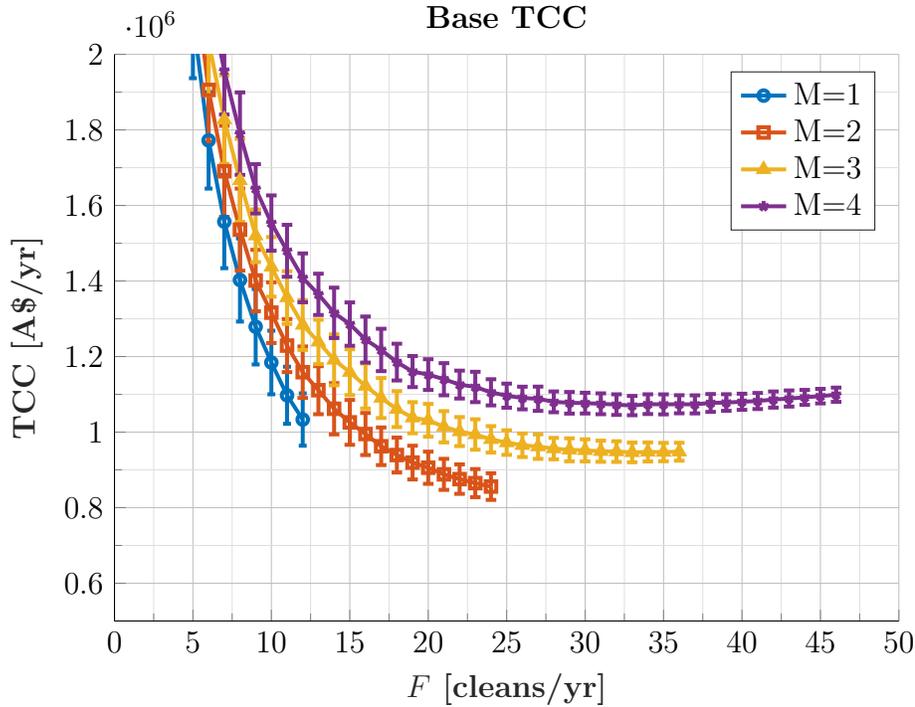

        The mean total cleaning cost are shown in Fig~\ref{fig:basetcc}. Using the cleaning optimizer the optimal solution occurs with two crews cleaning all heliostats 24 times per year with a total cleaning cost of $\ta{tcc}(2,24)=\text{A}\$ \SI{856000}{\per\year}$.  

\begin{figure}[h!]
            \centering
\definecolor{mycolor1}{rgb}{0.00000,0.44700,0.74100}%
\definecolor{mycolor2}{rgb}{0.85000,0.32500,0.09800}%
\definecolor{mycolor3}{rgb}{0.92900,0.69400,0.12500}%
\definecolor{mycolor4}{rgb}{0.49400,0.18400,0.55600}%
\definecolor{mycolor5}{rgb}{0.46600,0.67400,0.18800}%
\begin{tikzpicture}[trim axis left, trim axis right]
\begin{axis}[%
width=0.951\fwidth,
height=\fheight,
at={(0\fwidth,0\fheight)},
scale only axis,
xmin=0.000,
xmax=60.000,
grid=both,
axis y line*=left,
axis x line*=none,
ylabel={Cost [A\$/yr]},
ylabel style={font=\color{white!15!black}},
xlabel={\ta{ncleans} [cleans/yr]},
xlabel style={font=\color{white!15!black}},
title style={font=\bfseries},
title={Base Scenario \ta{cdeg} \& \ta{ccl}},
%every outer y axis line/.append style={mycolor5},
%every y tick label/.append style={font=\color{mycolor5}},
%every y tick/.append style={mycolor5},
ymin=0.000,
ymax=2000000.000,
%axis background/.style={fill=white},
grid style={line width=.1pt, draw=gray!25},
major grid style={line width=.2pt,draw=gray!50},
minor tick num=1,
legend pos=north east,
legend style={legend cell align=left, align=left, draw=white!15!black}
]
\addplot [color=mycolor4, line width=1.5pt]
 plot [error bars/.cd, y dir=both, y explicit, error bar style={line width=1.5pt}, error mark options={line width=1.5pt, mark size=1.5pt, rotate=90}]
 table[row sep=crcr, y error plus index=2, y error minus index=3]{%
1.000	9178003.685	491692.388	491692.388\\
2.000	4637549.295	448049.537	448049.537\\
3.000	3274318.617	287860.974	287860.974\\
4.000	2417161.653	258177.783	258177.783\\
5.000	1954002.471	187571.224	187571.224\\
6.000	1613093.374	146336.900	146336.900\\
7.000	1398142.579	116414.489	116414.489\\
8.000	1222802.301	109131.120	109131.120\\
9.000	1068267.319	65226.527	65226.527\\
10.000	969217.195	73156.746	73156.746\\
11.000	887358.469	68685.944	68685.944\\
12.000	807298.733	64669.489	64669.489\\
13.000	755157.885	54691.507	54691.507\\
14.000	697789.094	66691.925	66691.925\\
15.000	659564.868	57367.321	57367.321\\
16.000	610280.751	60959.410	60959.410\\
17.000	574317.594	56075.742	56075.742\\
18.000	533714.036	48643.118	48643.118\\
19.000	500256.314	41139.005	41139.005\\
20.000	483787.873	40669.408	40669.408\\
21.000	463109.749	42570.638	42570.638\\
22.000	440509.029	36948.237	36948.237\\
23.000	425802.289	40140.849	40140.849\\
24.000	403177.848	34647.464	34647.464\\
25.000	385877.338	31868.011	31868.011\\
26.000	371227.818	29988.315	29988.315\\
27.000	360880.485	32071.712	32071.712\\
28.000	344238.357	27517.905	27517.905\\
29.000	333332.244	28644.810	28644.810\\
30.000	324051.497	27476.465	27476.465\\
31.000	313846.684	27608.237	27608.237\\
32.000	303177.017	27304.493	27304.493\\
33.000	292217.813	25545.828	25545.828\\
34.000	287574.114	26074.014	26074.014\\
35.000	278585.016	26489.333	26489.333\\
36.000	270470.258	23932.348	23932.348\\
37.000	261731.502	23942.036	23942.036\\
38.000	257059.562	23647.680	23647.680\\
39.000	249766.904	21960.998	21960.998\\
40.000	243314.481	21348.312	21348.312\\
41.000	236936.104	21289.480	21289.480\\
42.000	232340.069	21398.378	21398.378\\
43.000	226649.028	21360.860	21360.860\\
44.000	221273.932	19749.164	19749.164\\
45.000	216137.675	19265.708	19265.708\\
46.000	211439.428	18843.601	18843.601\\
};
\addlegendentry{$\ta{cdeg}[_{4}]$}

\addplot [color=mycolor3, line width=1.5pt]
 plot [error bars/.cd, y dir=both, y explicit, error bar style={line width=1.5pt}, error mark options={line width=1.5pt, mark size=1.5pt, rotate=90}]
 table[row sep=crcr, y error plus index=2, y error minus index=3]{%
1.000	9168409.300	499573.991	499573.991\\
2.000	4648197.657	462113.906	462113.906\\
3.000	3272197.419	296781.884	296781.884\\
4.000	2424001.381	265527.424	265527.424\\
5.000	1956041.769	188854.786	188854.786\\
6.000	1609754.279	143845.593	143845.593\\
7.000	1393514.743	117196.621	117196.621\\
8.000	1224069.507	109879.232	109879.232\\
9.000	1069052.014	69708.070	69708.070\\
10.000	978260.202	78917.402	78917.402\\
11.000	888523.518	69676.090	69676.090\\
12.000	807425.199	65992.274	65992.274\\
13.000	754302.428	58797.175	58797.175\\
14.000	698276.716	67457.783	67457.783\\
15.000	657320.240	60626.826	60626.826\\
16.000	612246.511	59599.498	59599.498\\
17.000	572026.464	53004.502	53004.502\\
18.000	534258.431	47575.997	47575.997\\
19.000	503072.283	41177.121	41177.121\\
20.000	487856.908	43397.882	43397.882\\
21.000	462218.477	41458.202	41458.202\\
22.000	441126.464	38383.541	38383.541\\
23.000	424939.032	40296.177	40296.177\\
24.000	403747.718	34930.911	34930.911\\
25.000	386720.517	32508.064	32508.064\\
26.000	371257.075	31255.973	31255.973\\
27.000	359122.632	32453.452	32453.452\\
28.000	345488.064	28405.130	28405.130\\
29.000	333816.351	30475.526	30475.526\\
30.000	323985.540	29280.042	29280.042\\
31.000	313212.884	27985.094	27985.094\\
32.000	303885.424	27352.678	27352.678\\
33.000	293288.100	26050.486	26050.486\\
34.000	286069.583	25216.062	25216.062\\
35.000	277273.142	24284.802	24284.802\\
36.000	269786.713	23781.647	23781.647\\
37.000	nan	nan	nan\\
38.000	nan	nan	nan\\
39.000	nan	nan	nan\\
40.000	nan	nan	nan\\
41.000	nan	nan	nan\\
42.000	nan	nan	nan\\
43.000	nan	nan	nan\\
44.000	nan	nan	nan\\
45.000	nan	nan	nan\\
46.000	nan	nan	nan\\
};
\addlegendentry{$\ta{cdeg}[_{3}]$}
\addplot [color=mycolor2, line width=1.5pt]
 plot [error bars/.cd, y dir=both, y explicit, error bar style={line width=1.5pt}, error mark options={line width=1.5pt, mark size=1.5pt, rotate=90}]
 table[row sep=crcr, y error plus index=2, y error minus index=3]{%
1.000	9125942.534	515365.879	515365.879\\
2.000	4652376.345	473986.984	473986.984\\
3.000	3267379.972	305456.605	305456.605\\
4.000	2419675.767	267007.851	267007.851\\
5.000	1947191.987	183434.775	183434.775\\
6.000	1605264.875	137381.433	137381.433\\
7.000	1381503.876	120458.823	120458.823\\
8.000	1218709.493	108608.267	108608.267\\
9.000	1075464.888	81279.286	81279.286\\
10.000	981977.291	80428.397	80428.397\\
11.000	886406.119	70189.190	70189.190\\
12.000	807522.823	68144.682	68144.682\\
13.000	751472.949	63551.079	63551.079\\
14.000	693846.486	67829.556	67829.556\\
15.000	649465.816	59361.826	59361.826\\
16.000	610194.706	55765.915	55765.915\\
17.000	569325.362	49805.654	49805.654\\
18.000	537663.826	46235.119	46235.119\\
19.000	509273.060	44678.292	44678.292\\
20.000	487221.738	42599.253	42599.253\\
21.000	461311.539	41097.337	41097.337\\
22.000	440256.498	39001.561	39001.561\\
23.000	421156.979	37218.480	37218.480\\
24.000	403685.118	35321.124	35321.124\\
25.000	nan	nan	nan\\
26.000	nan	nan	nan\\
27.000	nan	nan	nan\\
28.000	nan	nan	nan\\
29.000	nan	nan	nan\\
30.000	nan	nan	nan\\
31.000	nan	nan	nan\\
32.000	nan	nan	nan\\
33.000	nan	nan	nan\\
34.000	nan	nan	nan\\
35.000	nan	nan	nan\\
36.000	nan	nan	nan\\
37.000	nan	nan	nan\\
38.000	nan	nan	nan\\
39.000	nan	nan	nan\\
40.000	nan	nan	nan\\
41.000	nan	nan	nan\\
42.000	nan	nan	nan\\
43.000	nan	nan	nan\\
44.000	nan	nan	nan\\
45.000	nan	nan	nan\\
46.000	nan	nan	nan\\
};
\addlegendentry{$\ta{cdeg}[_{2}]$}

\addplot [color=mycolor1, line width=1.5pt]
 plot [error bars/.cd, y dir=both, y explicit, error bar style={line width=1.5pt}, error mark options={line width=1.5pt, mark size=1.5pt, rotate=90}]
 table[row sep=crcr, y error plus index=2, y error minus index=3]{%
1.000	9044443.241	561164.845	561164.845\\
2.000	4669204.747	482536.083	482536.083\\
3.000	3265237.120	308769.739	308769.739\\
4.000	2416355.447	257327.032	257327.032\\
5.000	1931059.970	161240.675	161240.675\\
6.000	1596782.904	127643.077	127643.077\\
7.000	1373150.596	123176.418	123176.418\\
8.000	1210343.869	110156.133	110156.133\\
9.000	1078176.647	99877.706	99877.706\\
10.000	975038.602	84184.320	84184.320\\
11.000	879893.746	75573.591	75573.591\\
12.000	807190.453	69262.006	69262.006\\
13.000	nan	nan	nan\\
14.000	nan	nan	nan\\
15.000	nan	nan	nan\\
16.000	nan	nan	nan\\
17.000	nan	nan	nan\\
18.000	nan	nan	nan\\
19.000	nan	nan	nan\\
20.000	nan	nan	nan\\
21.000	nan	nan	nan\\
22.000	nan	nan	nan\\
23.000	nan	nan	nan\\
24.000	nan	nan	nan\\
25.000	nan	nan	nan\\
26.000	nan	nan	nan\\
27.000	nan	nan	nan\\
28.000	nan	nan	nan\\
29.000	nan	nan	nan\\
30.000	nan	nan	nan\\
31.000	nan	nan	nan\\
32.000	nan	nan	nan\\
33.000	nan	nan	nan\\
34.000	nan	nan	nan\\
35.000	nan	nan	nan\\
36.000	nan	nan	nan\\
37.000	nan	nan	nan\\
38.000	nan	nan	nan\\
39.000	nan	nan	nan\\
40.000	nan	nan	nan\\
41.000	nan	nan	nan\\
42.000	nan	nan	nan\\
43.000	nan	nan	nan\\
44.000	nan	nan	nan\\
45.000	nan	nan	nan\\
46.000	nan	nan	nan\\
};
\addlegendentry{$\ta{cdeg}[_{1}]$}

\addplot [color=mycolor1, dashed, line width=1.5pt]
  table[row sep=crcr]{%
1.000	133425.640\\
12.000	226107.678\\
};
\addlegendentry{$\ta{ccl}[_{1}]$}

\addplot [color=mycolor2, dotted, line width=1.5pt]
  table[row sep=crcr]{%
1.000	258425.640\\
24.000	452215.357\\
};
\addlegendentry{$\ta{ccl}[_{2}]$}

\addplot [color=mycolor3, dashdotted, line width=1.5pt]
  table[row sep=crcr]{%
1.000	383425.640\\
36.000	678323.035\\
};
\addlegendentry{$\ta{ccl}[_{3}]$}

\addplot [color=mycolor4, dash dot dot, line width=1.5pt]
  table[row sep=crcr]{%
1.000	508425.640\\
46.000	887579.434\\
};
\addlegendentry{$\ta{ccl}[_{4}]$}

\end{axis}
\end{tikzpicture}%
            \caption{Base scenario mean annual degradation ($\ta{cdeg}[_{M}]$) and cleaning costs ($\ta{ccl}[_{M}]$) for different amounts of cleaning crews (M)}
            \label{fig:basecdegccl}
        \end{figure}
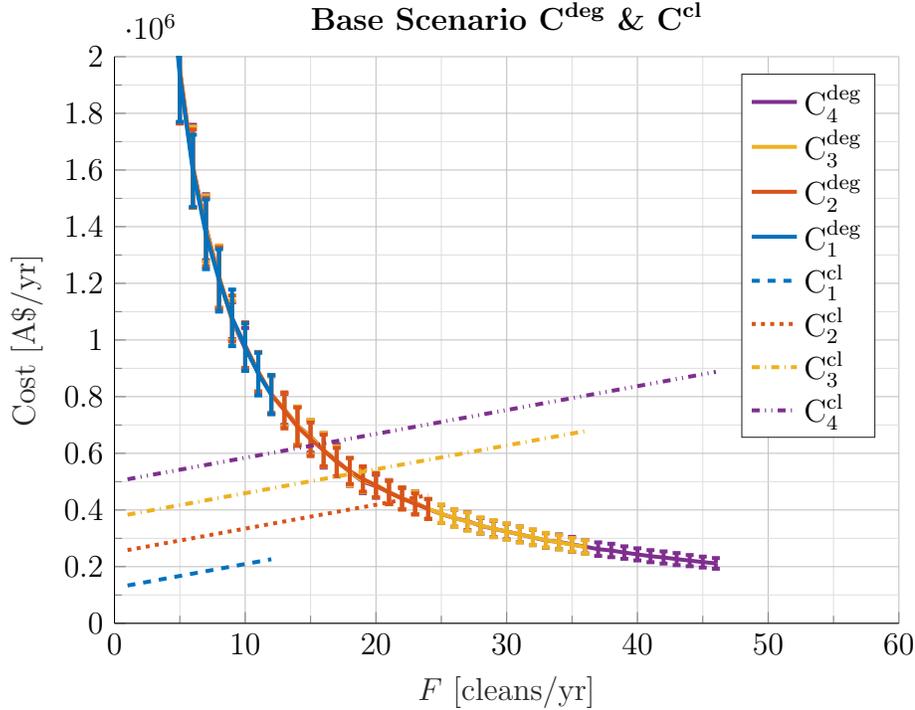

        Fig.~\ref{fig:basecdegccl} shows the total cleaning cost split between mean degradation and cleaning costs. Variations between degradation curves are minimal and increasing the number of field cleanings creates an exponentially decaying cost relation. Whereas, cleaning cost show a linear trend with each additional crew requiring a larger fixed cost and additional field cleanings will increase variable costs. As such surpassing the optimal cleaning resource setup will increase plant productivity but due to higher cleaning costs any production benefits are no longer economically efficient.

        \subsubsection{Cleaning Cost Breakdown}  
\begin{figure}[h!]
            \centering
\definecolor{mycolor1}{rgb}{0.00000,0.44700,0.74100}%
\definecolor{mycolor2}{rgb}{0.85000,0.32500,0.09800}%
\definecolor{mycolor3}{rgb}{0.92900,0.69400,0.12500}%
\definecolor{mycolor4}{rgb}{0.49400,0.18400,0.55600}%
\definecolor{mycolor5}{rgb}{0.46600,0.67400,0.18800}%
\begin{tikzpicture}[trim axis left, trim axis right]

\begin{axis}[%
width=0.951\fwidth,
height=\fheight,
at={(0\fwidth,0\fheight)},
scale only axis,
bar width=0.8,
xmin=-0.200,
xmax=3.200,
xtick={1.000,2.000},
xticklabels={{Fixed},{Variable}},
ymin=0.000,
ymax=150000.000,
ylabel style={font=\color{white!15!black}},
ylabel={Cost [A\$/yr/crew]},
title style={font=\bfseries},
title={Cleaning Cost Breakdown},
axis background/.style={fill=white},
ymajorgrids=true,
yminorgrids=true,
minor y tick num=1,
legend pos=north east,
legend style={legend cell align=left, align=left, draw=white!15!black}
]
\addplot[ybar stacked, fill=mycolor1, draw=none, area legend] table[row sep=crcr] {%
1.000	80000.000\\
2.000	0.000\\
};
\addplot[forget plot, color=white!15!black, draw=none, only marks] table[row sep=crcr] {%
-0.200	0.000\\
3.200	0.000\\
};
\addlegendentry{\ta{costop}}

\addplot[ybar stacked, fill=mycolor2, draw=none, area legend] table[row sep=crcr] {%
1.000	30000.000\\
2.000	0.000\\
};
\addplot[forget plot, color=white!15!black, draw=none, only marks] table[row sep=crcr] {%
-0.200	0.000\\
3.200	0.000\\
};
\addlegendentry{\ta{costtruck}}

\addplot[ybar stacked, fill=mycolor3, draw=none, area legend] table[row sep=crcr] {%
1.000	15000.000\\
2.000	0.000\\
};
\addplot[forget plot, color=white!15!black, draw=none, only marks] table[row sep=crcr] {%
-0.200	0.000\\
3.200	0.000\\
};
\addlegendentry{\ta{costmaint}}

\addplot[ybar stacked, fill=mycolor4, draw=none, area legend] table[row sep=crcr] {%
1.000	0.000\\
2.000	94500.000\\
};
\addplot[forget plot, color=white!15!black, draw=none, only marks] table[row sep=crcr] {%
-0.200	0.000\\
3.200	0.000\\
};
\addlegendentry{$\text{C}^{\text{fuel}}$}

\addplot[ybar stacked, fill=mycolor5, draw=none, area legend] table[row sep=crcr] {%
1.000	0.000\\
2.000	6605.107\\
};
\addplot[forget plot, color=white!15!black, draw=none, only marks] table[row sep=crcr] {%
-0.200	0.000\\
3.200	0.000\\
};
\addlegendentry{$\text{C}^{\text{water}}$}

\end{axis}
\end{tikzpicture}%
            \caption{Cleaning cost breakdown for one cleaning crew operating at the maximum cleaning rate}
            \label{fig:cleaningcostbar}
        \end{figure}
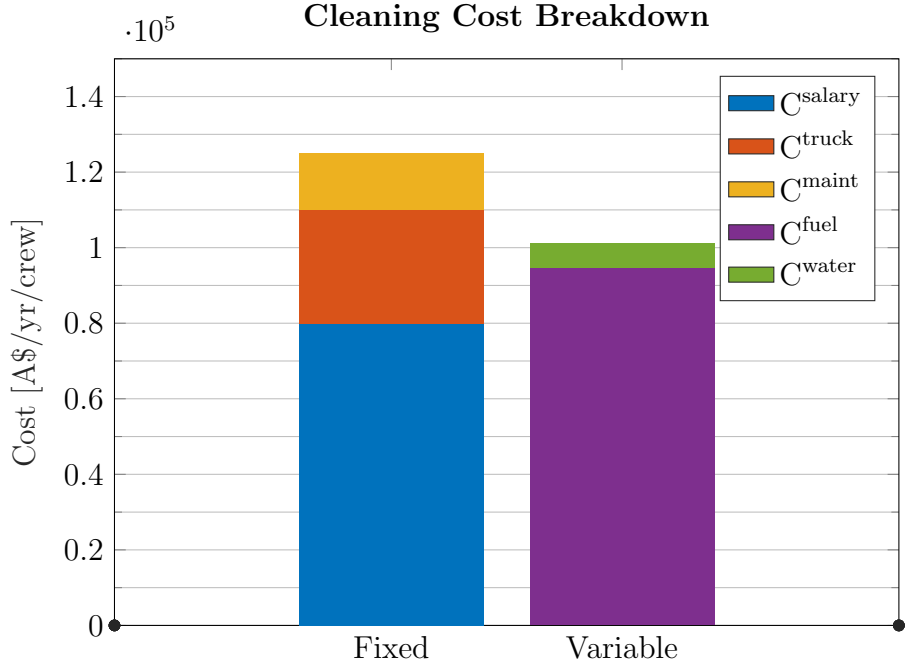

        The breakdown of costs associated with one cleaning crew operating at the maximum cleaning rate\footnote{The maximum cleaning rate corresponds to one crew cleaning each heliostat 12 times per year, with two crews able to clean each heliostat 24 times per year, and so on.} is depicted in Fig.~\ref{fig:cleaningcostbar}. It is evident that fixed costs outweigh variable costs, with salaries constituting the majority of yearly fixed expenses, accounting for \SI{35}{\percent} of the total cleaning costs. A smaller proportion of fixed costs is attributed to capital investment and maintenance of the cleaning vehicle itself.

        Notably, with the ongoing advancements in automated cleaning vehicles~\cite{Supcon2020,Hardt2011},  a reduction in salary costs per cleaning crew is expected, as a single supervisor may oversee a fleet of automated vehicles. This potential shift would come at the expense of higher vehicle costs. However, automated vehicles can work over a longer shift time and which may even reduce the total number of cleaning crews needed. As such this cost-trade off may be advantageous if the total fixed cost proves to be lower than that of the staffed cleaning crew presented in this study.
        
        Variable costs contribute nearly as much towards the cleaning cost. Primarily due to the \SI{67.5}{\kilo\litre\per\year} of fuel required to operate each truck leading to the the annual fuel cost of A\$\SI{94500}{\per\year} per crew or \SI{42}{\percent} of the cleaning costs; dominating all cost factors listed in both categories. Whereas, the \SI{2.2}{\mega\litre\per\year} of water contributes about A\$\SI{6600}{\per\year} per crew (\SI{3}{\percent} of cleaning costs) of the variable costs and is the lowest reported for all cleaning cost factors. The results suggest the need to reduce fuel usage within cleaning vehicles. The possibility of hybrid or fully electric vehicles could prove to be a solution as cheap electrical energy could be supplied by the plant.
        
        \subsubsection{Horizontal Stowing Evaluation}
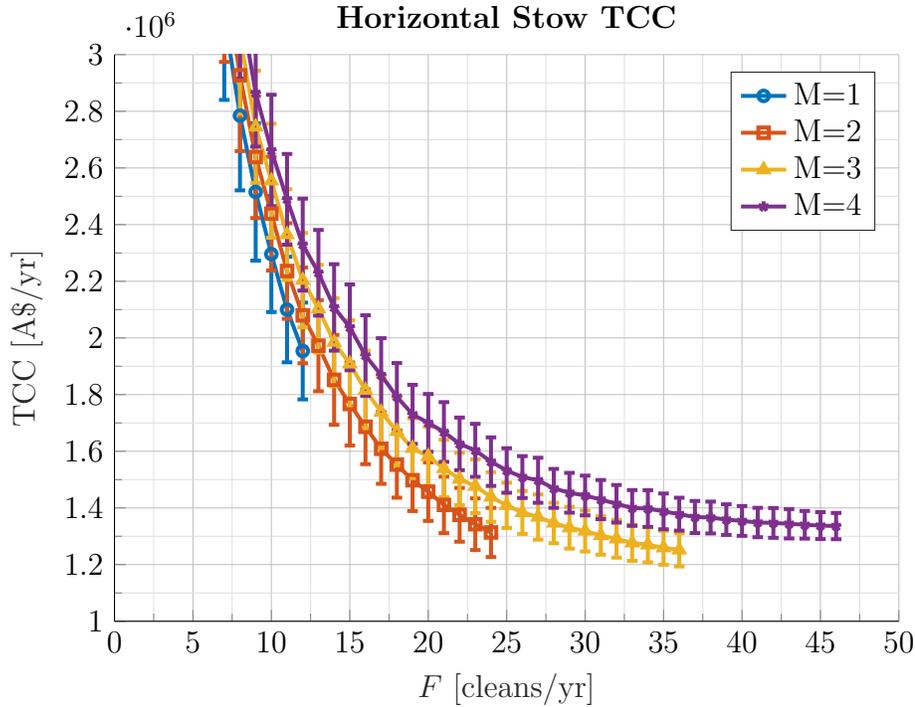
\begin{figure}[h!]
            \centering
\definecolor{mycolor1}{rgb}{0.00000,0.44700,0.74100}%
\definecolor{mycolor2}{rgb}{0.85000,0.32500,0.09800}%
\definecolor{mycolor3}{rgb}{0.92900,0.69400,0.12500}%
\definecolor{mycolor4}{rgb}{0.49400,0.18400,0.55600}%
\begin{tikzpicture}[trim axis left, trim axis right]

\begin{axis}[%
width=0.951\fwidth,
height=\fheight,
at={(0\fwidth,0\fheight)},
scale only axis,
unbounded coords=jump,
xmin=0.000,
xmax=50.000,
xlabel style={font=\color{white!15!black}},
xlabel={\ta{ncleans} [cleans/yr]},
ymin=1000000.000,
ymax=3000000.000,
ylabel style={font=\color{white!15!black}},
ylabel={TCC [A\$/yr]},
title style={font=\bfseries},
title={Horizontal Stow \ta{tcc}},
axis background/.style={fill=white},
axis x line*=bottom,
axis y line*=left,
grid=both,
grid style={line width=.1pt, draw=gray!25},
major grid style={line width=.2pt,draw=gray!50},
minor tick num=1,
legend pos=north east,
legend style={legend cell align=left, align=left, draw=white!15!black}
]
\addplot [color=mycolor1, line width=1.5pt ,mark=o, mark options={solid, mycolor1}]
 plot [error bars/.cd, y dir=both, y explicit, error bar style={line width=1.5pt}, error mark options={line width=1.5pt, mark size=2.0pt, rotate=90}]
 table[row sep=crcr, y error plus index=2, y error minus index=3]{%
1.000	17547265.781	1528735.916	1528735.916\\
2.000	10003361.341	1027418.282	1027418.282\\
3.000	7124338.412	719499.634	719499.634\\
4.000	5297217.596	567399.643	567399.643\\
5.000	4300873.114	373740.719	373740.719\\
6.000	3593474.687	322833.309	322833.309\\
7.000	3136736.380	296493.142	296493.142\\
8.000	2784750.169	263878.535	263878.535\\
9.000	2515368.597	242378.287	242378.287\\
10.000	2295980.203	204713.828	204713.828\\
11.000	2100265.687	186356.457	186356.457\\
12.000	1953921.151	170757.783	170757.783\\
13.000	nan	nan	nan\\
14.000	nan	nan	nan\\
15.000	nan	nan	nan\\
16.000	nan	nan	nan\\
17.000	nan	nan	nan\\
18.000	nan	nan	nan\\
19.000	nan	nan	nan\\
20.000	nan	nan	nan\\
21.000	nan	nan	nan\\
22.000	nan	nan	nan\\
23.000	nan	nan	nan\\
24.000	nan	nan	nan\\
25.000	nan	nan	nan\\
26.000	nan	nan	nan\\
27.000	nan	nan	nan\\
28.000	nan	nan	nan\\
29.000	nan	nan	nan\\
30.000	nan	nan	nan\\
31.000	nan	nan	nan\\
32.000	nan	nan	nan\\
33.000	nan	nan	nan\\
34.000	nan	nan	nan\\
35.000	nan	nan	nan\\
36.000	nan	nan	nan\\
37.000	nan	nan	nan\\
38.000	nan	nan	nan\\
39.000	nan	nan	nan\\
40.000	nan	nan	nan\\
41.000	nan	nan	nan\\
42.000	nan	nan	nan\\
43.000	nan	nan	nan\\
44.000	nan	nan	nan\\
45.000	nan	nan	nan\\
46.000	nan	nan	nan\\
};
\addlegendentry{M=1}

\addplot [color=mycolor2, line width=1.5pt ,mark=square, mark options={solid, mycolor2}]
 plot [error bars/.cd, y dir=both, y explicit, error bar style={line width=1.5pt}, error mark options={line width=1.5pt, mark size=2.0pt, rotate=90}]
 table[row sep=crcr, y error plus index=2, y error minus index=3]{%
1.000	17782323.842	1477745.135	1477745.135\\
2.000	10091412.875	1001376.613	1001376.613\\
3.000	7230912.175	696900.345	696900.345\\
4.000	5416937.852	567130.193	567130.193\\
5.000	4471789.674	411060.375	411060.375\\
6.000	3724923.896	323303.310	323303.310\\
7.000	3283431.822	309247.210	309247.210\\
8.000	2927223.230	267605.827	267605.827\\
9.000	2638392.881	215221.590	215221.590\\
10.000	2438699.197	200842.282	200842.282\\
11.000	2236151.556	168738.696	168738.696\\
12.000	2079599.239	169061.930	169061.930\\
13.000	1972571.039	160740.471	160740.471\\
14.000	1852058.836	158499.136	158499.136\\
15.000	1766961.727	146546.617	146546.617\\
16.000	1686355.118	131932.981	131932.981\\
17.000	1608743.457	123588.676	123588.676\\
18.000	1552853.467	116616.000	116616.000\\
19.000	1497682.093	108692.528	108692.528\\
20.000	1457292.286	103177.770	103177.770\\
21.000	1410711.901	99683.265	99683.265\\
22.000	1375679.251	94975.470	94975.470\\
23.000	1342466.763	91119.966	91119.966\\
24.000	1313435.422	86748.548	86748.548\\
25.000	nan	nan	nan\\
26.000	nan	nan	nan\\
27.000	nan	nan	nan\\
28.000	nan	nan	nan\\
29.000	nan	nan	nan\\
30.000	nan	nan	nan\\
31.000	nan	nan	nan\\
32.000	nan	nan	nan\\
33.000	nan	nan	nan\\
34.000	nan	nan	nan\\
35.000	nan	nan	nan\\
36.000	nan	nan	nan\\
37.000	nan	nan	nan\\
38.000	nan	nan	nan\\
39.000	nan	nan	nan\\
40.000	nan	nan	nan\\
41.000	nan	nan	nan\\
42.000	nan	nan	nan\\
43.000	nan	nan	nan\\
44.000	nan	nan	nan\\
45.000	nan	nan	nan\\
46.000	nan	nan	nan\\
};
\addlegendentry{M=2}

\addplot [color=mycolor3, line width=1.5pt ,mark=triangle, mark options={solid, mycolor3}]
 plot [error bars/.cd, y dir=both, y explicit, error bar style={line width=1.5pt}, error mark options={line width=1.5pt, mark size=2.0pt, rotate=90}]
 table[row sep=crcr, y error plus index=2, y error minus index=3]{%
1.000	17958830.145	1472127.010	1472127.010\\
2.000	10206109.333	983291.990	983291.990\\
3.000	7355228.031	684275.166	684275.166\\
4.000	5546347.096	558386.771	558386.771\\
5.000	4615455.009	428899.485	428899.485\\
6.000	3856368.587	332711.375	332711.375\\
7.000	3429491.105	307969.446	307969.446\\
8.000	3064123.466	273351.448	273351.448\\
9.000	2745121.592	198122.629	198122.629\\
10.000	2554812.878	201140.734	201140.734\\
11.000	2365056.823	160054.273	160054.273\\
12.000	2203905.966	166539.648	166539.648\\
13.000	2103400.370	154821.815	154821.815\\
14.000	1984957.702	155425.569	155425.569\\
15.000	1908535.966	154267.378	154267.378\\
16.000	1815701.683	139171.465	139171.465\\
17.000	1738080.472	126486.314	126486.314\\
18.000	1670438.724	120035.209	120035.209\\
19.000	1610582.187	105563.418	105563.418\\
20.000	1582132.008	104162.205	104162.205\\
21.000	1539325.346	101451.862	101451.862\\
22.000	1502111.422	92364.238	92364.238\\
23.000	1476662.515	94843.574	94843.574\\
24.000	1438588.275	87636.675	87636.675\\
25.000	1409197.380	79902.852	79902.852\\
26.000	1383971.476	75650.197	75650.197\\
27.000	1367888.637	79736.218	79736.218\\
28.000	1346775.873	71454.801	71454.801\\
29.000	1330082.309	73692.617	73692.617\\
30.000	1318421.691	72134.305	72134.305\\
31.000	1302655.894	67766.109	67766.109\\
32.000	1291159.370	66622.120	66622.120\\
33.000	1276729.191	63569.102	63569.102\\
34.000	1270032.113	61990.688	61990.688\\
35.000	1259449.208	59958.666	59958.666\\
36.000	1252069.576	58617.756	58617.756\\
37.000	nan	nan	nan\\
38.000	nan	nan	nan\\
39.000	nan	nan	nan\\
40.000	nan	nan	nan\\
41.000	nan	nan	nan\\
42.000	nan	nan	nan\\
43.000	nan	nan	nan\\
44.000	nan	nan	nan\\
45.000	nan	nan	nan\\
46.000	nan	nan	nan\\
};
\addlegendentry{M=3}

\addplot [color=mycolor4, line width=1.5pt ,mark=star, mark options={solid, mycolor4}]
 plot [error bars/.cd, y dir=both, y explicit, error bar style={line width=1.5pt}, error mark options={line width=1.5pt, mark size=2.0pt, rotate=90}]
 table[row sep=crcr, y error plus index=2, y error minus index=3]{%
1.000	18094225.521	1472318.633	1472318.633\\
2.000	10312267.500	968772.315	968772.315\\
3.000	7480776.149	669524.053	669524.053\\
4.000	5659629.288	544686.528	544686.528\\
5.000	4737064.236	431862.748	431862.748\\
6.000	3985889.098	332577.495	332577.495\\
7.000	3562496.941	309496.670	309496.670\\
8.000	3188277.709	273017.279	273017.279\\
9.000	2864695.209	189327.836	189327.836\\
10.000	2661767.982	196614.961	196614.961\\
11.000	2489309.994	160156.508	160156.508\\
12.000	2329464.878	162260.199	162260.199\\
13.000	2229948.135	151297.194	151297.194\\
14.000	2107775.362	152614.047	152614.047\\
15.000	2037422.982	151434.569	151434.569\\
16.000	1937767.259	142239.239	142239.239\\
17.000	1868677.420	130448.180	130448.180\\
18.000	1792252.358	119155.811	119155.811\\
19.000	1730344.285	104282.143	104282.143\\
20.000	1700177.101	102060.845	102060.845\\
21.000	1668042.073	105385.318	105385.318\\
22.000	1626133.303	92852.832	92852.832\\
23.000	1603230.309	93607.526	93607.526\\
24.000	1563383.816	85669.801	85669.801\\
25.000	1532141.783	78614.673	78614.673\\
26.000	1508954.383	73952.049	73952.049\\
27.000	1497648.726	79491.326	79491.326\\
28.000	1468956.435	68699.420	68699.420\\
29.000	1453562.783	70181.657	70181.657\\
30.000	1443848.486	70078.634	70078.634\\
31.000	1429462.556	69185.139	69185.139\\
32.000	1414654.547	66356.592	66356.592\\
33.000	1399865.455	63059.914	63059.914\\
34.000	1397787.027	64647.841	64647.841\\
35.000	1387798.880	63090.370	63090.370\\
36.000	1377884.597	58053.348	58053.348\\
37.000	1367913.912	56680.871	56680.871\\
38.000	1366126.376	56197.104	56197.104\\
39.000	1358737.323	54059.233	54059.233\\
40.000	1354058.800	52659.339	52659.339\\
41.000	1348597.647	51874.076	51874.076\\
42.000	1347063.340	52574.516	52574.516\\
43.000	1344143.850	51852.383	51852.383\\
44.000	1340505.421	48781.261	48781.261\\
45.000	1337688.182	47452.935	47452.935\\
46.000	1335978.995	46403.822	46403.822\\
};
\addlegendentry{M=4}

\end{axis}
\end{tikzpicture}%
            \caption{Horizontal stowing scenario mean total cleaning cost for various amounts of cleaning crews (M)}
            \label{fig:horztcc}
        \end{figure}

        Stowing heliostats in a horizontal position will lead to higher soiling rates compared to vertical heliostats due to the increased soil deposition exposure and lower removal forces (see Fig.~\ref{fig:tiltcleanlinessmeasures}). Table~\ref{table:optimalresultssummary} shows the total cleaning cost increased around A\$\SI{400000}{\per\year} or \SI{51}{\percent} relative to the vertical stow results of the base scenario. Higher soiling rates even motivate the use of a third cleaning crew to reach the optimal cleaning schedule (see Fig.~\ref{fig:horztcc}). Although, two cleaning crews is A\$\SI{61000}{\per\year} more expensive and within the standard deviation of the three crew solution and may be adequate for low soiling trajectories. 
        
        The additional degradation cost associated with horizontally stowing the heliostats is approximately A\$166 per heliostat or A\$\SI{25.9}{\per\metre\squared} of reflective area, assuming a lifespan of 30 years. Considering heliostat prices of A\$\SI{100}{\per\metre\squared} \citep{Solar2016}, this results in a cost increase of \SI{26}{\percent} of the total heliostat price. Consequently, solar field designers should carefully weigh the economic benefits of reduced structural and maintenance costs due to wind loading against the additional expenses related to cleaning and productivity losses caused by soiling.
        
\begin{figure}[h!]
            \centering
\begin{tikzpicture}[%
trim axis left, trim axis right
]

\begin{axis}[%
width=0.951\fwidth,
height=\fheight,
at={(0\fwidth,0\fheight)},
scale only axis,
point meta min=0.000,
point meta max=0.200,
colormap/jet,
xmin=-100.000,
xmax=100.000,
xlabel style={font=\color{white!15!black}},
xlabel={X Position [m]},
ymin=-220.000,
ymax=0.000,
ylabel style={font=\color{white!15!black}},
ylabel={Y Position [m]},
axis background/.style={fill=white},
title style={font=\bfseries},
title={Vertical Stow Daily Soiling Rate (pp/day)},
xmajorgrids,
ymajorgrids
]
\addplot[contour prepared = {contour label style={/pgf/number format/fixed,/pgf/number format/precision=3}}, contour prepared format=matlab] table[row sep=crcr] {%
0.060	31.000\\
-97.500	-180.512\\
-90.169	-181.992\\
-87.237	-182.538\\
-76.974	-183.249\\
-70.461	-183.744\\
-66.711	-184.025\\
-56.447	-184.756\\
-51.822	-185.496\\
-46.184	-186.351\\
-37.373	-187.247\\
-35.921	-187.394\\
-25.658	-188.266\\
-17.639	-188.999\\
-15.395	-189.202\\
-5.132	-190.105\\
1.942	-190.751\\
5.132	-191.041\\
15.395	-191.967\\
21.265	-192.502\\
25.658	-192.903\\
35.921	-193.861\\
40.174	-194.254\\
46.184	-194.813\\
56.447	-195.817\\
58.608	-196.005\\
66.711	-196.718\\
76.974	-197.505\\
79.080	-197.757\\
87.237	-198.738\\
94.458	-199.509\\
97.500	-199.847\\
0.062	29.000\\
-97.500	-147.653\\
-88.321	-148.711\\
-87.237	-148.824\\
-76.974	-149.745\\
-68.641	-150.463\\
-66.711	-150.637\\
-56.447	-151.825\\
-52.674	-152.215\\
-46.184	-153.078\\
-39.066	-153.966\\
-35.921	-154.359\\
-25.658	-155.604\\
-24.710	-155.718\\
-15.395	-156.826\\
-9.480	-157.470\\
-5.132	-157.939\\
5.132	-158.997\\
7.516	-159.221\\
15.395	-159.961\\
25.658	-160.848\\
27.213	-160.973\\
35.921	-161.673\\
46.184	-162.414\\
50.644	-162.724\\
56.447	-163.132\\
66.711	-163.764\\
76.974	-163.552\\
87.237	-163.828\\
97.500	-163.362\\
0.064	31.000\\
-97.500	-121.527\\
-95.276	-122.437\\
-90.831	-124.189\\
-87.237	-125.522\\
-84.918	-125.940\\
-76.974	-127.314\\
-74.697	-127.692\\
-66.711	-128.967\\
-64.072	-129.443\\
-56.447	-130.773\\
-53.304	-131.195\\
-46.184	-132.130\\
-39.929	-132.947\\
-35.921	-133.441\\
-28.951	-134.698\\
-25.658	-135.273\\
-15.395	-136.316\\
-13.683	-136.450\\
-5.132	-137.113\\
5.132	-137.782\\
13.632	-138.202\\
15.395	-138.289\\
25.658	-138.609\\
35.921	-138.904\\
46.184	-138.612\\
56.447	-138.898\\
66.711	-138.547\\
76.974	-138.374\\
84.712	-138.202\\
87.237	-138.143\\
97.500	-136.809\\
0.066	37.000\\
-97.500	-99.785\\
-91.275	-101.417\\
-87.237	-102.671\\
-85.434	-103.169\\
-78.865	-104.921\\
-76.974	-105.408\\
-71.923	-106.672\\
-66.711	-107.923\\
-64.698	-108.424\\
-57.447	-110.176\\
-56.447	-110.411\\
-48.476	-111.927\\
-46.184	-112.348\\
-40.995	-113.679\\
-35.921	-114.897\\
-33.424	-115.430\\
-25.658	-117.069\\
-24.850	-117.182\\
-15.395	-118.477\\
-10.764	-118.934\\
-5.132	-119.482\\
5.132	-120.219\\
15.395	-120.651\\
19.085	-120.685\\
25.658	-120.748\\
33.617	-120.685\\
35.921	-120.667\\
46.184	-119.646\\
56.447	-119.406\\
61.417	-118.934\\
66.711	-118.416\\
76.974	-117.251\\
77.745	-117.182\\
87.237	-116.305\\
90.199	-115.430\\
96.139	-113.679\\
97.500	-113.285\\
0.068	40.000\\
-97.500	-79.464\\
-96.209	-80.398\\
-93.522	-82.149\\
-90.404	-83.901\\
-87.237	-85.432\\
-86.533	-85.653\\
-80.647	-87.404\\
-76.974	-88.417\\
-74.229	-89.156\\
-67.196	-90.908\\
-66.711	-91.022\\
-59.272	-92.659\\
-56.447	-93.259\\
-48.476	-94.411\\
-46.184	-94.741\\
-38.640	-96.162\\
-35.921	-96.825\\
-30.821	-97.914\\
-25.658	-99.057\\
-22.125	-99.666\\
-15.395	-100.803\\
-10.268	-101.417\\
-5.132	-102.025\\
5.132	-102.826\\
15.156	-103.169\\
15.395	-103.177\\
16.140	-103.169\\
25.658	-103.063\\
35.921	-102.628\\
44.605	-101.417\\
46.184	-101.196\\
56.447	-99.998\\
58.836	-99.666\\
66.711	-98.599\\
71.221	-97.914\\
76.974	-97.130\\
86.190	-96.162\\
87.237	-96.060\\
96.172	-94.411\\
97.500	-94.134\\
0.070	45.000\\
-97.500	-60.401\\
-96.637	-61.130\\
-94.613	-62.881\\
-92.593	-64.633\\
-90.536	-66.385\\
-88.402	-68.136\\
-87.237	-69.055\\
-85.389	-69.888\\
-81.298	-71.640\\
-76.974	-73.365\\
-76.909	-73.391\\
-72.368	-75.143\\
-67.466	-76.895\\
-66.711	-77.152\\
-61.099	-78.646\\
-56.447	-79.840\\
-53.413	-80.398\\
-46.184	-81.655\\
-44.729	-82.149\\
-39.222	-83.901\\
-35.921	-84.855\\
-32.849	-85.653\\
-25.658	-87.253\\
-24.664	-87.404\\
-15.395	-88.726\\
-9.879	-89.156\\
-5.132	-89.513\\
5.132	-90.016\\
15.395	-90.125\\
25.658	-89.774\\
32.858	-89.156\\
35.921	-88.864\\
46.184	-87.483\\
47.319	-87.404\\
56.447	-86.716\\
62.871	-85.653\\
66.711	-84.971\\
70.560	-83.901\\
76.159	-82.149\\
76.974	-81.873\\
87.092	-80.398\\
87.237	-80.376\\
92.083	-78.646\\
96.520	-76.895\\
97.500	-76.485\\
0.072	52.000\\
-96.309	-46.914\\
-95.758	-47.117\\
-92.706	-48.868\\
-90.023	-50.620\\
-87.643	-52.372\\
-87.237	-52.687\\
-84.797	-54.123\\
-81.679	-55.875\\
-78.390	-57.627\\
-76.974	-58.352\\
-75.000	-59.378\\
-71.466	-61.130\\
-67.713	-62.881\\
-66.711	-63.332\\
-62.886	-64.633\\
-57.452	-66.385\\
-56.447	-66.697\\
-52.514	-68.136\\
-46.753	-69.888\\
-46.184	-70.043\\
-42.226	-71.640\\
-37.644	-73.391\\
-35.921	-74.025\\
-32.841	-75.143\\
-27.742	-76.895\\
-25.658	-77.569\\
-20.229	-78.646\\
-15.395	-79.563\\
-7.718	-80.398\\
-5.132	-80.669\\
5.132	-81.267\\
15.395	-81.223\\
25.658	-80.484\\
26.160	-80.398\\
35.358	-78.646\\
35.921	-78.528\\
44.066	-76.895\\
46.184	-76.444\\
53.305	-75.143\\
56.447	-74.504\\
60.994	-73.391\\
66.711	-71.891\\
67.323	-71.640\\
71.392	-69.888\\
75.157	-68.136\\
76.974	-67.256\\
80.820	-66.385\\
87.237	-64.781\\
87.544	-64.633\\
91.105	-62.881\\
94.522	-61.130\\
97.500	-59.555\\
0.074	50.000\\
-79.058	-43.969\\
-76.974	-44.675\\
-74.971	-45.365\\
-71.060	-47.117\\
-68.083	-48.868\\
-66.711	-49.639\\
-64.835	-50.620\\
-61.086	-52.372\\
-56.754	-54.123\\
-56.447	-54.240\\
-53.393	-55.875\\
-49.989	-57.627\\
-46.298	-59.378\\
-46.184	-59.429\\
-42.533	-61.130\\
-38.572	-62.881\\
-35.921	-64.006\\
-34.396	-64.633\\
-30.025	-66.385\\
-25.658	-68.065\\
-25.365	-68.136\\
-17.951	-69.888\\
-15.395	-70.467\\
-7.162	-71.640\\
-5.132	-71.918\\
5.132	-72.559\\
15.395	-72.315\\
21.797	-71.640\\
25.658	-71.216\\
30.602	-69.888\\
35.921	-68.334\\
36.716	-68.136\\
43.444	-66.385\\
46.184	-65.625\\
49.632	-64.633\\
55.157	-62.881\\
56.447	-62.440\\
60.471	-61.130\\
65.510	-59.378\\
66.711	-58.941\\
69.329	-57.627\\
72.735	-55.875\\
76.054	-54.123\\
76.974	-53.630\\
80.207	-52.372\\
84.280	-50.620\\
87.237	-49.214\\
87.883	-48.868\\
90.937	-47.117\\
92.704	-46.298\\
0.076	40.000\\
-63.483	-41.311\\
-61.798	-41.862\\
-56.447	-43.535\\
-56.228	-43.614\\
-52.830	-45.365\\
-49.306	-47.117\\
-46.184	-48.799\\
-46.063	-48.868\\
-42.919	-50.620\\
-39.591	-52.372\\
-36.056	-54.123\\
-35.921	-54.188\\
-32.137	-55.875\\
-28.082	-57.627\\
-25.658	-58.647\\
-23.053	-59.378\\
-16.574	-61.130\\
-15.395	-61.439\\
-7.149	-62.881\\
-5.132	-63.222\\
5.132	-63.881\\
15.395	-63.405\\
19.147	-62.881\\
25.658	-61.934\\
28.103	-61.130\\
33.191	-59.378\\
35.921	-58.394\\
38.346	-57.627\\
43.597	-55.875\\
46.184	-54.964\\
48.304	-54.123\\
52.424	-52.372\\
56.186	-50.620\\
56.447	-50.493\\
60.374	-48.868\\
64.338	-47.117\\
66.711	-46.000\\
67.874	-45.365\\
71.095	-43.614\\
72.433	-42.838\\
0.078	29.000\\
-44.800	-39.874\\
-44.380	-40.110\\
-40.644	-41.862\\
-37.499	-43.614\\
-35.921	-44.491\\
-34.104	-45.365\\
-30.401	-47.117\\
-26.602	-48.868\\
-25.658	-49.296\\
-21.510	-50.620\\
-15.767	-52.372\\
-15.395	-52.482\\
-7.367	-54.123\\
-5.132	-54.566\\
5.132	-55.227\\
15.395	-54.494\\
17.494	-54.123\\
25.658	-52.608\\
26.283	-52.372\\
30.806	-50.620\\
35.157	-48.868\\
35.921	-48.554\\
39.683	-47.117\\
44.052	-45.365\\
46.184	-44.479\\
47.932	-43.614\\
51.363	-41.862\\
55.024	-40.110\\
55.419	-39.935\\
0.080	18.000\\
-30.921	-37.505\\
-28.925	-38.359\\
-25.658	-39.982\\
-25.312	-40.110\\
-20.493	-41.862\\
-15.419	-43.614\\
-15.395	-43.622\\
-7.714	-45.365\\
-5.132	-45.937\\
5.132	-46.580\\
15.395	-45.581\\
16.382	-45.365\\
24.040	-43.614\\
25.658	-43.226\\
28.914	-41.862\\
33.010	-40.110\\
35.921	-39.048\\
37.729	-38.359\\
0.082	5.000\\
-10.177	-36.607\\
-5.132	-37.558\\
5.132	-38.040\\
15.395	-37.086\\
17.603	-36.607\\
};
\end{axis}

\begin{axis}[%
width=1.227\fwidth,
height=1.227\fheight,
at={(-0.16\fwidth,-0.135\fheight)},
scale only axis,
xmin=0.000,
xmax=1.000,
ymin=0.000,
ymax=1.000,
axis line style={draw=none},
ticks=none,
axis x line*=bottom,
axis y line*=left
]
\end{axis}
\end{tikzpicture}%
            \caption{Mean daily heliostat field soiling rate (\SI{}{pp\per\day}) interpolated in-between sector representatives for vertical stowing, receiver at (0,0)}
            \label{fig:soilingmapvertical}
        \end{figure}
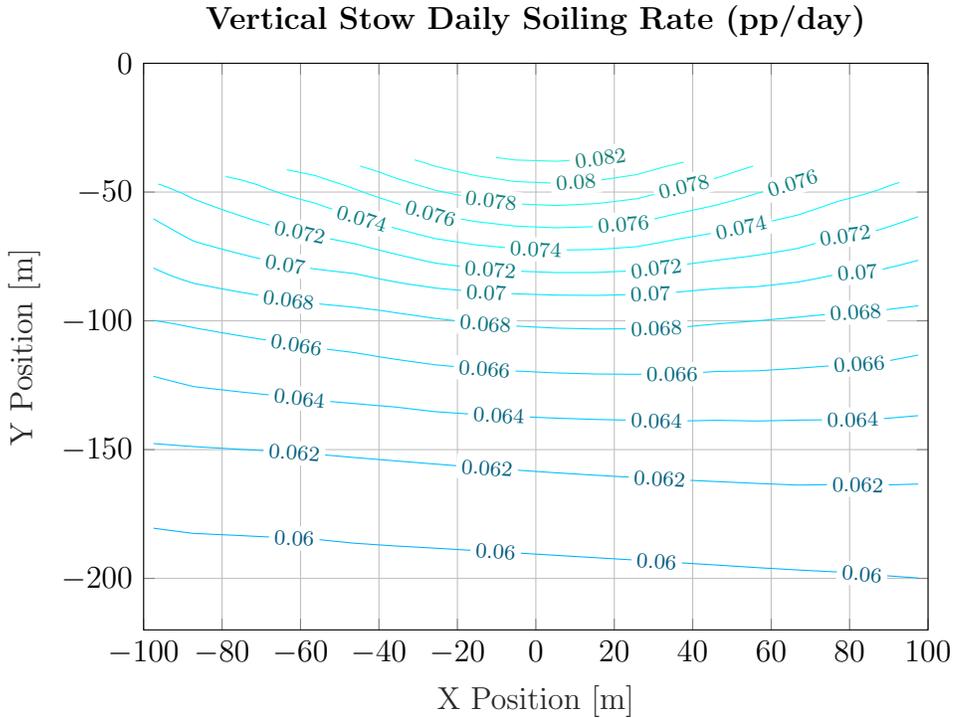

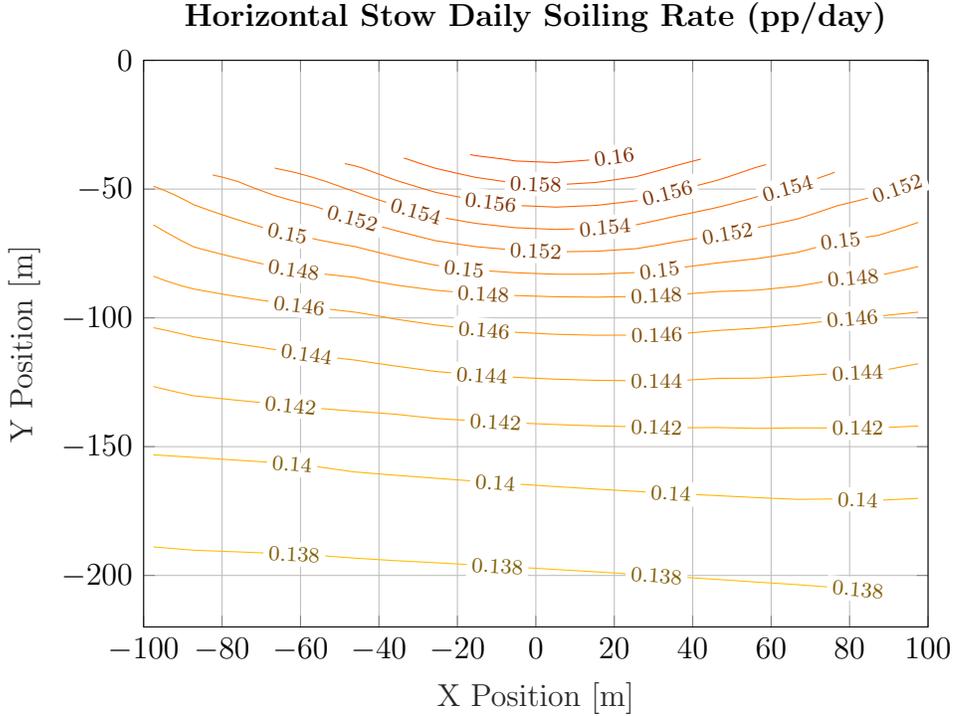
\begin{figure}[h!]
            \centering
\begin{tikzpicture}[%
trim axis left, trim axis right
]

\begin{axis}[%
width=0.951\fwidth,
height=\fheight,
at={(0\fwidth,0\fheight)},
scale only axis,
point meta min=0.000,
point meta max=0.200,
colormap/jet,
xmin=-100.000,
xmax=100.000,
xlabel style={font=\color{white!15!black}},
xlabel={X Position [m]},
ymin=-220.000,
ymax=0.000,
ylabel style={font=\color{white!15!black}},
ylabel={Y Position [m]},
axis background/.style={fill=white},
title style={font=\bfseries},
title={Horizontal Stow Daily Soiling Rate (pp/day)},
xmajorgrids,
ymajorgrids
]
\addplot[contour prepared, contour prepared format=matlab] table[row sep=crcr] {%
0.138	30.000\\
-97.500	-188.934\\
-97.028	-188.999\\
-87.237	-190.214\\
-77.754	-190.751\\
-76.974	-190.794\\
-66.711	-191.453\\
-56.447	-192.085\\
-52.995	-192.502\\
-46.184	-193.277\\
-35.921	-194.251\\
-35.888	-194.254\\
-25.658	-195.051\\
-15.395	-195.925\\
-14.446	-196.005\\
-5.132	-196.786\\
5.132	-197.696\\
5.815	-197.757\\
15.395	-198.615\\
25.092	-199.509\\
25.658	-199.561\\
35.921	-200.545\\
43.308	-201.260\\
46.184	-201.541\\
56.447	-202.602\\
60.778	-203.012\\
66.711	-203.587\\
76.974	-204.756\\
77.039	-204.764\\
87.237	-206.030\\
88.891	-206.233\\
0.140	30.000\\
-97.500	-153.189\\
-89.061	-153.966\\
-87.237	-154.145\\
-76.974	-155.173\\
-71.245	-155.718\\
-66.711	-156.180\\
-57.525	-157.470\\
-56.447	-157.625\\
-48.997	-159.221\\
-46.184	-159.829\\
-36.849	-160.973\\
-35.921	-161.086\\
-25.658	-162.254\\
-21.529	-162.724\\
-15.395	-163.418\\
-5.240	-164.476\\
-5.132	-164.487\\
5.132	-165.521\\
12.717	-166.228\\
15.395	-166.477\\
25.658	-167.375\\
32.944	-167.979\\
35.921	-168.228\\
46.184	-169.011\\
55.678	-169.731\\
56.447	-169.790\\
66.711	-170.466\\
76.974	-170.299\\
87.237	-170.848\\
97.500	-170.087\\
0.142	29.000\\
-97.500	-126.644\\
-94.719	-127.692\\
-89.710	-129.443\\
-87.237	-130.243\\
-81.253	-131.195\\
-76.974	-131.850\\
-69.479	-132.947\\
-66.711	-133.338\\
-58.324	-134.698\\
-56.447	-134.992\\
-46.184	-136.251\\
-44.422	-136.450\\
-35.921	-137.380\\
-30.977	-138.202\\
-25.658	-139.035\\
-15.882	-139.953\\
-15.395	-139.998\\
-5.132	-140.753\\
5.132	-141.408\\
10.978	-141.705\\
15.395	-141.931\\
25.658	-142.297\\
35.921	-142.667\\
46.184	-142.572\\
56.447	-142.905\\
66.711	-142.711\\
76.974	-142.768\\
87.237	-142.697\\
97.500	-141.985\\
0.144	36.000\\
-97.500	-103.809\\
-94.003	-104.921\\
-89.036	-106.672\\
-87.237	-107.351\\
-83.009	-108.424\\
-76.974	-109.892\\
-75.771	-110.176\\
-68.056	-111.927\\
-66.711	-112.222\\
-60.423	-113.679\\
-56.447	-114.570\\
-51.617	-115.430\\
-46.184	-116.366\\
-42.692	-117.182\\
-35.921	-118.662\\
-34.624	-118.934\\
-26.151	-120.685\\
-25.658	-120.786\\
-15.395	-122.123\\
-12.072	-122.437\\
-5.132	-123.084\\
5.132	-123.808\\
13.871	-124.189\\
15.395	-124.255\\
25.658	-124.397\\
35.921	-124.395\\
38.348	-124.189\\
46.184	-123.501\\
56.447	-123.398\\
66.711	-122.527\\
67.695	-122.437\\
76.974	-121.549\\
87.237	-120.756\\
87.491	-120.685\\
93.585	-118.934\\
97.500	-117.773\\
0.146	40.000\\
-97.500	-83.810\\
-97.355	-83.901\\
-94.296	-85.653\\
-90.646	-87.404\\
-87.237	-88.791\\
-85.954	-89.156\\
-79.405	-90.908\\
-76.974	-91.513\\
-72.206	-92.659\\
-66.711	-93.877\\
-64.216	-94.411\\
-56.447	-96.014\\
-55.412	-96.162\\
-46.184	-97.777\\
-45.610	-97.914\\
-38.553	-99.666\\
-35.921	-100.466\\
-31.632	-101.417\\
-25.658	-102.740\\
-23.066	-103.169\\
-15.395	-104.416\\
-11.029	-104.921\\
-5.132	-105.594\\
5.132	-106.382\\
13.248	-106.672\\
15.395	-106.749\\
25.658	-106.678\\
25.830	-106.672\\
35.921	-106.310\\
46.184	-104.923\\
46.211	-104.921\\
56.447	-103.970\\
62.486	-103.169\\
66.711	-102.597\\
73.334	-101.417\\
76.974	-100.759\\
84.455	-99.666\\
87.237	-99.337\\
96.729	-97.914\\
97.500	-97.787\\
0.148	43.000\\
-97.500	-63.893\\
-96.663	-64.633\\
-94.688	-66.385\\
-92.662	-68.136\\
-90.545	-69.888\\
-88.292	-71.640\\
-87.237	-72.417\\
-84.915	-73.391\\
-80.495	-75.143\\
-76.974	-76.449\\
-75.775	-76.895\\
-70.810	-78.646\\
-66.711	-79.991\\
-65.144	-80.398\\
-58.210	-82.149\\
-56.447	-82.583\\
-48.517	-83.901\\
-46.184	-84.275\\
-41.663	-85.653\\
-35.921	-87.196\\
-35.019	-87.404\\
-26.285	-89.156\\
-25.658	-89.266\\
-15.395	-90.614\\
-11.303	-90.908\\
-5.132	-91.336\\
5.132	-91.817\\
15.395	-91.956\\
25.658	-91.688\\
35.921	-90.993\\
36.723	-90.908\\
46.184	-89.803\\
56.447	-89.238\\
57.023	-89.156\\
66.711	-87.679\\
67.872	-87.404\\
74.585	-85.653\\
76.974	-84.952\\
84.899	-83.901\\
87.237	-83.579\\
91.706	-82.149\\
96.628	-80.398\\
97.500	-80.066\\
0.150	50.000\\
-97.500	-48.870\\
-94.541	-50.620\\
-91.957	-52.372\\
-89.620	-54.123\\
-87.438	-55.875\\
-87.237	-56.041\\
-84.398	-57.627\\
-81.094	-59.378\\
-77.595	-61.130\\
-76.974	-61.431\\
-74.007	-62.881\\
-70.218	-64.633\\
-66.711	-66.162\\
-66.037	-66.385\\
-60.538	-68.136\\
-56.447	-69.372\\
-54.881	-69.888\\
-48.746	-71.640\\
-46.184	-72.275\\
-43.356	-73.391\\
-38.711	-75.143\\
-35.921	-76.145\\
-33.797	-76.895\\
-28.513	-78.646\\
-25.658	-79.533\\
-21.119	-80.398\\
-15.395	-81.437\\
-8.436	-82.149\\
-5.132	-82.476\\
5.132	-83.058\\
15.395	-83.049\\
25.658	-82.387\\
27.188	-82.149\\
35.921	-80.641\\
37.270	-80.398\\
46.184	-78.688\\
46.467	-78.646\\
56.447	-76.998\\
56.912	-76.895\\
64.387	-75.143\\
66.711	-74.561\\
69.676	-73.391\\
73.812	-71.640\\
76.974	-70.195\\
78.494	-69.888\\
86.566	-68.136\\
87.237	-67.980\\
90.649	-66.385\\
94.212	-64.633\\
97.500	-62.944\\
0.152	52.000\\
-82.374	-44.535\\
-79.923	-45.365\\
-76.974	-46.624\\
-75.863	-47.117\\
-73.039	-48.868\\
-70.016	-50.620\\
-66.816	-52.372\\
-66.711	-52.428\\
-62.955	-54.123\\
-58.495	-55.875\\
-56.447	-56.624\\
-54.567	-57.627\\
-51.133	-59.378\\
-47.296	-61.130\\
-46.184	-61.597\\
-43.351	-62.881\\
-39.312	-64.633\\
-35.921	-66.039\\
-35.064	-66.385\\
-30.624	-68.136\\
-25.982	-69.888\\
-25.658	-70.007\\
-18.533	-71.640\\
-15.395	-72.328\\
-7.558	-73.391\\
-5.132	-73.708\\
5.132	-74.343\\
15.395	-74.144\\
22.998	-73.391\\
25.658	-73.117\\
31.480	-71.640\\
35.921	-70.407\\
38.089	-69.888\\
45.029	-68.136\\
46.184	-67.830\\
51.539	-66.385\\
56.447	-64.904\\
57.345	-64.633\\
62.896	-62.881\\
66.711	-61.594\\
67.664	-61.130\\
71.188	-59.378\\
74.613	-57.627\\
76.974	-56.392\\
78.479	-55.875\\
83.166	-54.123\\
87.237	-52.401\\
87.291	-52.372\\
90.566	-50.620\\
93.827	-48.868\\
97.088	-47.117\\
97.199	-47.065\\
0.154	42.000\\
-66.597	-41.842\\
-66.538	-41.862\\
-61.205	-43.614\\
-57.062	-45.365\\
-56.447	-45.640\\
-53.388	-47.117\\
-50.171	-48.868\\
-46.878	-50.620\\
-46.184	-50.985\\
-43.630	-52.372\\
-40.218	-54.123\\
-36.599	-55.875\\
-35.921	-56.193\\
-32.655	-57.627\\
-28.544	-59.378\\
-25.658	-60.573\\
-23.621	-61.130\\
-16.980	-62.881\\
-15.395	-63.286\\
-7.404	-64.633\\
-5.132	-65.002\\
5.132	-65.660\\
15.395	-65.233\\
19.921	-64.633\\
25.658	-63.841\\
28.674	-62.881\\
33.903	-61.130\\
35.921	-60.423\\
39.349	-59.378\\
44.751	-57.627\\
46.184	-57.139\\
49.526	-55.875\\
53.779	-54.123\\
56.447	-52.932\\
57.892	-52.372\\
62.213	-50.620\\
66.242	-48.868\\
66.711	-48.656\\
69.557	-47.117\\
72.738	-45.365\\
75.968	-43.614\\
76.197	-43.481\\
0.156	31.000\\
-48.701	-40.110\\
-46.184	-40.953\\
-44.184	-41.862\\
-41.048	-43.614\\
-37.929	-45.365\\
-35.921	-46.470\\
-34.563	-47.117\\
-30.822	-48.868\\
-26.978	-50.620\\
-25.658	-51.210\\
-21.924	-52.372\\
-16.049	-54.123\\
-15.395	-54.313\\
-7.544	-55.875\\
-5.132	-56.339\\
5.132	-57.001\\
15.395	-56.323\\
18.046	-55.875\\
25.658	-54.527\\
26.753	-54.123\\
31.366	-52.372\\
35.786	-50.620\\
35.921	-50.566\\
40.483	-48.868\\
44.936	-47.117\\
46.184	-46.609\\
48.768	-45.365\\
52.217	-43.614\\
55.545	-41.862\\
56.447	-41.437\\
58.732	-40.500\\
0.158	20.000\\
-33.663	-37.973\\
-32.761	-38.359\\
-29.370	-40.110\\
-25.722	-41.862\\
-25.658	-41.893\\
-20.805	-43.614\\
-15.603	-45.365\\
-15.395	-45.433\\
-7.842	-47.117\\
-5.132	-47.706\\
5.132	-48.355\\
15.395	-47.410\\
16.789	-47.117\\
24.730	-45.365\\
25.658	-45.151\\
29.376	-43.614\\
33.507	-41.862\\
35.921	-40.820\\
37.587	-40.110\\
42.070	-38.359\\
0.160	8.000\\
-16.709	-36.607\\
-15.395	-37.009\\
-8.244	-38.359\\
-5.132	-39.094\\
5.132	-39.707\\
15.395	-38.513\\
16.009	-38.359\\
24.014	-36.607\\
};
\end{axis}

\begin{axis}[%
width=1.227\fwidth,
height=1.227\fheight,
at={(-0.16\fwidth,-0.135\fheight)},
scale only axis,
xmin=0.000,
xmax=1.000,
ymin=0.000,
ymax=1.000,
axis line style={draw=none},
ticks=none,
axis x line*=bottom,
axis y line*=left
]
\end{axis}
\end{tikzpicture}%
            \caption{Mean daily heliostat field soiling rate (\SI{}{pp\per\day}) interpolated in-between sector representatives for horizontal stowing, receiver at (0,0)}
            \label{fig:soilingmaphorizontal}
        \end{figure}

        Soiling rates predicted from the \gls{dsm} for all simulation periods are summed daily and averaged across each heliostat sector\footnote{All solar field arrays will experience similar soiling fluxes but differ due to there tilt angles} and shown in Figs~\ref{fig:soilingmapvertical} and \ref{fig:soilingmaphorizontal} for both vertical and horizontal stow respectively. The plots suggest that horizontally stowed heliostats undergo higher daily soiling rates (0.15 pp/day) with respect to those vertically stowed (0.07 pp/day). Thus, an increase in daily soiling rates of \SI{114}{\percent}.

        Both soiling rate maps show a relation between soiling rate and relative position within the solar field. Soiling rates tend to decrease with distance from the receiver; since heliostats close to the receiver have horizontally inclined tilt angles corresponding to higher soiling rates. Although mirrors closer to the receiver will have higher soiling rates the resultant impact on optical performance is slightly compensated due to lower incidence angles and thus lower geometry factor values, as expressed in \eqref{eq:geometryfactor}. 
        
        Additionally, for heliostats with the same Y position, daily soiling rates increase with increasing X position (i.e. further East). This is due to the intraday time synchronous average \gls{tsp} (see Fig.~\ref{fig:seasonalcomponentcheck} right) being slightly higher during the morning when eastern heliostats have lower tilt angles with respect to western heliostats aiming at the receiver.
        
        \subsubsection{Night Time Dispatching Only Evaluation}
        \label{subsec:nighttimeresults}
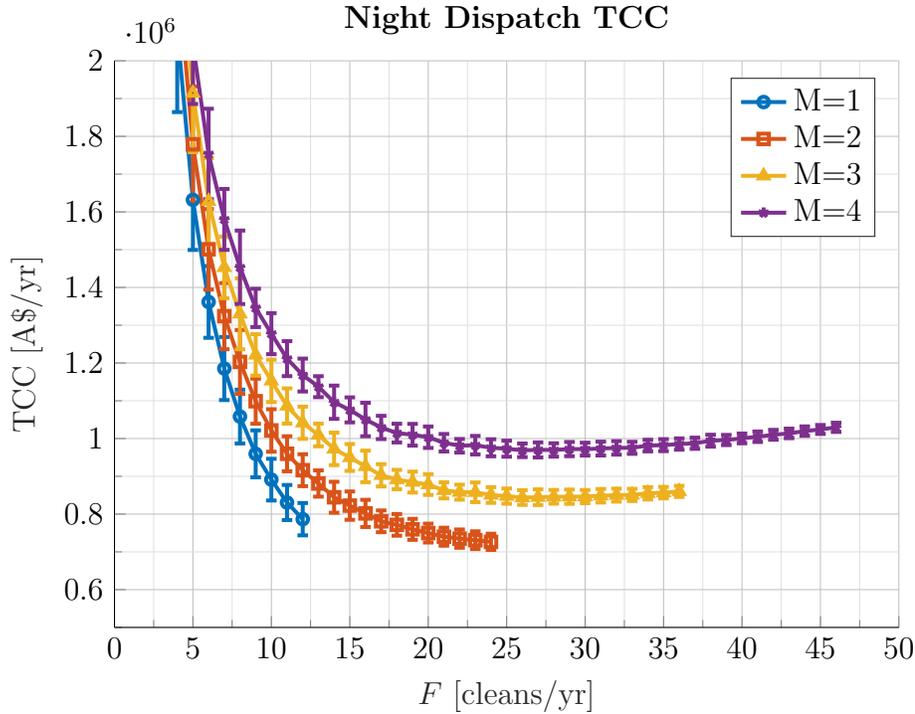
\begin{figure}[h!]
            \centering
\definecolor{mycolor1}{rgb}{0.00000,0.44700,0.74100}%
\definecolor{mycolor2}{rgb}{0.85000,0.32500,0.09800}%
\definecolor{mycolor3}{rgb}{0.92900,0.69400,0.12500}%
\definecolor{mycolor4}{rgb}{0.49400,0.18400,0.55600}%
\begin{tikzpicture}[%
trim axis left, trim axis right
]

\begin{axis}[%
width=0.951\fwidth,
height=\fheight,
at={(0\fwidth,0\fheight)},
scale only axis,
unbounded coords=jump,
xmin=0.000,
xmax=50.000,
xlabel style={font=\color{white!15!black}},
xlabel={\ta{ncleans} [cleans/yr]},
ymin=500000.000,
ymax=2000000.000,
ylabel style={font=\color{white!15!black}},
ylabel={TCC [A\$/yr]},
title style={font=\bfseries},
title={Night Dispatch \ta{tcc}},
axis background/.style={fill=white},
axis x line*=bottom,
axis y line*=left,
grid=both,
grid style={line width=.1pt, draw=gray!25},
major grid style={line width=.2pt,draw=gray!50},
minor tick num=1,
legend pos=north east,
legend style={legend cell align=left, align=left, draw=white!15!black}
]
\addplot [color=mycolor1, line width=1.5pt ,mark=o, mark options={solid, mycolor1}]
 plot [error bars/.cd, y dir=both, y explicit, error bar style={line width=1.5pt}, error mark options={line width=1.5pt, mark size=2.0pt, rotate=90}]
 table[row sep=crcr, y error plus index=2, y error minus index=3]{%
1.000	8562790.877	508988.193	508988.193\\
2.000	4215108.751	451892.856	451892.856\\
3.000	2886580.196	271518.277	271518.277\\
4.000	2083681.760	219487.080	219487.080\\
5.000	1631856.232	132573.471	132573.471\\
6.000	1361600.177	94937.134	94937.134\\
7.000	1185355.807	83456.029	83456.029\\
8.000	1058246.240	71263.705	71263.705\\
9.000	959500.254	62211.585	62211.585\\
10.000	891218.433	55235.649	55235.649\\
11.000	830665.413	46410.633	46410.633\\
12.000	786552.289	42781.252	42781.252\\
13.000	nan	nan	nan\\
14.000	nan	nan	nan\\
15.000	nan	nan	nan\\
16.000	nan	nan	nan\\
17.000	nan	nan	nan\\
18.000	nan	nan	nan\\
19.000	nan	nan	nan\\
20.000	nan	nan	nan\\
21.000	nan	nan	nan\\
22.000	nan	nan	nan\\
23.000	nan	nan	nan\\
24.000	nan	nan	nan\\
25.000	nan	nan	nan\\
26.000	nan	nan	nan\\
27.000	nan	nan	nan\\
28.000	nan	nan	nan\\
29.000	nan	nan	nan\\
30.000	nan	nan	nan\\
31.000	nan	nan	nan\\
32.000	nan	nan	nan\\
33.000	nan	nan	nan\\
34.000	nan	nan	nan\\
35.000	nan	nan	nan\\
36.000	nan	nan	nan\\
37.000	nan	nan	nan\\
38.000	nan	nan	nan\\
39.000	nan	nan	nan\\
40.000	nan	nan	nan\\
41.000	nan	nan	nan\\
42.000	nan	nan	nan\\
43.000	nan	nan	nan\\
44.000	nan	nan	nan\\
45.000	nan	nan	nan\\
46.000	nan	nan	nan\\
};
\addlegendentry{M=1}

\addplot [color=mycolor2, line width=1.5pt ,mark=square, mark options={solid, mycolor2}]
 plot [error bars/.cd, y dir=both, y explicit, error bar style={line width=1.5pt}, error mark options={line width=1.5pt, mark size=2.0pt, rotate=90}]
 table[row sep=crcr, y error plus index=2, y error minus index=3]{%
1.000	8792586.285	462913.959	462913.959\\
2.000	4347176.269	447614.398	447614.398\\
3.000	3030321.373	278475.384	278475.384\\
4.000	2245401.164	227351.059	227351.059\\
5.000	1778792.363	153476.409	153476.409\\
6.000	1500950.671	106652.281	106652.281\\
7.000	1323951.722	87468.967	87468.967\\
8.000	1202862.472	84590.235	84590.235\\
9.000	1098795.636	59697.736	59697.736\\
10.000	1021436.730	56268.872	56268.872\\
11.000	959650.219	46387.683	46387.683\\
12.000	916191.592	42268.236	42268.236\\
13.000	880923.857	34717.236	34717.236\\
14.000	844557.022	41022.833	41022.833\\
15.000	822428.769	37457.940	37457.940\\
16.000	801821.572	36008.787	36008.787\\
17.000	780912.675	28770.140	28770.140\\
18.000	771521.259	28730.424	28730.424\\
19.000	759680.944	27793.422	27793.422\\
20.000	749756.535	24911.574	24911.574\\
21.000	740161.080	23816.014	23816.014\\
22.000	735079.758	24235.643	24235.643\\
23.000	730933.401	23503.898	23503.898\\
24.000	726576.646	21525.138	21525.138\\
25.000	nan	nan	nan\\
26.000	nan	nan	nan\\
27.000	nan	nan	nan\\
28.000	nan	nan	nan\\
29.000	nan	nan	nan\\
30.000	nan	nan	nan\\
31.000	nan	nan	nan\\
32.000	nan	nan	nan\\
33.000	nan	nan	nan\\
34.000	nan	nan	nan\\
35.000	nan	nan	nan\\
36.000	nan	nan	nan\\
37.000	nan	nan	nan\\
38.000	nan	nan	nan\\
39.000	nan	nan	nan\\
40.000	nan	nan	nan\\
41.000	nan	nan	nan\\
42.000	nan	nan	nan\\
43.000	nan	nan	nan\\
44.000	nan	nan	nan\\
45.000	nan	nan	nan\\
46.000	nan	nan	nan\\
};
\addlegendentry{M=2}

\addplot [color=mycolor3, line width=1.5pt ,mark=triangle, mark options={solid, mycolor3}]
 plot [error bars/.cd, y dir=both, y explicit, error bar style={line width=1.5pt}, error mark options={line width=1.5pt, mark size=2.0pt, rotate=90}]
 table[row sep=crcr, y error plus index=2, y error minus index=3]{%
1.000	8959689.544	444896.597	444896.597\\
2.000	4465272.656	434656.391	434656.391\\
3.000	3153606.304	268963.974	268963.974\\
4.000	2379300.050	230850.687	230850.687\\
5.000	1914626.615	156499.053	156499.053\\
6.000	1628330.320	113838.933	113838.933\\
7.000	1453103.636	81593.768	81593.768\\
8.000	1330164.824	93917.975	93917.975\\
9.000	1221340.127	54910.877	54910.877\\
10.000	1152757.827	56189.130	56189.130\\
11.000	1086224.444	46614.009	46614.009\\
12.000	1041713.971	42670.624	42670.624\\
13.000	1009438.965	30043.261	30043.261\\
14.000	972642.888	43000.525	43000.525\\
15.000	949858.367	35580.104	35580.104\\
16.000	926233.381	41797.926	41797.926\\
17.000	902709.127	29490.988	29490.988\\
18.000	891534.007	25601.478	25601.478\\
19.000	884699.333	27739.034	27739.034\\
20.000	878233.720	27521.373	27521.373\\
21.000	862793.458	21312.077	21312.077\\
22.000	858102.325	19828.424	19828.424\\
23.000	857720.524	26040.193	26040.193\\
24.000	850543.039	20820.730	20820.730\\
25.000	847588.316	20142.253	20142.253\\
26.000	843883.836	19449.371	19449.371\\
27.000	844722.027	20788.881	20788.881\\
28.000	845754.435	18147.898	18147.898\\
29.000	846030.491	18922.613	18922.613\\
30.000	846391.384	17515.794	17515.794\\
31.000	848541.856	17940.439	17940.439\\
32.000	850252.564	17167.374	17167.374\\
33.000	850716.764	16195.881	16195.881\\
34.000	855027.592	16262.472	16262.472\\
35.000	856716.648	15150.324	15150.324\\
36.000	860137.843	14709.907	14709.907\\
37.000	nan	nan	nan\\
38.000	nan	nan	nan\\
39.000	nan	nan	nan\\
40.000	nan	nan	nan\\
41.000	nan	nan	nan\\
42.000	nan	nan	nan\\
43.000	nan	nan	nan\\
44.000	nan	nan	nan\\
45.000	nan	nan	nan\\
46.000	nan	nan	nan\\
};
\addlegendentry{M=3}

\addplot [color=mycolor4, line width=1.5pt ,mark=star, mark options={solid, mycolor4}]
 plot [error bars/.cd, y dir=both, y explicit, error bar style={line width=1.5pt}, error mark options={line width=1.5pt, mark size=2.0pt, rotate=90}]
 table[row sep=crcr, y error plus index=2, y error minus index=3]{%
1.000	9092763.180	435489.874	435489.874\\
2.000	4579296.147	421518.345	421518.345\\
3.000	3276822.049	261514.136	261514.136\\
4.000	2501899.741	228448.462	228448.462\\
5.000	2044195.901	158630.139	158630.139\\
6.000	1753331.705	119772.085	119772.085\\
7.000	1580149.607	80653.733	80653.733\\
8.000	1453347.630	97288.003	97288.003\\
9.000	1345792.670	50737.658	50737.658\\
10.000	1277695.211	54293.929	54293.929\\
11.000	1211518.715	46275.886	46275.886\\
12.000	1168106.631	43531.712	43531.712\\
13.000	1137301.676	27763.403	27763.403\\
14.000	1096161.296	44019.917	44019.917\\
15.000	1075892.022	33077.365	33077.365\\
16.000	1050419.401	44387.019	44387.019\\
17.000	1029387.345	30791.876	30791.876\\
18.000	1014484.427	24770.244	24770.244\\
19.000	1010046.788	28771.771	28771.771\\
20.000	1003532.901	28354.769	28354.769\\
21.000	988926.595	23187.348	23187.348\\
22.000	980915.635	17978.007	17978.007\\
23.000	982435.139	24544.243	24544.243\\
24.000	975594.349	22587.440	22587.440\\
25.000	973136.661	21248.000	21248.000\\
26.000	968891.410	17975.311	17975.311\\
27.000	969952.590	19923.083	19923.083\\
28.000	969833.497	17681.009	17681.009\\
29.000	972130.896	18851.226	18851.226\\
30.000	972028.841	17044.529	17044.529\\
31.000	973501.403	18318.188	18318.188\\
32.000	975308.863	17860.276	17860.276\\
33.000	975511.049	15970.087	15970.087\\
34.000	981428.166	15390.279	15390.279\\
35.000	982542.215	15900.503	15900.503\\
36.000	985471.889	14923.455	14923.455\\
37.000	987563.601	14857.262	14857.262\\
38.000	994343.345	15798.518	15798.518\\
39.000	996188.205	13483.490	13483.490\\
40.000	1000367.609	12960.091	12960.091\\
41.000	1004813.927	13863.316	13863.316\\
42.000	1009926.121	13061.638	13061.638\\
43.000	1013630.721	12663.811	12663.811\\
44.000	1019716.345	12458.750	12458.750\\
45.000	1024089.278	11850.812	11850.812\\
46.000	1029699.561	12219.908	12219.908\\
};
\addlegendentry{M=4}

\end{axis}
\end{tikzpicture}%
            \caption{Total cleaning cost for night-time dispatching scenario}
            \label{fig:nightdtcc}
        \end{figure}

        The night-time only dispatch policy introduces a night-time only \gls{csp} power block operation mode. Where \gls{pv} energy provides day-time power while the \gls{csp} charges the \gls{tes}; during the night the power block is run using the charged \gls{tes}. Comparing the always cleaned power generation of the night-time dispatch and base scenarios of table~\ref{table:optimalresultssummary} shows that the night-time dispatch scenario produces less electricity then the base scenario. The difference is due to the receiver system turning off when the \gls{tes} is filled, whose size is kept equal to the base case. Total cleaning costs plotted in Fig.~\ref{fig:nightdtcc} show an optimal cleaning setup with two cleaning crews cleaning the field 24 times per year.
        
\begin{figure}[h!]
            \centering
\definecolor{mycolor1}{rgb}{0.00000,0.44700,0.74100}%
\definecolor{mycolor2}{rgb}{0.85000,0.32500,0.09800}%
\begin{tikzpicture}[%
trim axis left, trim axis right
]

\begin{axis}[%
width=0.951\fwidth,
height=\fheight,
at={(0\fwidth,0\fheight)},
scale only axis,
bar width=0.8,
xmin=-0.200,
xmax=5.200,
xtick={1.000,2.000,3.000,4.000},
xticklabels={{Base},{Base C.},{Night},{Night C.}},
ymin=0.000,
ymax=15000.000,
ylabel style={font=\color{white!15!black}},
ylabel={$\text{Energy Loss [\SI{}{\mega\watt\hour\per\year}}^{\text{-1}}\text{]}$},
title style={font=\bfseries},
title={Energy Loss Comparison},
axis background/.style={fill=white},
ymajorgrids=true,
yminorgrids=true,
minor y tick num=1,
legend pos=north west,
legend style={legend cell align=left, align=left, draw=white!15!black}
]
\addplot[ybar stacked, fill=mycolor1, draw=none, area legend] table[row sep=crcr] {%
1.000	227.678\\
2.000	393.107\\
3.000	227.678\\
4.000	393.107\\
};
\addplot[forget plot, color=white!15!black, draw=none, only marks] table[row sep=crcr] {%
-0.200	0.000\\
5.200	0.000\\
};
\addlegendentry{Saturation}

\addplot[ybar stacked, fill=mycolor2, draw=none, area legend] table[row sep=crcr] {%
1.000	0.000\\
2.000	0.000\\
3.000	10316.839\\
4.000	13004.196\\
};
\addplot[forget plot, color=white!15!black, draw=none, only marks] table[row sep=crcr] {%
-0.200	0.000\\
5.200	0.000\\
};
\addlegendentry{Capacity}

\end{axis}

\begin{axis}[%
width=1.227\fwidth,
height=1.227\fheight,
at={(-0.16\fwidth,-0.135\fheight)},
scale only axis,
xmin=0.000,
xmax=1.000,
ymin=0.000,
ymax=1.000,
axis line style={draw=none},
ticks=none,
axis x line*=bottom,
axis y line*=left
]
\end{axis}
\end{tikzpicture}%
            \caption{Thermal energy loss due to receiver and \gls{tes} operational constraints during the base and night time dispatch scenarios for a soiled and cleaned (C) heliostat field}
            \label{fig:energyloss}
        \end{figure}
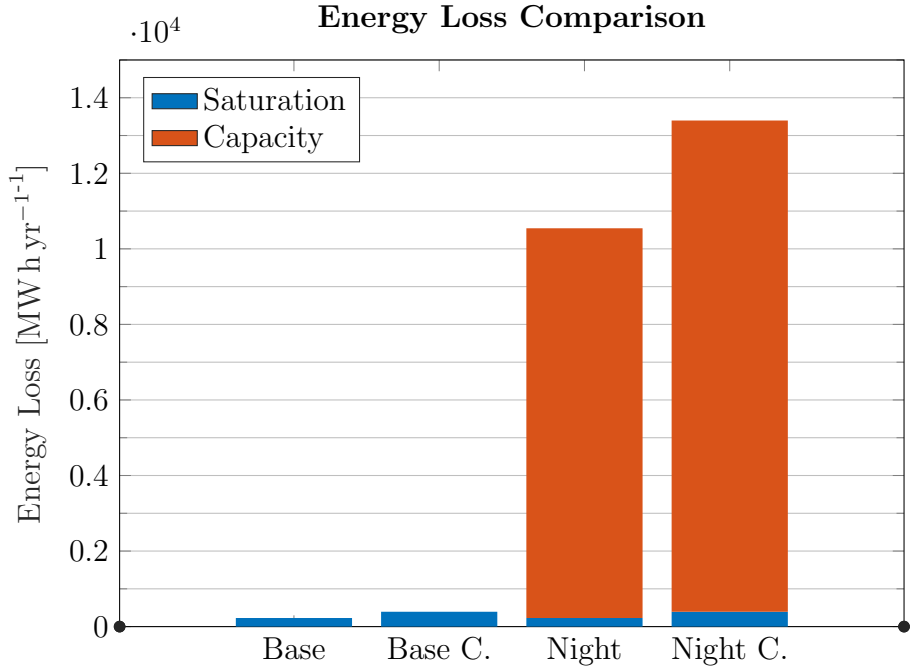

        In the night dispatch scenario, degradation costs are lower compared to the base case due to the introduced constraints on Thermal Energy Storage (TES) capacity. These constraints result in a reduced need for a perfectly clean solar field. The concept of "free" soiling also applies when the receiver operates in a saturated condition, where heliostats are defocused to prevent overheating of the receiver system. Figure~\ref{fig:energyloss} illustrates the excess energy lost due to operational constraints on the receiver and TES for both the base and night-time dispatch scenarios. Comparing the perfectly clean and soiled scenarios reveals that some of the energy reflected from a perfectly clean solar field is partially lost due to the storage and receiver saturation constraints. Consequently, the plant's average yearly cleanliness, while meeting night-time production requirements, can be approximated as... [add the relevant approximation or further discussion here].
        
        Degradation costs are notably lower in the night dispatch scenario compared to the base case, primarily due to the implementation of \gls{tes} capacity constraints. These constraints effectively reduce the necessity for maintaining a perfectly clean solar field. Interestingly, the concept of "free" soiling also extends to instances when the receiver operates in a saturated condition, leading to heliostats being defocused to avoid overheating. Fig.~\ref{fig:energyloss} illustrates the excess energy lost due to operational constraints on the receiver and \gls{tes} for both the base and night-time dispatch scenarios. By comparing the perfectly clean and soiled scenarios, it becomes evident that some of the energy reflected from a perfectly clean solar field is partially lost due to the storage and receiver saturation constraints. Consequently, it is possible to approximate the average yearly cleanliness that the plant can maintain and still fully charge the storage as:
        \begin{equation}
            \label{eq:freesoiling}
            \zeta_{\text{max}} = 1 - \frac{ w_{\text{sat}}^{\text{clean}} + w_{\text{cap}}^{\text{clean}} }{w_{\text{rated}}^{\text{clean}} } \cdot 100 \\
        \end{equation}
        where, $\zeta_{\text{max}}$ is the yearly field average cleanliness corresponding to maximum production, $w_{\text{sat}}^{\text{clean}}$ is the work loss due to receiver saturation and $w_{\text{cap}}^{\text{clean}}$ is the work loss due to \gls{tes} capacity constraints with an always clean solar field. As such, when operating with a night-time dispatch, a yearly field average cleanliness of \SI{97.6}{\percent} would allow for the \gls{tes} to completely charge on most days. 

        \subsubsection{Day Time Cleaning Evaluation}
        Results from Table~\ref{table:optimalresultssummary} show an optimal cleaning schedule occurs with two crews cleaning the solar fields 24 times per year (see Fig.~\ref{fig:dayctcc}). The effects of cleaning during the day can be seen when comparing the work produced by an always clean solar field,  where the base scenario produces \SI{495}{\mega\watt\hour} more then the day cleaning policy. Interestingly, degradation costs are slightly lower due to the solar field being cleaner during production time. However the reduced heliostat up-time leads to higher operation costs that exceed any benefit of having a cleaner solar field during the day; incurring a total cleaning cost increase of \SI{7}{\percent} relative to a night-time cleaning policy. As such each crew contributes around A\$\SI{32000}{\per\year} to the total cleaning cost making it a minor cost relative to other cleaning expenditures listed in Fig.~\ref{fig:cleaningcostbar}.
        
        \begin{figure}[h!]
            \centering
\definecolor{mycolor1}{rgb}{0.00000,0.44700,0.74100}%
\definecolor{mycolor2}{rgb}{0.85000,0.32500,0.09800}%
\definecolor{mycolor3}{rgb}{0.92900,0.69400,0.12500}%
\definecolor{mycolor4}{rgb}{0.49400,0.18400,0.55600}%
\begin{tikzpicture}[%
trim axis left, trim axis right
]

\begin{axis}[%
width=0.951\fwidth,
height=\fheight,
at={(0\fwidth,0\fheight)},
scale only axis,
unbounded coords=jump,
xmin=0.000,
xmax=50.000,
xlabel style={font=\color{white!15!black}},
xlabel={\ta{ncleans} [cleans/yr]},
ymin=500000.000,
ymax=2000000.000,
ylabel style={font=\color{white!15!black}},
ylabel={TCC [A\$/yr]},
title style={font=\bfseries},
title={Day Cleaning \ta{tcc}},
axis background/.style={fill=white},
axis x line*=bottom,
axis y line*=left,
grid=both,
grid style={line width=.1pt, draw=gray!25},
major grid style={line width=.2pt,draw=gray!50},
minor tick num=1,
legend pos=south west,
legend style={legend cell align=left, align=left, draw=white!15!black}
]
\addplot [color=mycolor1, line width=1.5pt ,mark=o, mark options={solid, mycolor1}]
 plot [error bars/.cd, y dir=both, y explicit, error bar style={line width=1.5pt}, error mark options={line width=1.5pt, mark size=2.0pt, rotate=90}]
 table[row sep=crcr, y error plus index=2, y error minus index=3]{%
1.000	9176686.742	563753.039	563753.039\\
2.000	4820240.043	484024.034	484024.034\\
3.000	3427504.420	308784.484	308784.484\\
4.000	2594165.655	256830.104	256830.104\\
5.000	2121289.484	159692.485	159692.485\\
6.000	1801056.935	127413.609	127413.609\\
7.000	1592008.603	124678.657	124678.657\\
8.000	1443200.299	110412.005	110412.005\\
9.000	1325629.354	100529.089	100529.089\\
10.000	1236257.485	84295.361	84295.361\\
11.000	1155120.512	76075.906	76075.906\\
12.000	1096546.482	69522.846	69522.846\\
13.000	nan	nan	nan\\
14.000	nan	nan	nan\\
15.000	nan	nan	nan\\
16.000	nan	nan	nan\\
17.000	nan	nan	nan\\
18.000	nan	nan	nan\\
19.000	nan	nan	nan\\
20.000	nan	nan	nan\\
21.000	nan	nan	nan\\
22.000	nan	nan	nan\\
23.000	nan	nan	nan\\
24.000	nan	nan	nan\\
25.000	nan	nan	nan\\
26.000	nan	nan	nan\\
27.000	nan	nan	nan\\
28.000	nan	nan	nan\\
29.000	nan	nan	nan\\
30.000	nan	nan	nan\\
31.000	nan	nan	nan\\
32.000	nan	nan	nan\\
33.000	nan	nan	nan\\
34.000	nan	nan	nan\\
35.000	nan	nan	nan\\
36.000	nan	nan	nan\\
37.000	nan	nan	nan\\
38.000	nan	nan	nan\\
39.000	nan	nan	nan\\
40.000	nan	nan	nan\\
41.000	nan	nan	nan\\
42.000	nan	nan	nan\\
43.000	nan	nan	nan\\
44.000	nan	nan	nan\\
45.000	nan	nan	nan\\
46.000	nan	nan	nan\\
};
\addlegendentry{M=1}

\addplot [color=mycolor2, line width=1.5pt ,mark=square, mark options={solid, mycolor2}]
 plot [error bars/.cd, y dir=both, y explicit, error bar style={line width=1.5pt}, error mark options={line width=1.5pt, mark size=2.0pt, rotate=90}]
 table[row sep=crcr, y error plus index=2, y error minus index=3]{%
1.000	9379935.592	517661.345	517661.345\\
2.000	4927145.801	475144.137	475144.137\\
3.000	3553886.640	305787.457	305787.457\\
4.000	2721142.413	266990.319	266990.319\\
5.000	2262261.059	182128.977	182128.977\\
6.000	1933557.484	135614.044	135614.044\\
7.000	1723556.902	121118.965	121118.965\\
8.000	1574566.868	108769.187	108769.187\\
9.000	1448572.195	83319.265	83319.265\\
10.000	1369617.206	81023.097	81023.097\\
11.000	1286425.166	70501.738	70501.738\\
12.000	1221771.021	68413.077	68413.077\\
13.000	1178556.207	64240.992	64240.992\\
14.000	1135024.718	67376.689	67376.689\\
15.000	1104718.097	59593.073	59593.073\\
16.000	1080276.536	56170.098	56170.098\\
17.000	1053784.290	50243.035	50243.035\\
18.000	1037002.955	46337.734	46337.734\\
19.000	1024455.501	45146.144	45146.144\\
20.000	1014523.171	42821.178	42821.178\\
21.000	1002992.035	41762.012	41762.012\\
22.000	995768.218	39551.227	39551.227\\
23.000	990249.345	37726.859	37726.859\\
24.000	987218.386	35863.398	35863.398\\
25.000	nan	nan	nan\\
26.000	nan	nan	nan\\
27.000	nan	nan	nan\\
28.000	nan	nan	nan\\
29.000	nan	nan	nan\\
30.000	nan	nan	nan\\
31.000	nan	nan	nan\\
32.000	nan	nan	nan\\
33.000	nan	nan	nan\\
34.000	nan	nan	nan\\
35.000	nan	nan	nan\\
36.000	nan	nan	nan\\
37.000	nan	nan	nan\\
38.000	nan	nan	nan\\
39.000	nan	nan	nan\\
40.000	nan	nan	nan\\
41.000	nan	nan	nan\\
42.000	nan	nan	nan\\
43.000	nan	nan	nan\\
44.000	nan	nan	nan\\
45.000	nan	nan	nan\\
46.000	nan	nan	nan\\
};
\addlegendentry{M=2}

\addplot [color=mycolor3, line width=1.5pt ,mark=triangle, mark options={solid, mycolor3}]
 plot [error bars/.cd, y dir=both, y explicit, error bar style={line width=1.5pt}, error mark options={line width=1.5pt, mark size=2.0pt, rotate=90}]
 table[row sep=crcr, y error plus index=2, y error minus index=3]{%
1.000	9550297.662	503459.322	503459.322\\
2.000	5050761.618	467211.045	467211.045\\
3.000	3684380.019	297861.098	297861.098\\
4.000	2852987.708	268213.691	268213.691\\
5.000	2397823.616	187568.130	187568.130\\
6.000	2063368.543	141943.171	141943.171\\
7.000	1859709.449	119255.744	119255.744\\
8.000	1706213.773	111442.160	111442.160\\
9.000	1566326.732	71458.254	71458.254\\
10.000	1493425.439	79343.233	79343.233\\
11.000	1414862.444	70362.609	70362.609\\
12.000	1347422.346	66305.362	66305.362\\
13.000	1306454.031	60628.644	60628.644\\
14.000	1264213.244	68704.153	68704.153\\
15.000	1238167.144	61219.000	61219.000\\
16.000	1208260.345	60938.911	60938.911\\
17.000	1180538.524	53370.194	53370.194\\
18.000	1157460.062	47276.947	47276.947\\
19.000	1144806.978	41805.811	41805.811\\
20.000	1142227.938	43757.761	43757.761\\
21.000	1128908.020	42305.467	42305.467\\
22.000	1122197.551	39828.568	39828.568\\
23.000	1119149.990	40366.082	40366.082\\
24.000	1112890.986	35365.478	35365.478\\
25.000	1109808.006	32990.928	32990.928\\
26.000	1108000.577	32617.189	32617.189\\
27.000	1108837.119	33075.708	33075.708\\
28.000	1110663.620	29524.530	29524.530\\
29.000	1112677.172	31651.711	31651.711\\
30.000	1116386.753	29490.266	29490.266\\
31.000	1120609.536	28557.760	28557.760\\
32.000	1124785.212	28197.483	28197.483\\
33.000	1128725.105	26997.964	26997.964\\
34.000	1135346.126	26002.148	26002.148\\
35.000	1140824.308	25135.348	25135.348\\
36.000	1147426.979	24680.257	24680.257\\
37.000	nan	nan	nan\\
38.000	nan	nan	nan\\
39.000	nan	nan	nan\\
40.000	nan	nan	nan\\
41.000	nan	nan	nan\\
42.000	nan	nan	nan\\
43.000	nan	nan	nan\\
44.000	nan	nan	nan\\
45.000	nan	nan	nan\\
46.000	nan	nan	nan\\
};
\addlegendentry{M=3}

\addplot [color=mycolor4, line width=1.5pt ,mark=star, mark options={solid, mycolor4}]
 plot [error bars/.cd, y dir=both, y explicit, error bar style={line width=1.5pt}, error mark options={line width=1.5pt, mark size=2.0pt, rotate=90}]
 table[row sep=crcr, y error plus index=2, y error minus index=3]{%
1.000	9688400.369	495028.768	495028.768\\
2.000	5167446.115	453389.172	453389.172\\
3.000	3812845.073	289302.497	289302.497\\
4.000	2973631.831	260858.130	260858.130\\
5.000	2522934.103	187348.387	187348.387\\
6.000	2191915.746	144723.288	144723.288\\
7.000	1991464.037	118318.507	118318.507\\
8.000	1831245.542	110824.133	110824.133\\
9.000	1690624.172	67030.458	67030.458\\
10.000	1610670.644	74386.478	74386.478\\
11.000	1539391.787	69802.239	69802.239\\
12.000	1473308.142	64653.966	64653.966\\
13.000	1431902.039	55645.112	55645.112\\
14.000	1389791.604	67884.064	67884.064\\
15.000	1366660.740	58965.025	58965.025\\
16.000	1332747.384	62443.382	62443.382\\
17.000	1308528.174	56131.505	56131.505\\
18.000	1281847.339	48297.128	48297.128\\
19.000	1265825.804	41190.688	41190.688\\
20.000	1263690.945	41265.583	41265.583\\
21.000	1256295.216	43504.480	43504.480\\
22.000	1247106.156	38307.068	38307.068\\
23.000	1246266.509	40851.104	40851.104\\
24.000	1237887.208	34991.451	34991.451\\
25.000	1234399.506	32903.092	32903.092\\
26.000	1232823.293	30721.730	30721.730\\
27.000	1235554.361	32532.284	32532.284\\
28.000	1234893.765	27980.080	27980.080\\
29.000	1234876.893	30681.025	30681.025\\
30.000	1241692.788	28395.141	28395.141\\
31.000	1245497.154	28732.822	28732.822\\
32.000	1250070.732	28039.750	28039.750\\
33.000	1252191.276	27257.039	27257.039\\
34.000	1261277.660	27010.348	27010.348\\
35.000	1268672.604	27082.546	27082.546\\
36.000	1274367.203	24699.257	24699.257\\
37.000	1280194.015	25177.421	25177.421\\
38.000	1288720.763	24180.439	24180.439\\
39.000	1293617.966	22788.925	22788.925\\
40.000	1301466.320	21469.865	21469.865\\
41.000	1310124.309	21948.653	21948.653\\
42.000	1320064.845	22498.677	22498.677\\
43.000	1328159.442	22400.125	22400.125\\
44.000	1336698.431	20645.220	20645.220\\
45.000	1345348.598	20543.928	20543.928\\
46.000	1354917.426	20229.739	20229.739\\
};
\addlegendentry{M=4}

\end{axis}
\end{tikzpicture}%
            \caption{Total cleaning cost for day-time cleaning scenario}
            \label{fig:dayctcc}
        \end{figure}
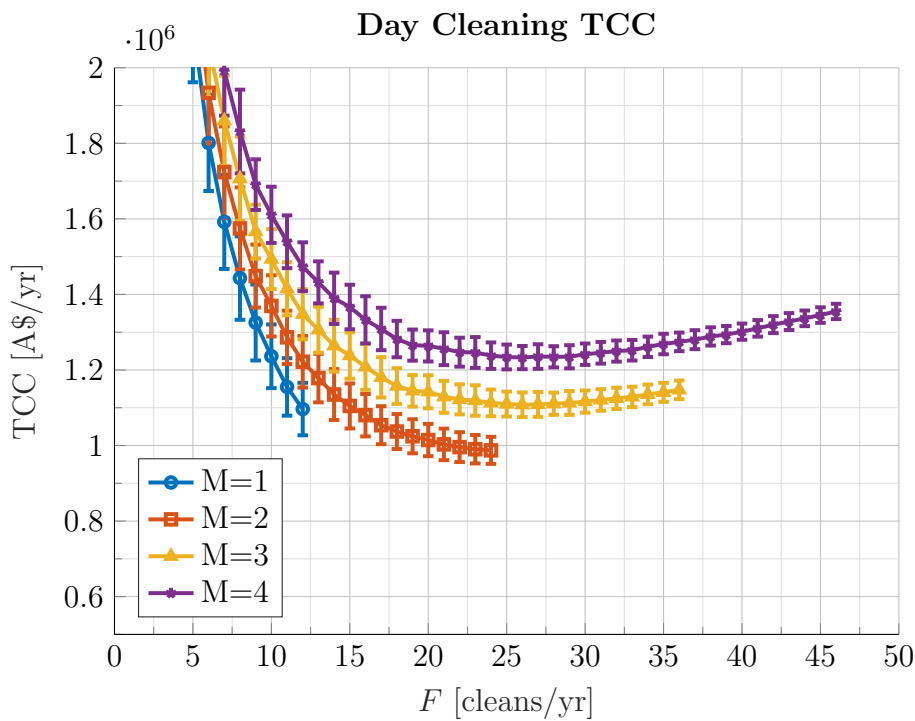
        \begin{figure}[h!]
            \centering
\definecolor{mycolor1}{rgb}{0.00000,0.44700,0.74100}%
\definecolor{mycolor2}{rgb}{0.85000,0.32500,0.09800}%
\definecolor{mycolor3}{rgb}{0.92900,0.69400,0.12500}%
\definecolor{mycolor4}{rgb}{0.49400,0.18400,0.55600}%
\begin{tikzpicture}[%
trim axis left, trim axis right
]

\begin{axis}[%
width=0.951\fwidth,
height=\fheight,
at={(0\fwidth,0\fheight)},
scale only axis,
unbounded coords=jump,
xmin=0.000,
xmax=50.000,
xlabel style={font=\color{white!15!black}},
xlabel={\ta{ncleans} [cleans/yr]},
ymin=0.000,
ymax=150000.000,
ylabel style={font=\color{white!15!black}},
ylabel={Operation Cost [A\$/yr]},
title style={font=\bfseries},
title={Day Cleaning \ta{costoperational}},
axis background/.style={fill=white},
axis x line*=bottom,
axis y line*=left,
xmajorgrids,
ymajorgrids,
legend pos=north west,
legend style={legend cell align=left, align=left, draw=white!15!black}
]
\addplot [color=mycolor4, line width=1.5pt]
 plot [error bars/.cd, y dir=both, y explicit, error bar style={line width=1.5pt}, error mark options={line width=1.5pt, mark size=2.0pt, rotate=90}]
 table[row sep=crcr, y error plus index=2, y error minus index=3]{%
1.000	2302.355	435.815	435.815\\
2.000	5350.574	565.581	565.581\\
3.000	8299.987	578.606	578.606\\
4.000	11604.471	996.227	996.227\\
5.000	14566.828	710.185	710.185\\
6.000	16500.198	880.814	880.814\\
7.000	19824.613	1017.352	1017.352\\
8.000	22337.575	1251.103	1251.103\\
9.000	25027.273	960.924	960.924\\
10.000	29187.074	760.095	760.095\\
11.000	31446.685	1298.978	1298.978\\
12.000	34386.671	1679.633	1679.633\\
13.000	36274.921	1524.034	1524.034\\
14.000	39110.940	1477.831	1477.831\\
15.000	42558.201	1650.416	1650.416\\
16.000	45430.245	1766.989	1766.989\\
17.000	47854.851	1178.697	1178.697\\
18.000	50322.231	1521.966	1521.966\\
19.000	54363.070	1596.045	1596.045\\
20.000	57074.703	1427.066	1427.066\\
21.000	60324.047	1608.131	1608.131\\
22.000	62645.103	1764.165	1764.165\\
23.000	65454.349	1951.585	1951.585\\
24.000	68385.042	1766.845	1766.845\\
25.000	71204.557	1577.542	1577.542\\
26.000	73567.149	2448.168	2448.168\\
27.000	75916.912	1234.382	1234.382\\
28.000	79314.678	2532.998	2532.998\\
29.000	81183.554	2487.758	2487.758\\
30.000	84518.949	1537.154	1537.154\\
31.000	87477.601	2632.939	2632.939\\
32.000	90826.268	1662.816	1662.816\\
33.000	93146.983	2146.372	2146.372\\
34.000	95882.045	2139.706	2139.706\\
35.000	99660.557	1940.107	1940.107\\
36.000	102427.998	1963.804	1963.804\\
37.000	105569.262	1777.387	1777.387\\
38.000	107841.632	2041.902	2041.902\\
39.000	110055.944	2142.899	2142.899\\
40.000	112912.842	2050.788	2050.788\\
41.000	116190.245	2068.297	2068.297\\
42.000	119105.997	2020.668	2020.668\\
43.000	121888.529	2392.536	2392.536\\
44.000	124459.750	2185.984	2185.984\\
45.000	127311.421	2207.524	2207.524\\
46.000	130266.523	2518.883	2518.883\\
};
\addlegendentry{M=4}

\addplot [color=mycolor3, line width=1.5pt]
 plot [error bars/.cd, y dir=both, y explicit, error bar style={line width=1.5pt}, error mark options={line width=1.5pt, mark size=2.0pt, rotate=90}]
 table[row sep=crcr, y error plus index=2, y error minus index=3]{%
1.000	2203.853	339.129	339.129\\
2.000	5187.420	533.675	533.675\\
3.000	8232.380	560.158	560.158\\
4.000	11400.499	936.720	936.720\\
5.000	14349.269	571.605	571.605\\
6.000	16498.757	836.908	836.908\\
7.000	19178.457	960.545	960.545\\
8.000	22046.363	1067.484	1067.484\\
9.000	24983.144	931.528	931.528\\
10.000	29052.150	488.775	488.775\\
11.000	31341.406	1265.366	1265.366\\
12.000	34088.934	1514.257	1514.257\\
13.000	36339.359	1147.859	1147.859\\
14.000	38775.282	1126.161	1126.161\\
15.000	42216.294	1575.649	1575.649\\
16.000	45070.681	1667.259	1667.259\\
17.000	47735.261	1042.221	1042.221\\
18.000	50200.883	1352.052	1352.052\\
19.000	54929.143	1169.061	1169.061\\
20.000	57024.375	1210.312	1210.312\\
21.000	59699.066	1534.653	1534.653\\
22.000	62535.230	1790.941	1790.941\\
23.000	64980.869	1731.593	1731.593\\
24.000	68286.285	1489.189	1489.189\\
25.000	71066.694	1443.823	1443.823\\
26.000	73617.860	2291.647	2291.647\\
27.000	75895.387	1380.115	1380.115\\
28.000	79078.594	2007.287	2007.287\\
29.000	82204.861	2048.547	2048.547\\
30.000	84460.272	1527.616	1527.616\\
31.000	87969.831	1862.270	1862.270\\
32.000	90511.258	1809.908	1809.908\\
33.000	93336.176	1650.473	1650.473\\
34.000	96245.141	1591.466	1591.466\\
35.000	99132.187	1730.896	1730.896\\
36.000	101913.647	1814.499	1814.499\\
37.000	nan	nan	nan\\
38.000	nan	nan	nan\\
39.000	nan	nan	nan\\
40.000	nan	nan	nan\\
41.000	nan	nan	nan\\
42.000	nan	nan	nan\\
43.000	nan	nan	nan\\
44.000	nan	nan	nan\\
45.000	nan	nan	nan\\
46.000	nan	nan	nan\\
};
\addlegendentry{M=3}

\addplot [color=mycolor2, line width=1.5pt]
 plot [error bars/.cd, y dir=both, y explicit, error bar style={line width=1.5pt}, error mark options={line width=1.5pt, mark size=2.0pt, rotate=90}]
 table[row sep=crcr, y error plus index=2, y error minus index=3]{%
1.000	2085.382	401.015	401.015\\
2.000	5037.213	466.017	466.017\\
3.000	8166.592	528.164	528.164\\
4.000	11162.646	597.469	597.469\\
5.000	14015.721	701.129	701.129\\
6.000	16571.568	687.643	687.643\\
7.000	19416.524	732.312	732.312\\
8.000	21860.550	841.028	841.028\\
9.000	25595.777	789.596	789.596\\
10.000	28551.462	430.341	430.341\\
11.000	31140.597	1020.719	1020.719\\
12.000	33804.881	1023.012	1023.012\\
13.000	36559.463	610.647	610.647\\
14.000	39075.212	895.028	895.028\\
15.000	42040.888	991.362	991.362\\
16.000	44795.223	1274.537	1274.537\\
17.000	47900.934	1176.561	1176.561\\
18.000	51012.042	897.498	897.498\\
19.000	54435.847	793.775	793.775\\
20.000	56677.396	1072.927	1072.927\\
21.000	59581.297	1123.874	1123.874\\
22.000	62276.561	1191.199	1191.199\\
23.000	65039.383	1213.463	1213.463\\
24.000	67942.431	1209.666	1209.666\\
25.000	nan	nan	nan\\
26.000	nan	nan	nan\\
27.000	nan	nan	nan\\
28.000	nan	nan	nan\\
29.000	nan	nan	nan\\
30.000	nan	nan	nan\\
31.000	nan	nan	nan\\
32.000	nan	nan	nan\\
33.000	nan	nan	nan\\
34.000	nan	nan	nan\\
35.000	nan	nan	nan\\
36.000	nan	nan	nan\\
37.000	nan	nan	nan\\
38.000	nan	nan	nan\\
39.000	nan	nan	nan\\
40.000	nan	nan	nan\\
41.000	nan	nan	nan\\
42.000	nan	nan	nan\\
43.000	nan	nan	nan\\
44.000	nan	nan	nan\\
45.000	nan	nan	nan\\
46.000	nan	nan	nan\\
};
\addlegendentry{M=2}

\addplot [color=mycolor1, line width=1.5pt]
 plot [error bars/.cd, y dir=both, y explicit, error bar style={line width=1.5pt}, error mark options={line width=1.5pt, mark size=2.0pt, rotate=90}]
 table[row sep=crcr, y error plus index=2, y error minus index=3]{%
1.000	2212.156	335.113	335.113\\
2.000	5282.651	386.715	386.715\\
3.000	8254.344	449.178	449.178\\
4.000	11470.533	389.278	389.278\\
5.000	14017.032	491.951	491.951\\
6.000	16736.545	606.654	606.654\\
7.000	19774.599	518.771	518.771\\
8.000	22519.696	479.988	479.988\\
9.000	25460.971	507.449	507.449\\
10.000	28240.487	590.520	590.520\\
11.000	31102.684	640.748	640.748\\
12.000	33971.216	604.833	604.833\\
13.000	nan	nan	nan\\
14.000	nan	nan	nan\\
15.000	nan	nan	nan\\
16.000	nan	nan	nan\\
17.000	nan	nan	nan\\
18.000	nan	nan	nan\\
19.000	nan	nan	nan\\
20.000	nan	nan	nan\\
21.000	nan	nan	nan\\
22.000	nan	nan	nan\\
23.000	nan	nan	nan\\
24.000	nan	nan	nan\\
25.000	nan	nan	nan\\
26.000	nan	nan	nan\\
27.000	nan	nan	nan\\
28.000	nan	nan	nan\\
29.000	nan	nan	nan\\
30.000	nan	nan	nan\\
31.000	nan	nan	nan\\
32.000	nan	nan	nan\\
33.000	nan	nan	nan\\
34.000	nan	nan	nan\\
35.000	nan	nan	nan\\
36.000	nan	nan	nan\\
37.000	nan	nan	nan\\
38.000	nan	nan	nan\\
39.000	nan	nan	nan\\
40.000	nan	nan	nan\\
41.000	nan	nan	nan\\
42.000	nan	nan	nan\\
43.000	nan	nan	nan\\
44.000	nan	nan	nan\\
45.000	nan	nan	nan\\
46.000	nan	nan	nan\\
};
\addlegendentry{M=1}

\end{axis}

\begin{axis}[%
width=1.227\fwidth,
height=1.227\fheight,
at={(-0.16\fwidth,-0.135\fheight)},
scale only axis,
xmin=0.000,
xmax=1.000,
ymin=0.000,
ymax=1.000,
axis line style={draw=none},
ticks=none,
axis x line*=bottom,
axis y line*=left
]
\end{axis}
\end{tikzpicture}%
            \caption{Day cleaning mean operational costs}
            \label{fig:operationalcost}
        \end{figure}
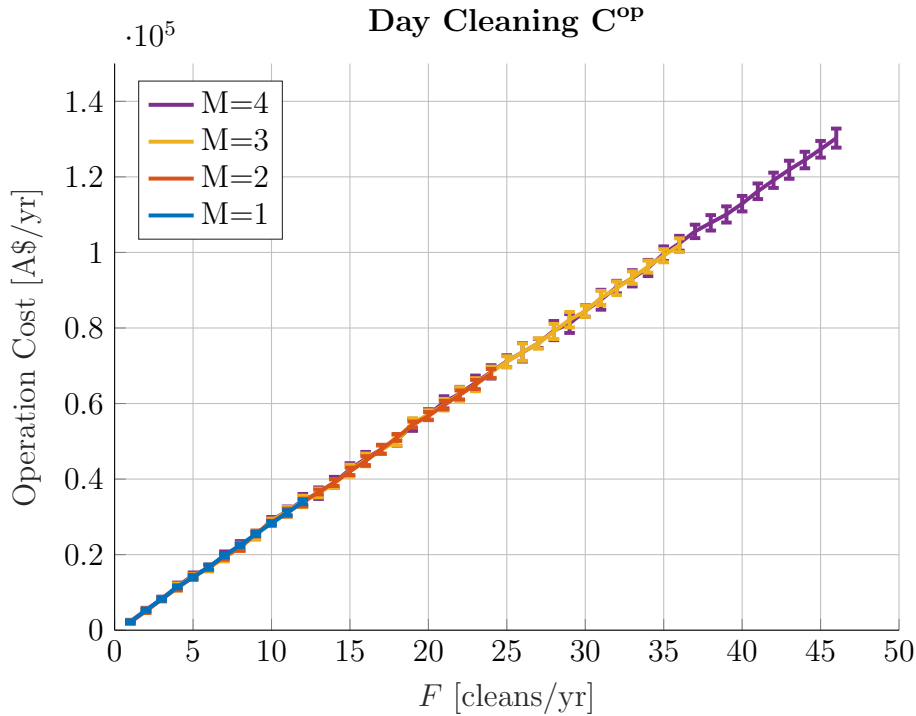

        Operational downtime costs could be reduced through opportunistic maintenance such as cleaning during periods of downtime from insufficient solar energy, when the receiver is at saturated conditions or when the \gls{tes} is fully charged and operating with a night time dispatch policy. Cleaning times could be adjusted to coincide with probable hours where a heliostat would not be producing energy to reduce total impact on productivity.
\section{Conclusion}
\label{sec:conclusion}
Cleaning resources for a modular \SI{50}{\mega\watt} \gls{csp}+\gls{tes} were optimized early in the design phase using the developed soiling prediction and cleaning resource methodology. By calibrating the soiling model to the site through experimental measurements, mean soiling rates of \SI{0.12}{pp\per\day} and \SI{0.22}{pp\per\day} were predicted for the low and high dust season respectively. An autoregressive time-series model was used to extend two years of meteorological measurements into ten years allowing for multiple soiling trajectories to be evaluated. Additionally, usage of a second surface geometry model and updated receiver and thermal storage models allowed for an assessment of various \gls{csp} operation and design philosophies on soiling induced costs. The proposed methodology ultimately reduces financial uncertainty associated to soiling early in the design and site selection phases of a \gls{csp} plant.

For the base scenario utilizing vertically stowed heliostats and a continuous energy dispatch policy, an optimal cleaning schedule was found with two crews cleaning the solar field 24 times per annum. A break down of cleaning costs showed that fuel and operator salaries contribute \SI{42}{\percent} and \SI{35}{\percent} of cleaning costs with a high pressure wash and brush cleaning crew. 

The second scenario introduced a horizontally night-time stow operational policy. By stowing horizontally instead of vertically at night the soiling rate increased by \SI{114}{\percent}. This led to hiring a third cleaning crew to reach the optimal cleaning schedule. Ultimately, the total cleaning cost and total lifetime cost of each heliostat increased by \SI{51}{\percent} and \SI{26}{\percent} relative to the base vertical stow scenario.

A night-time dispatch policy was explored, where during the day the modular CSP field would charge the TES for night time electrical production. Simulations showed that the oversized solar fields led to a large amount of energy loss due to maximum storage constraints on the TES system. The excess energy suggests that a daily fully-charged thermal storage could occur with a yearly field average cleanliness of \SI{97.6}{\percent}, reducing the need to maintain perfectly clean solar fields.

A day-time cleaning scenario was setup to analyse the effect of shutting down groups of heliostats for cleaning during the day. Results showed that day-time cleaning incurred a productivity loss that increases linearly with the number of field cleanings per year. Resulting in a total cleaning cost increase of \SI{7}{\percent} relative to the night-time cleaning policy. To minimize these costs it may be possible to perform opportunistic cleaning during periods where the receiver is operating in a saturated/depraved condition or when the thermal energy storage is full (assuming a night time only dispatch policy).
\section*{Acknowledgements}
C.B. Anderson, G. Picotti, M.E. Cholette, and T.A. Steinberg acknowledge the support of the Australian Government for this study, through the Australian Renewable Energy Agency (ARENA) and within the framework of the Australian Solar Thermal Research Initiative (ASTRI Project ID P54). Additionally, the ground soil composition work was enabled by use of the Central Analytical Research Facility hosted by the Institute for Future Environments at QUT.

%% Appendices
\appendix
\section{Overlap derivation}
\label{apx:a}
Heliostats operating at near normal incident angles can experience an overlapping of shading and blocking areas with the given second surface geometry model. This overlapping of areas causes a reduction in the total obstructed area and can be corrected as a function of how overlapped the two regions are. As the blocking and shading areas are ellipses of the same size along one axis it is possible to use the \als{overlaplength} to determine how overlapped the shaded and blocked areas are and analytically solve for the overlapping areas.

\begin{figure}[ht]
    \centering
    \includegraphics[width=0.951\columnwidth]{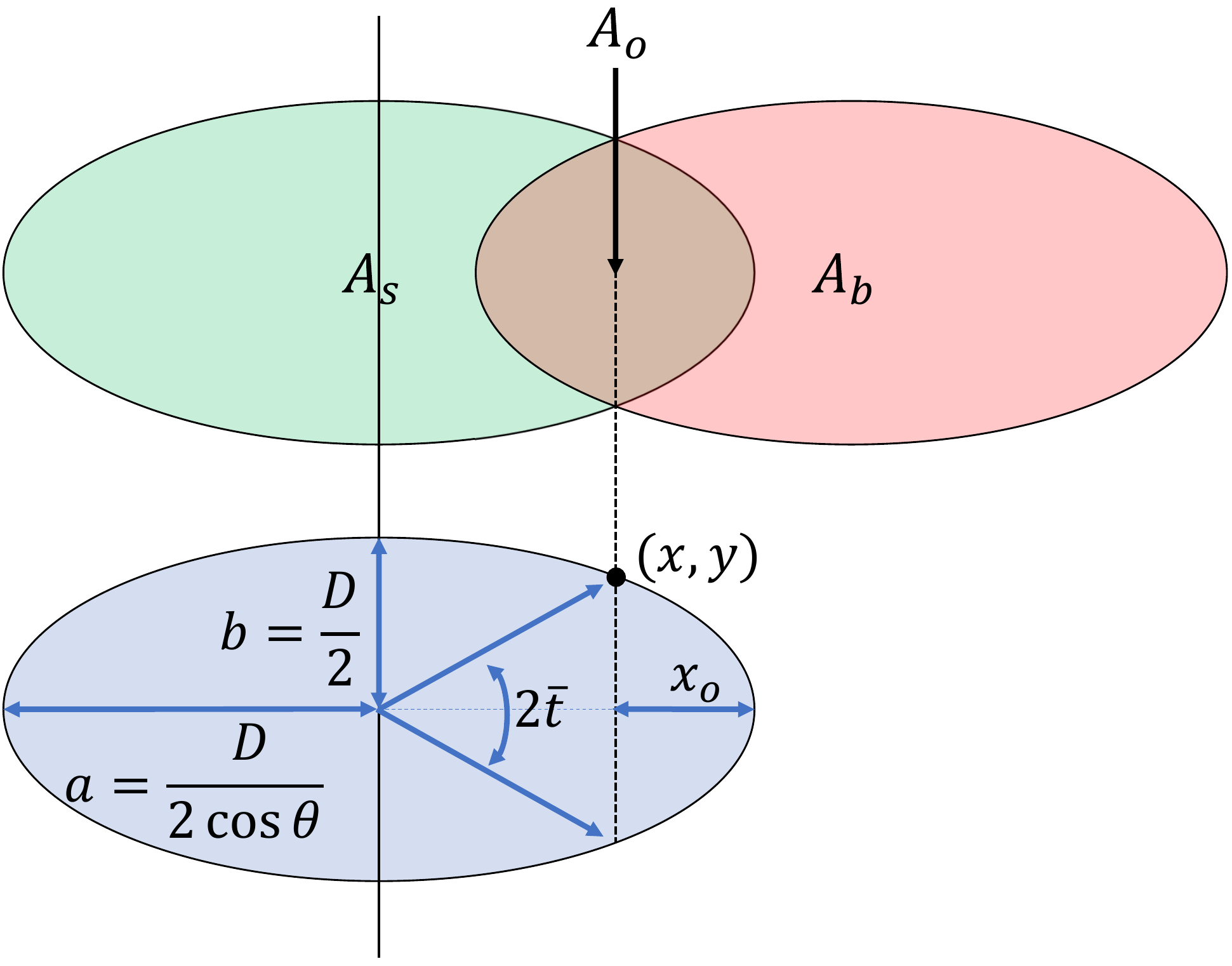}
    \caption{Schematic of particle overlapping, shading and blocking areas}
    \label{fig:shadeblockoverlaparea}
\end{figure}

Consider the two ellipses shown in Fig.~\ref{fig:shadeblockoverlaparea} that are overlapped by length \glsname{overlaplength}, with a shared area denoted by \glsname{areaoverlap}. Both shading and blocking ellipses can be parameterised with $(x,y) = (a\cos{t}, b\sin{t}), 0 \leq t \leq 2\pi$, where $2t = \ta{eccentricanomaly}$. The \als{eccentricanomaly} associated to the position of the overlap position relative to the ellipse length $a-\ta{overlaplength}$ is derived as:
\begin{subequations}
    \label{eq:eccentricanomaly}
    \begin{align}
    dx & = -a \sin{t}\text{ }dt\\
    a-\ta{overlaplength} & = a \cos{\ta{eccentricanomaly}}\\
    \ta{eccentricanomaly} & = \cos^{-1} \left( \frac{ a - \ta{overlaplength}} {a} \right)\\
    \ta{eccentricanomaly} & = \cos^{-1}{ \left( \frac{ \ta{particlediameter} - 2 \ta{overlaplength}\cos{\ta{incidenceangle}}}{\ta{particlediameter}} \right)}
    \end{align}
\end{subequations}
so therefore, the overlapping area can be calculated as:
\begin{subequations}
    \label{eq:overlapareaderivation}
    \begin{align}
        \frac{A_{o}}{2} & = 2 \int^{a}_{a-\ta{overlaplength}} y\text{ }dx\\
        & = -2ab\int^{0}_{\ta{eccentricanomaly}}\sin^{2}{t}\text{ }dt\\
        & = 2ab\int^{\ta{eccentricanomaly}}_{0} \sin^{2}{t}\text{ }dt\\
        & = ab \left( \ta{eccentricanomaly}-\sin{\ta{eccentricanomaly}}\cdot \cos{\ta{eccentricanomaly}} \right)\\
        A_{o} & = 2ab\left( \ta{eccentricanomaly}-\sin{\ta{eccentricanomaly}}\cdot \cos{\ta{eccentricanomaly}} \right)\\
        A_{o} & = \frac{\ta{particlediameter}^2}{2\cos{\ta{incidenceangle}}} \cdot \left( \ta{eccentricanomaly}-\sin{\ta{eccentricanomaly}}\cdot \cos{\ta{eccentricanomaly}} \right)
    \end{align}
\end{subequations}

\begin{figure}[h!]
    \centering
    \includegraphics[width=0.951\columnwidth]{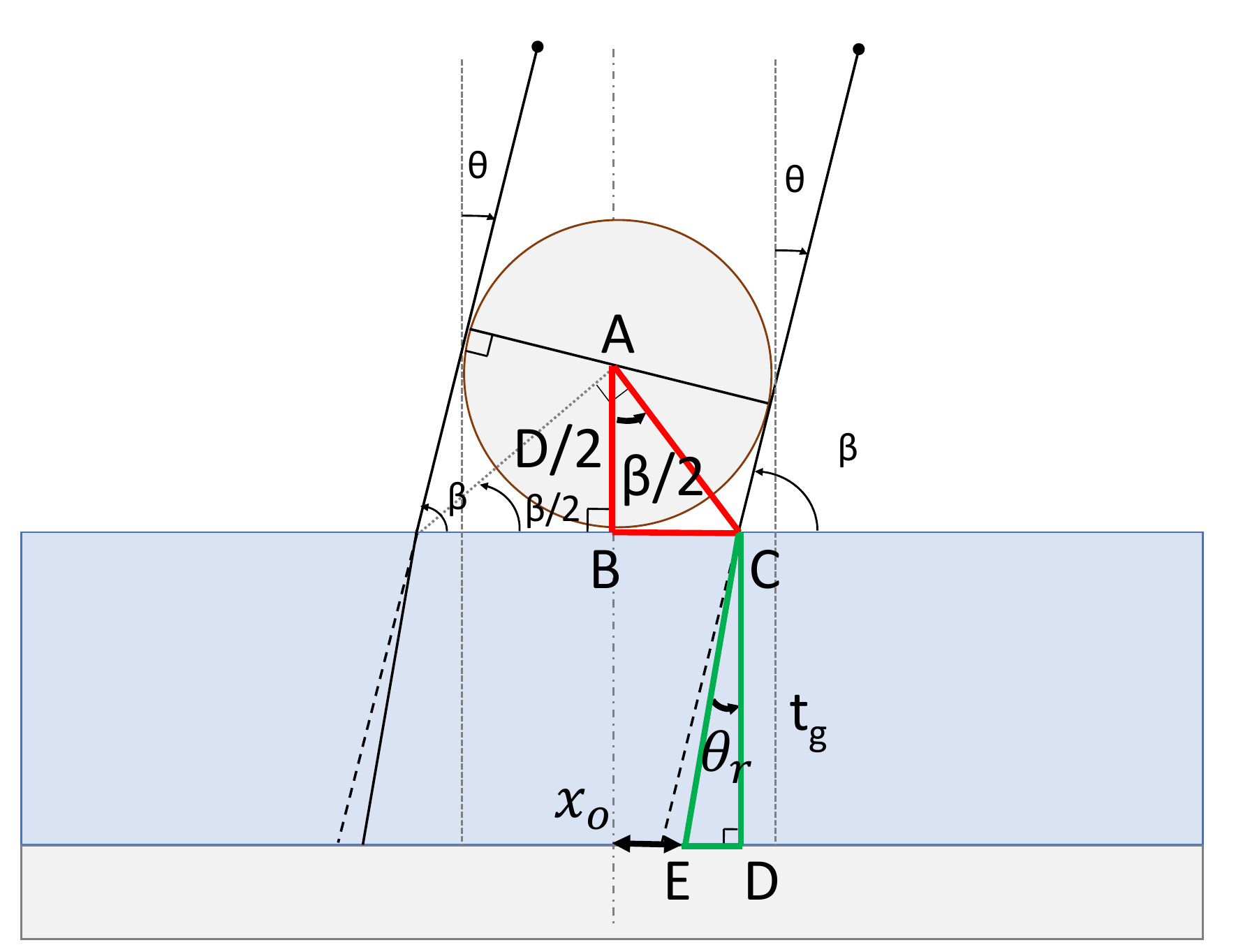}
    \caption{Geometry diagram of shaded area boundary positions relative to a particle}
    \label{fig:geometryangle}
\end{figure}

The overlap length is equivalent to the difference between length $\overline{BC}$ and $\overline{ED}$ depicted in Fig.~\ref{fig:geometryangle}. Where length $\overline{BC}$ denotes the distance between the centre line of a spherical particle and the intersecting point of a \gls{dni} ray skimming the particle into the glass, and $\overline{ED}$ is the length between the intersecting \gls{dni} point as projected onto the reflector vertically below and the final position of the ray before reflection. Overlapping will therefore only occur when $\overline{BC}>\overline{ED}$.

As \gls{dni} passes through the glass with thickness, \glsname{glassthick} it will be refracted at a new angle. This \gls{refractedangle} is calculated using Snell's law assuming a fixed refractive index for air and glass ($n_{air}=1$ and $n_{glass} = 1.517$), $\overline{ED}$ can be calculated using the refractive angle and thus the overlapping length is solved using Eq.~\ref{eq:xro}.
\begin{subequations}
    \label{eq:xro}
    \begin{align}
        \ta{overlaplength} & = \overline{BC}-\overline{ED}\\
        & = \frac{\ta{particlediameter}}{2} \tan{\left(\frac{\beta}{2}\right)} - \ta{glassthick} \tan{\left( \ta{refractedangle} \right)}\\
        & = \frac{\ta{particlediameter}}{2} \tan{ \left( \frac{90-\ta{incidenceangle}}{2} \right)} - \ta{glassthick} \tan{ \left( \sin^{-1}{ \left(\frac{n_{\text{air}}}{n_{\text{glass}}}\cdot \sin{ \ta{incidenceangle}} \right)} \right)}
    \end{align}
\end{subequations}

%% Bibliography
\bibliographystyle{elsarticle-num-names} 
\biboptions{sort&compress}
\bibliography{SPCM_Main} % This is an updated Mendeley Group bibliography from Mendeley see Cody for invite 

\begin{thebibliography}{41}
\expandafter\ifx\csname natexlab\endcsname\relax\def\natexlab#1{#1}\fi
\providecommand{\url}[1]{\texttt{#1}}
\providecommand{\href}[2]{#2}
\providecommand{\path}[1]{#1}
\providecommand{\DOIprefix}{doi:}
\providecommand{\ArXivprefix}{arXiv:}
\providecommand{\URLprefix}{URL: }
\providecommand{\Pubmedprefix}{pmid:}
\providecommand{\doi}[1]{\href{http://dx.doi.org/#1}{\path{#1}}}
\providecommand{\Pubmed}[1]{\href{pmid:#1}{\path{#1}}}
\providecommand{\bibinfo}[2]{#2}
\ifx\xfnm\relax \def\xfnm[#1]{\unskip,\space#1}\fi
%Type = Techreport
\bibitem[{{IRENA}(2020)}]{IRENA2020}
\bibinfo{author}{{IRENA}}, \bibinfo{title}{{Renewable Capacity Statistics
  2020}}, \bibinfo{type}{Technical Report}, International Renewable Energy
  Agency, \bibinfo{address}{Abu Dhabi}, \bibinfo{year}{2020}. \URLprefix
  \url{https://irena.org/-/media/Files/IRENA/Agency/Publication/2020/Mar/IRENA_RE_Capacity_Statistics_2020.pdf}.
%Type = Article
\bibitem[{Bahadori and Nwaoha(2013)}]{Bahadori2013}
\bibinfo{author}{A.~Bahadori}, \bibinfo{author}{C.~Nwaoha},
\newblock \bibinfo{title}{{A review on solar energy utilisation in Australia}},
\newblock \bibinfo{journal}{Renewable and Sustainable Energy Reviews}
  \bibinfo{volume}{18} (\bibinfo{year}{2013}) \bibinfo{pages}{1--5}.
  \DOIprefix\doi{10.1016/j.rser.2012.10.003}.
%Type = Techreport
\bibitem[{Feldman et~al.(2016)Feldman, Margolis, Denholm, and
  Stekli}]{Feldman2016}
\bibinfo{author}{D.~Feldman}, \bibinfo{author}{R.~Margolis},
  \bibinfo{author}{P.~Denholm}, \bibinfo{author}{J.~Stekli},
  \bibinfo{title}{{Exploring the Potential Competitiveness of Utility-Scale
  Photovoltaics plus Batteries with Concentrating Solar Power , 2015 –
  2030}}, \bibinfo{type}{Technical Report}, NREL, \bibinfo{address}{Denver,
  CO}, \bibinfo{year}{2016}. \DOIprefix\doi{10.2172/1321487}.
%Type = Article
\bibitem[{Picotti et~al.(2018)Picotti, Borghesani, Cholette, and
  Manzolini}]{Picotti2018c}
\bibinfo{author}{G.~Picotti}, \bibinfo{author}{P.~Borghesani},
  \bibinfo{author}{M.~E. Cholette}, \bibinfo{author}{G.~Manzolini},
\newblock \bibinfo{title}{{Soiling of solar collectors – Modelling approaches
  for airborne dust and its interactions with surfaces}},
\newblock \bibinfo{journal}{Renewable and Sustainable Energy Reviews}
  \bibinfo{volume}{81} (\bibinfo{year}{2018}) \bibinfo{pages}{2343--2357}.
  \URLprefix \url{http://dx.doi.org/10.1016/j.rser.2017.06.043}.
  \DOIprefix\doi{10.1016/j.rser.2017.06.043}.
%Type = Techreport
\bibitem[{Kolb et~al.(2011)Kolb, Ho, Mancini, and Gary}]{Kolb2012}
\bibinfo{author}{G.~J. Kolb}, \bibinfo{author}{C.~K. Ho},
  \bibinfo{author}{T.~R. Mancini}, \bibinfo{author}{J.~A. Gary},
  \bibinfo{title}{{Power tower technology roadmap and cost reduction plan}},
  \bibinfo{type}{Technical Report}, Sandia Labs, \bibinfo{address}{Albuquerque,
  NM, and Livermore, CA (United States)}, \bibinfo{year}{2011}.
  \DOIprefix\doi{10.2172/1011644}.
%Type = Techreport
\bibitem[{Berg(1978)}]{Berg1978}
\bibinfo{author}{R.~S. Berg}, \bibinfo{title}{{Heliostat dust buildup and
  cleaning studies}}, \bibinfo{type}{Technical Report}
  \bibinfo{number}{SAND-78-0432C}, Sandia Labs, \bibinfo{address}{Albuquerque,
  NM (United States)}, \bibinfo{year}{1978}. \URLprefix
  \url{http://www.osti.gov/servlets/purl/6867834/}.
  \DOIprefix\doi{10.2172/6867834}.
%Type = Techreport
\bibitem[{Roth and Pettit(1980)}]{Roth1980}
\bibinfo{author}{E.~Roth}, \bibinfo{author}{R.~Pettit}, \bibinfo{title}{{The
  effect of soiling on solar mirrors and techniques used to maintain high
  reflectivity}}, \bibinfo{type}{Technical Report}, Sandia Labs,
  \bibinfo{address}{Albuquerque, NM, and Livermore, CA (United States)},
  \bibinfo{year}{1980}. \DOIprefix\doi{10.2172/5249717}.
%Type = Article
\bibitem[{Heimsath and Nitz(2019)}]{Heimsath2019}
\bibinfo{author}{A.~Heimsath}, \bibinfo{author}{P.~Nitz},
\newblock \bibinfo{title}{{The effect of soiling on the reflectance of solar
  reflector materials - Model for prediction of incidence angle dependent
  reflectance and attenuation due to dust deposition}},
\newblock \bibinfo{journal}{Solar Energy Materials and Solar Cells}
  \bibinfo{volume}{195} (\bibinfo{year}{2019}) \bibinfo{pages}{258--268}.
  \DOIprefix\doi{10.1016/j.solmat.2019.03.015}.
%Type = Article
\bibitem[{Picotti et~al.(2021)Picotti, Simonetti, Schmidt, Cholette, Heimsath,
  Ernst, and Manzolini}]{Picotti2021a}
\bibinfo{author}{G.~Picotti}, \bibinfo{author}{R.~Simonetti},
  \bibinfo{author}{T.~Schmidt}, \bibinfo{author}{M.~E. Cholette},
  \bibinfo{author}{A.~Heimsath}, \bibinfo{author}{S.~J. Ernst},
  \bibinfo{author}{G.~Manzolini},
\newblock \bibinfo{title}{{Evaluation of reflectance measurement techniques for
  artificially soiled solar reflectors: Experimental campaign and model
  assessment}},
\newblock \bibinfo{journal}{Solar Energy Materials and Solar Cells}
  \bibinfo{volume}{231} (\bibinfo{year}{2021}) \bibinfo{pages}{111321}.
  \DOIprefix\doi{10.1016/j.solmat.2021.111321}.
%Type = Techreport
\bibitem[{Zhu et~al.(2022)Zhu, Augustine, Mitchell, Muller, Kurup, Zolan,
  Yellapantula, Brost, Armijo, Sment, Schaller, Gordon, Collins, Coventry, Pye,
  Cholette, Picotti, Arjomandi, Emes, Potter, Rae, Zhu, Augustine, Mitchell,
  Muller, Kurup, Zolan, Yellapantula, Brost, Armijo, Sment, Schaller, Gordon,
  Collins, Coventry, Pye, Cholette, Picotti, Arjomandi, Emes, Potter, and
  Rae}]{Zhu2022}
\bibinfo{author}{G.~Zhu}, \bibinfo{author}{C.~Augustine},
  \bibinfo{author}{R.~Mitchell}, \bibinfo{author}{M.~Muller},
  \bibinfo{author}{P.~Kurup}, \bibinfo{author}{A.~Zolan},
  \bibinfo{author}{S.~Yellapantula}, \bibinfo{author}{R.~Brost},
  \bibinfo{author}{K.~Armijo}, \bibinfo{author}{J.~Sment},
  \bibinfo{author}{R.~Schaller}, \bibinfo{author}{M.~Gordon},
  \bibinfo{author}{M.~Collins}, \bibinfo{author}{J.~Coventry},
  \bibinfo{author}{J.~Pye}, \bibinfo{author}{M.~Cholette},
  \bibinfo{author}{G.~Picotti}, \bibinfo{author}{M.~Arjomandi},
  \bibinfo{author}{M.~Emes}, \bibinfo{author}{D.~Potter},
  \bibinfo{author}{M.~Rae}, \bibinfo{author}{G.~Zhu},
  \bibinfo{author}{C.~Augustine}, \bibinfo{author}{R.~Mitchell},
  \bibinfo{author}{M.~Muller}, \bibinfo{author}{P.~Kurup},
  \bibinfo{author}{A.~Zolan}, \bibinfo{author}{S.~Yellapantula},
  \bibinfo{author}{R.~Brost}, \bibinfo{author}{K.~Armijo},
  \bibinfo{author}{J.~Sment}, \bibinfo{author}{R.~Schaller},
  \bibinfo{author}{M.~Gordon}, \bibinfo{author}{M.~Collins},
  \bibinfo{author}{J.~Coventry}, \bibinfo{author}{J.~Pye},
  \bibinfo{author}{M.~Cholette}, \bibinfo{author}{G.~Picotti},
  \bibinfo{author}{M.~Arjomandi}, \bibinfo{author}{M.~Emes},
  \bibinfo{author}{D.~Potter}, \bibinfo{author}{M.~Rae},
  \bibinfo{title}{{Roadmap to Advance Heliostat Technologies for Concentrating
  Solar-Thermal Power}}, \bibinfo{type}{Technical Report}
  \bibinfo{number}{NREL/TP-5700-83041}, NREL, \bibinfo{year}{2022}. \URLprefix
  \url{https://www.nrel.gov/docs/fy22osti/83041.pdf}.
%Type = Article
\bibitem[{Hachicha et~al.(2019)Hachicha, Al-Sawafta, and
  Ben~Hamadou}]{Hachicha2019}
\bibinfo{author}{A.~A. Hachicha}, \bibinfo{author}{I.~Al-Sawafta},
  \bibinfo{author}{D.~Ben~Hamadou},
\newblock \bibinfo{title}{{Numerical and experimental investigations of dust
  effect on CSP performance under United Arab Emirates weather conditions}},
\newblock \bibinfo{journal}{Renewable Energy} \bibinfo{volume}{143}
  (\bibinfo{year}{2019}) \bibinfo{pages}{263--276}.
  \DOIprefix\doi{10.1016/j.renene.2019.04.144}.
%Type = Article
\bibitem[{Azouzoute et~al.(2020)Azouzoute, Merrouni, Garoum, and
  Bennouna}]{Azouzoute2020}
\bibinfo{author}{A.~Azouzoute}, \bibinfo{author}{A.~A. Merrouni},
  \bibinfo{author}{M.~Garoum}, \bibinfo{author}{E.~G. Bennouna},
\newblock \bibinfo{title}{{Soiling loss of solar glass and mirror samples in
  the region with arid climate}},
\newblock \bibinfo{journal}{Energy Reports} \bibinfo{volume}{6}
  (\bibinfo{year}{2020}) \bibinfo{pages}{693--698}.
  \DOIprefix\doi{10.1016/j.egyr.2019.09.051}.
%Type = Article
\bibitem[{Concei{\c{c}}{\~{a}}o et~al.(2023)Concei{\c{c}}{\~{a}}o,
  Mart{\'{i}}nez~Hern{\'{a}}ndez, Romero, and
  Gonz{\'{a}}lez-Aguilar}]{Conceicao2023}
\bibinfo{author}{R.~Concei{\c{c}}{\~{a}}o},
  \bibinfo{author}{A.~Mart{\'{i}}nez~Hern{\'{a}}ndez},
  \bibinfo{author}{M.~Romero}, \bibinfo{author}{J.~Gonz{\'{a}}lez-Aguilar},
\newblock \bibinfo{title}{{Experimental soiling assessment, characterization
  and modelling of a highly-compact heliostat field in an urban environment}},
\newblock \bibinfo{journal}{Solar Energy} \bibinfo{volume}{262}
  (\bibinfo{year}{2023}) \bibinfo{pages}{111812}.
  \DOIprefix\doi{10.1016/j.solener.2023.111812}.
%Type = Article
\bibitem[{Bouaddi et~al.(2015)Bouaddi, Ihlal, and
  Fern{\'{a}}ndez-Garc{\'{i}}a}]{Bouaddi2015}
\bibinfo{author}{S.~Bouaddi}, \bibinfo{author}{A.~Ihlal},
  \bibinfo{author}{A.~Fern{\'{a}}ndez-Garc{\'{i}}a},
\newblock \bibinfo{title}{{Soiled CSP solar reflectors modeling using dynamic
  linear models}},
\newblock \bibinfo{journal}{Solar Energy} \bibinfo{volume}{122}
  (\bibinfo{year}{2015}) \bibinfo{pages}{847--863}.
  \DOIprefix\doi{10.1016/j.solener.2015.09.044}.
%Type = Article
\bibitem[{Concei{\c{c}}{\~{a}}o et~al.(2018)Concei{\c{c}}{\~{a}}o, Silva, and
  Collares-Pereira}]{Conceicao2018a}
\bibinfo{author}{R.~Concei{\c{c}}{\~{a}}o}, \bibinfo{author}{H.~G. Silva},
  \bibinfo{author}{M.~Collares-Pereira},
\newblock \bibinfo{title}{{CSP mirror soiling characterization and modeling}},
\newblock \bibinfo{journal}{Solar Energy Materials and Solar Cells}
  \bibinfo{volume}{185} (\bibinfo{year}{2018}) \bibinfo{pages}{233--239}.
  \DOIprefix\doi{10.1016/j.solmat.2018.05.035}.
%Type = Article
\bibitem[{Dehghan et~al.(2022)Dehghan, Rashidi, and Waqas}]{Dehghan2022}
\bibinfo{author}{M.~Dehghan}, \bibinfo{author}{S.~Rashidi},
  \bibinfo{author}{A.~Waqas},
\newblock \bibinfo{title}{{Modeling of soiling losses in solar energy
  systems}},
\newblock \bibinfo{journal}{Sustainable Energy Technologies and Assessments}
  \bibinfo{volume}{53} (\bibinfo{year}{2022}) \bibinfo{pages}{102435}.
  \DOIprefix\doi{10.1016/j.seta.2022.102435}.
%Type = Article
\bibitem[{Picotti et~al.(2018)Picotti, Borghesani, Manzolini, Cholette, and
  Wang}]{Picotti2018}
\bibinfo{author}{G.~Picotti}, \bibinfo{author}{P.~Borghesani},
  \bibinfo{author}{G.~Manzolini}, \bibinfo{author}{M.~Cholette},
  \bibinfo{author}{R.~Wang},
\newblock \bibinfo{title}{{Development and experimental validation of a
  physical model for the soiling of mirrors for CSP industry applications}},
\newblock \bibinfo{journal}{Solar Energy} \bibinfo{volume}{173}
  (\bibinfo{year}{2018}) \bibinfo{pages}{1287--1305}.
  \DOIprefix\doi{10.1016/j.solener.2018.08.066}.
%Type = Article
\bibitem[{Coello and Boyle(2019)}]{Coello2019}
\bibinfo{author}{M.~Coello}, \bibinfo{author}{L.~Boyle},
\newblock \bibinfo{title}{{Simple Model for Predicting Time Series Soiling of
  Photovoltaic Panels}},
\newblock \bibinfo{journal}{IEEE Journal of Photovoltaics} \bibinfo{volume}{9}
  (\bibinfo{year}{2019}) \bibinfo{pages}{1382--1387}.
  \DOIprefix\doi{10.1109/JPHOTOV.2019.2919628}.
%Type = Article
\bibitem[{Sengupta et~al.(2020)Sengupta, Sengupta, and Saha}]{Sengupta2020}
\bibinfo{author}{S.~Sengupta}, \bibinfo{author}{S.~Sengupta},
  \bibinfo{author}{H.~Saha},
\newblock \bibinfo{title}{{Comprehensive Modeling of Dust Accumulation on PV
  Modules Through Dry Deposition Processes}},
\newblock \bibinfo{journal}{IEEE Journal of Photovoltaics} \bibinfo{volume}{10}
  (\bibinfo{year}{2020}) \bibinfo{pages}{1--10}.
  \DOIprefix\doi{10.1109/jphotov.2020.2992352}.
%Type = Article
\bibitem[{El~Boujdaini et~al.(2022)El~Boujdaini, Mezrhab, Amine~Moussaoui,
  Antonio Carballo~Lopez, and Wolfertstetter}]{Elboujdaini2022}
\bibinfo{author}{L.~El~Boujdaini}, \bibinfo{author}{A.~Mezrhab},
  \bibinfo{author}{M.~Amine~Moussaoui}, \bibinfo{author}{J.~Antonio
  Carballo~Lopez}, \bibinfo{author}{F.~Wolfertstetter},
\newblock \bibinfo{title}{{The effect of soiling on the performance of solar
  mirror materials: Experimentation and modeling}},
\newblock \bibinfo{journal}{Sustainable Energy Technologies and Assessments}
  \bibinfo{volume}{53} (\bibinfo{year}{2022}) \bibinfo{pages}{102741}.
  \DOIprefix\doi{10.1016/j.seta.2022.102741}.
%Type = Article
\bibitem[{Wolfertstetter et~al.(2014)Wolfertstetter, Pottler, Geuder, Affolter,
  Merrouni, Mezrhab, and Pitz-Paal}]{Wolfertstetter2014}
\bibinfo{author}{F.~Wolfertstetter}, \bibinfo{author}{K.~Pottler},
  \bibinfo{author}{N.~Geuder}, \bibinfo{author}{R.~Affolter},
  \bibinfo{author}{A.~A. Merrouni}, \bibinfo{author}{A.~Mezrhab},
  \bibinfo{author}{R.~Pitz-Paal},
\newblock \bibinfo{title}{{Monitoring of mirror and sensor soiling with TraCS
  for improved quality of ground based irradiance measurements}},
\newblock \bibinfo{journal}{Energy Procedia} \bibinfo{volume}{49}
  (\bibinfo{year}{2014}) \bibinfo{pages}{2422--2432}.
  \DOIprefix\doi{10.1016/j.egypro.2014.03.257}.
%Type = Article
\bibitem[{Ballestr{\'{i}}n et~al.(2022)Ballestr{\'{i}}n, Polo,
  Mart{\'{i}}n-Chivelet, Barbero, Carra, Alonso-Montesinos, and
  Marzo}]{Ballestrin2022}
\bibinfo{author}{J.~Ballestr{\'{i}}n}, \bibinfo{author}{J.~Polo},
  \bibinfo{author}{N.~Mart{\'{i}}n-Chivelet}, \bibinfo{author}{J.~Barbero},
  \bibinfo{author}{E.~Carra}, \bibinfo{author}{J.~Alonso-Montesinos},
  \bibinfo{author}{A.~Marzo},
\newblock \bibinfo{title}{{Soiling forecasting of solar plants: A combined
  heuristic approach and autoregressive model}},
\newblock \bibinfo{journal}{Energy} \bibinfo{volume}{239}
  (\bibinfo{year}{2022}) \bibinfo{pages}{122442}.
  \DOIprefix\doi{10.1016/j.energy.2021.122442}.
%Type = Article
\bibitem[{Picotti et~al.(2020)Picotti, Moretti, Cholette, Binotti, Simonetti,
  Martelli, Steinberg, and Manzolini}]{Picotti2020}
\bibinfo{author}{G.~Picotti}, \bibinfo{author}{L.~Moretti},
  \bibinfo{author}{M.~E. Cholette}, \bibinfo{author}{M.~Binotti},
  \bibinfo{author}{R.~Simonetti}, \bibinfo{author}{E.~Martelli},
  \bibinfo{author}{T.~A. Steinberg}, \bibinfo{author}{G.~Manzolini},
\newblock \bibinfo{title}{{Optimization of cleaning strategies for heliostat
  fields in solar tower plants}},
\newblock \bibinfo{journal}{Solar Energy} \bibinfo{volume}{204}
  (\bibinfo{year}{2020}) \bibinfo{pages}{501--514}.
  \DOIprefix\doi{10.1016/j.solener.2020.04.032}.
%Type = Article
\bibitem[{Wales et~al.(2021)Wales, Zolan, Newman, Wagner, Newmans, and
  Wagner}]{Wales2021}
\bibinfo{author}{J.~G. Wales}, \bibinfo{author}{A.~J. Zolan},
  \bibinfo{author}{A.~M. Newman}, \bibinfo{author}{M.~J. Wagner},
  \bibinfo{author}{A.~M. Newmans}, \bibinfo{author}{M.~J. Wagner},
\newblock \bibinfo{title}{{Optimizing Vehicle Fleet and Assignment for
  Concentrating Solar Power Plant Heliostat Washing}},
\newblock \bibinfo{journal}{IISE Transactions} \bibinfo{volume}{54}
  (\bibinfo{year}{2021}) \bibinfo{pages}{550--562}.
  \DOIprefix\doi{10.1080/24725854.2021.1966858}.
%Type = Article
\bibitem[{Wolfertstetter et~al.(2018)Wolfertstetter, Wilbert, Dersch,
  Dieckmann, Pitz-Paal, and Ghennioui}]{Wolfertstetter2018}
\bibinfo{author}{F.~Wolfertstetter}, \bibinfo{author}{S.~Wilbert},
  \bibinfo{author}{J.~Dersch}, \bibinfo{author}{S.~Dieckmann},
  \bibinfo{author}{R.~Pitz-Paal}, \bibinfo{author}{A.~Ghennioui},
\newblock \bibinfo{title}{{Integration of soiling-rate measurements and
  cleaning strategies in yield analysis of parabolic trough plants}},
\newblock \bibinfo{journal}{Journal of Solar Energy Engineering, Transactions
  of the ASME} \bibinfo{volume}{140} (\bibinfo{year}{2018})
  \bibinfo{pages}{4--8}. \DOIprefix\doi{10.1115/1.4039631}.
%Type = Article
\bibitem[{Truong~Ba et~al.(2017)Truong~Ba, Cholette, Wang, Borghesani, Ma, and
  Steinberg}]{Truong2017}
\bibinfo{author}{H.~Truong~Ba}, \bibinfo{author}{M.~E. Cholette},
  \bibinfo{author}{R.~Wang}, \bibinfo{author}{P.~Borghesani},
  \bibinfo{author}{L.~Ma}, \bibinfo{author}{T.~A. Steinberg},
\newblock \bibinfo{title}{{Optimal condition-based cleaning of solar power
  collectors}},
\newblock \bibinfo{journal}{Solar Energy} \bibinfo{volume}{157}
  (\bibinfo{year}{2017}) \bibinfo{pages}{762--777}.
  \DOIprefix\doi{10.1016/j.solener.2017.08.076}.
%Type = Article
\bibitem[{Truong-Ba et~al.(2020)Truong-Ba, Cholette, Picotti, Steinberg, and
  Manzolini}]{Truong-Ba2020}
\bibinfo{author}{H.~Truong-Ba}, \bibinfo{author}{M.~E. Cholette},
  \bibinfo{author}{G.~Picotti}, \bibinfo{author}{T.~A. Steinberg},
  \bibinfo{author}{G.~Manzolini},
\newblock \bibinfo{title}{{Sectorial reflectance-based cleaning policy of
  heliostats for Solar Tower power plants}},
\newblock \bibinfo{journal}{Renewable Energy} \bibinfo{volume}{166}
  (\bibinfo{year}{2020}) \bibinfo{pages}{176--189}.
  \DOIprefix\doi{10.1016/j.renene.2020.11.129}.
%Type = Article
\bibitem[{Sutter et~al.(2018)Sutter, Montecchi, von Dahlen,
  Fern{\'{a}}ndez-Garc{\'{i}}a, and R{\"{o}}ger}]{Sutter2018}
\bibinfo{author}{F.~Sutter}, \bibinfo{author}{M.~Montecchi},
  \bibinfo{author}{H.~von Dahlen},
  \bibinfo{author}{A.~Fern{\'{a}}ndez-Garc{\'{i}}a},
  \bibinfo{author}{M.~R{\"{o}}ger},
\newblock \bibinfo{title}{{The effect of incidence angle on the reflectance of
  solar mirrors}},
\newblock \bibinfo{journal}{Solar Energy Materials and Solar Cells}
  \bibinfo{volume}{176} (\bibinfo{year}{2018}) \bibinfo{pages}{119--133}.
  \DOIprefix\doi{10.1016/j.solmat.2017.11.029}.
%Type = Article
\bibitem[{Karasu et~al.(2017)Karasu, Altan, Sarac, and Hacioglu}]{Seckin2017}
\bibinfo{author}{S.~Karasu}, \bibinfo{author}{A.~Altan},
  \bibinfo{author}{Z.~Sarac}, \bibinfo{author}{R.~Hacioglu},
\newblock \bibinfo{title}{{Prediction of Solar Radiation Based on Machine}},
\newblock \bibinfo{journal}{The journal of cognitive systems}
  \bibinfo{volume}{2} (\bibinfo{year}{2017}) \bibinfo{pages}{16--20}.
  \DOIprefix\doi{10.1109/ICCONS.2018.8663110}.
%Type = Article
\bibitem[{Zhang et~al.(2018)Zhang, Lin, Qiu, Hu, Zhang, Chen, Tan, Lin, and
  Wang}]{Zhang2018}
\bibinfo{author}{L.~Zhang}, \bibinfo{author}{J.~Lin}, \bibinfo{author}{R.~Qiu},
  \bibinfo{author}{X.~Hu}, \bibinfo{author}{H.~Zhang},
  \bibinfo{author}{Q.~Chen}, \bibinfo{author}{H.~Tan},
  \bibinfo{author}{D.~Lin}, \bibinfo{author}{J.~Wang},
\newblock \bibinfo{title}{{Trend analysis and forecast of PM2.5 in Fuzhou,
  China using the ARIMA model}},
\newblock \bibinfo{journal}{Ecological Indicators} \bibinfo{volume}{95}
  (\bibinfo{year}{2018}) \bibinfo{pages}{702--710}.
  \DOIprefix\doi{10.1016/j.ecolind.2018.08.032}.
%Type = Article
\bibitem[{Altan et~al.(2021)Altan, Karasu, and Zio}]{Altan2021a}
\bibinfo{author}{A.~Altan}, \bibinfo{author}{S.~Karasu},
  \bibinfo{author}{E.~Zio},
\newblock \bibinfo{title}{{A new hybrid model for wind speed forecasting
  combining long short-term memory neural network, decomposition methods and
  grey wolf optimizer}},
\newblock \bibinfo{journal}{Applied Soft Computing} \bibinfo{volume}{100}
  (\bibinfo{year}{2021}) \bibinfo{pages}{106996}.
  \DOIprefix\doi{10.1016/j.asoc.2020.106996}.
%Type = Article
\bibitem[{Cleveland et~al.(1990)Cleveland, Cleveland, JE, and
  Terpenning}]{Cleveland1990}
\bibinfo{author}{R.~Cleveland}, \bibinfo{author}{W.~Cleveland},
  \bibinfo{author}{M.~JE}, \bibinfo{author}{I.~Terpenning},
\newblock \bibinfo{title}{{STL: A seasonal-trend decomposition}},
\newblock \bibinfo{journal}{Journal of Official Statistics} \bibinfo{volume}{6}
  (\bibinfo{year}{1990}) \bibinfo{pages}{3--73}.
  \DOIprefix\doi{10.1007/978-1-4613-4499-5{\_}24}.
%Type = Article
\bibitem[{McRae et~al.(1982)McRae, Goodin, and Seinfeld}]{McRae1982a}
\bibinfo{author}{G.~J. McRae}, \bibinfo{author}{W.~R. Goodin},
  \bibinfo{author}{J.~H. Seinfeld},
\newblock \bibinfo{title}{{Development of a second-generation mathematical
  model for Urban air pollution-I. Model formulation}},
\newblock \bibinfo{journal}{Atmospheric Environment (1967)}
  \bibinfo{volume}{16} (\bibinfo{year}{1982}) \bibinfo{pages}{679--696}.
  \DOIprefix\doi{10.1016/0004-6981(82)90386-9}.
%Type = Article
\bibitem[{Wagner and Wendelin(2018)}]{Wagner2018}
\bibinfo{author}{M.~J. Wagner}, \bibinfo{author}{T.~Wendelin},
\newblock \bibinfo{title}{{SolarPILOT: A power tower solar field layout and
  characterization tool}},
\newblock \bibinfo{journal}{Solar Energy} \bibinfo{volume}{171}
  (\bibinfo{year}{2018}) \bibinfo{pages}{185--196}.
  \DOIprefix\doi{10.1016/j.solener.2018.06.063}.
%Type = Article
\bibitem[{Michalsky(1988)}]{Michalsky1988}
\bibinfo{author}{J.~J. Michalsky},
\newblock \bibinfo{title}{{The Astronomical Almanac's algorithm for approximate
  solar position (1950-2050)}},
\newblock \bibinfo{journal}{Solar Energy} \bibinfo{volume}{40}
  (\bibinfo{year}{1988}) \bibinfo{pages}{227--235}.
  \DOIprefix\doi{10.1016/0038-092X(88)90045-X}.
%Type = Article
\bibitem[{Guo et~al.(2013)Guo, Sun, Wang, and Zhang}]{Guo2013}
\bibinfo{author}{M.~Guo}, \bibinfo{author}{F.~Sun}, \bibinfo{author}{Z.~Wang},
  \bibinfo{author}{J.~Zhang},
\newblock \bibinfo{title}{{Properties of a general azimuth-elevation tracking
  angle formula for a heliostat with a mirror-pivot offset and other angular
  errors}},
\newblock \bibinfo{journal}{Solar Energy} \bibinfo{volume}{96}
  (\bibinfo{year}{2013}) \bibinfo{pages}{159--167}.
  \DOIprefix\doi{10.1016/j.solener.2013.06.031}.
%Type = Article
\bibitem[{Grisso et~al.(2004)Grisso, Kocher, and Vaughan}]{Grisso2004}
\bibinfo{author}{R.~D. Grisso}, \bibinfo{author}{M.~F. Kocher},
  \bibinfo{author}{D.~H. Vaughan},
\newblock \bibinfo{title}{{Predicting tractor fuel consumption}},
\newblock \bibinfo{journal}{Applied Engineering in Agriculture}
  \bibinfo{volume}{20} (\bibinfo{year}{2004}) \bibinfo{pages}{553--561}.
  \DOIprefix\doi{10.13031/2013.13732}.
%Type = Techreport
\bibitem[{{Queensland Government}(2021)}]{QueenslandGovernment2021}
\bibinfo{author}{{Queensland Government}}, \bibinfo{title}{{Electricity supply
  options for the North West Minerals Province - CRIS}},
  \bibinfo{type}{Technical Report} \bibinfo{number}{December}, Queensland
  Government, \bibinfo{year}{2021}. \URLprefix
  \url{https://www.epw.qld.gov.au/__data/assets/pdf_file/0023/19715/north-west-electricity-province-cris.pdf}.
%Type = Techreport
\bibitem[{Xuzhong et~al.(2020)Xuzhong, Neng, Xiaoling, Kangli, Long, Yuchao,
  Gangqiang, and Xiaobo}]{Supcon2020}
\bibinfo{author}{Z.~Xuzhong}, \bibinfo{author}{X.~Neng},
  \bibinfo{author}{M.~Xiaoling}, \bibinfo{author}{C.~Kangli},
  \bibinfo{author}{H.~Long}, \bibinfo{author}{H.~Yuchao},
  \bibinfo{author}{X.~Gangqiang}, \bibinfo{author}{L.~Xiaobo},
  \bibinfo{title}{{Cleaning Vehicle}}, \bibinfo{type}{Technical Report}, SUPCON
  SOLAR, \bibinfo{year}{2020}. \URLprefix
  \url{https://www.solarpaces.org/wp-content/uploads/1-1-page-description-of-the-innovative-idea-and-its-impact-and-the-role-of-the-applicants.pdf}.
%Type = Article
\bibitem[{Hardt et~al.(2011)Hardt, Mart{\'{i}}nez, Gonz{\'{a}}lez, Garrido,
  Aladren, Villa, and Saenz}]{Hardt2011}
\bibinfo{author}{M.~Hardt}, \bibinfo{author}{D.~Mart{\'{i}}nez},
  \bibinfo{author}{A.~Gonz{\'{a}}lez}, \bibinfo{author}{C.~Garrido},
  \bibinfo{author}{S.~Aladren}, \bibinfo{author}{J.~R. Villa},
  \bibinfo{author}{J.~Saenz},
\newblock \bibinfo{title}{{HECTOR – heliostat cleaning team-oriented robot}},
\newblock \bibinfo{journal}{SolarPaces Conference}  (\bibinfo{year}{2011})
  \bibinfo{pages}{20--23}.
%Type = Misc
\bibitem[{{Vast Solar}(2016)}]{Solar2016}
\bibinfo{author}{{Vast Solar}}, \bibinfo{title}{{Presentation at ASTRI Annual
  Workshop Public Symposium}}, \bibinfo{year}{2016}. \URLprefix
  \url{https://www.astri.org.au/wp-content/uploads/2016/04/ASTRI-20160502-1450-James_Fisher-Vast_Solar.pdf}.

\end{thebibliography}

%% Nomenclature
\renewenvironment{theglossary}{% Removes glossary titles from 
  \begin{description}%
}{%
  \end{description}%
}

\section*{Nomenclature}
    \begin{singlespace*}
        \subsection*{Latin}
        \printnoidxglossary[type={latin},title={Latin}]
        \subsection*{Greek}
        \printnoidxglossary[type={greek},title={Greek}]
        \subsection*{Subscripts}
        \printnoidxglossary[type={subscript},title={Subscripts}]
        \subsection*{Acronyms}
        \printnoidxglossary[type={acronyms},title={Acronyms}]
    \end{singlespace*}
\end{document}